\documentclass[preprint,aps,prc,showpacs,nofootinbib]{revtex4-1}\usepackage{amsmath}
\usepackage{amssymb}
\usepackage{graphics}
\usepackage{epsfig}

\usepackage{subfigure}

\begin{document}
\title{Polarization effects in elastic deuteron-electron scattering}

\%
\author{\fbox{G. I. Gakh}}
\affiliation{\it National Science Centre, Kharkov Institute of
Physics and Technology, Akademicheskaya 1, and V. N. Karazin
Kharkov National University, Dept. of
Physics and Technology, 31 Kurchatov, 61108 Kharkov, Ukraine}

\author{M.I. Konchatnij}
 \email{konchatnij@kipt.kharkov.ua}
 \affiliation{\it National Science Centre, Kharkov Institute of
Physics and Technology, Akademicheskaya 1, and V. N. Karazin
Kharkov National University, Dept. of
Physics and Technology, 31 Kurchatov, 61108 Kharkov, Ukraine}

\author{N.P. Merenkov}
\email{merenkov@kipt.kharkov.ua}
\affiliation{\it National Science Centre, Kharkov Institute of
Physics and Technology, Akademicheskaya 1, and V. N. Karazin
Kharkov
National University, Dept. of
Physics and Technology, 31 Kurchatov, 61108 Kharkov, Ukraine}

\author{E. Tomasi--Gustafsson}
\email{egle.tomasi@cea.fr}
\affiliation{\it IRFU, CEA, Universit\'e Paris-Saclay, 91191
Gif-sur-Yvette, France}

\author{A. G. Gakh}
\email{agakh@karazin.ua}
\affiliation{\it  V. N. Karazin
Kharkov National University, Dept. of
Physics and Technology, 31 Kurchatov, 61108 Kharkov, Ukraine}

\hspace{0.7cm}

\begin{abstract}
The differential cross section and polarization observables for the
elastic reaction induced by deuteron scattering off electrons at
rest are calculated in the one-photon-exchange (Born) approximation.
Specific attention is given to the kinematical
conditions, that is, to the specific range of incident energy
and transferred momentum. The specific interest of this reaction is 
to access very small transferred momenta.
Numerical estimates are given for polarization
observables that describe the of single- and double-spin effects, provided that
the polarization components (both, vector and tensor) of each particle in the reaction are determined in the rest frame of the electron target.
\end{abstract}

\maketitle

\section{Introduction}
The use of the inverse kinematics in the experimental study of nuclear reactions has peculiar features. Inverse kinematics in hadron-
electron reactions (where the projectile is much heavier than the target) is characterized by an extremely small momentum transfer squared. Therefore, these reactions induced by proton and deuteron beams as well, may be viewed as a possibility to measure the electromagnetic hadron form factors at very small momenta, inachieavable by direct kinematics.

Inverse kinematics was previously used in a number of the experiments to measure the
pion or kaon radius in the elastic scattering of negative pions (kaons) from atomic  electrons in a liquid-hydrogen target. The first experiment was done at Serpukhov \cite{Adylov:1974sg} with pion beam
energy of 50 GeV. Later, a few experiments were done at Fermilab with pion beam energy
of 100 GeV \cite{Dally:1977vt} and 250 GeV \cite{Dally:1982zk}. At this laboratory, the
electromagnetic form factors of negative kaons were measured by direct scattering of 250 GeV
kaons from stationary electrons \cite{Dally:1980dj}. Later on, a measurement of the pion
electromagnetic form factor was done at the CERN SPS \cite{Amet86,Amet84}
by measuring the interaction of 300 GeV pions with the electrons in a liquid hydrogen target.

The use of the inverse kinematics is proposed in a new experiment at CERN \cite{Abbiendi:2016xup} to
measure the running of the fine-structure constant in the space-like region by scattering
high-energy muons (with energy 150 GeV) on atomic electrons, $\mu e \to \mu e$. The aim of the
experiment is the extraction of the hadron vacuum polarization contribution.

In this work  we consider the scattering of a polarized deuteron beam on a
polarized electron target. We assume that the electron target is at
rest and the deuteron beam interacts through the exchange of one
photon with four momentum squared $-k^2=Q^2>0$. We follow the
formalism from Ref. \cite{Gakh:2011sh}, which was developed for the
process of the elastic proton-electron scattering $p+e^-\to p+e^-$.

Polarization observables are essential to disentangle the hadron structure and the reaction mechanism so to be able to test the validity  and the predictions of hadron models .
In particular, low-$Q^2$ data are used to determine the hadron charge radius, $r_c$. 
In case of protons, renewed interest in the charge radius is due to the discrepancy between several experiments, in atomic physics. Replacing an electron with a muon in an hydrogen atom is expected to give a more precise measurement on the proton radius, that enter as a correction in the Lamb shift of atomic levels, as the muonic atom is much more compact and therefore,  corrective terms are more sensitive to the proton. 

An experiment on muonic hydrogen  \cite{Pohl:2010zza} gives a value
of $r_c=0.84184(67)$ fm. This value is one order of magnitude more
precise but smaller by five standard deviation compared to the values of A1 Collaboration (Mainz) $(r_c=0.879(15))$ fm \cite{Bernauer:2010wm} and CODATA  $r_c=0.8768(69)$ fm $r_c=0.8768(69)$ fm  \cite{RevModPhys.84.1527}. The problem of {\it the charge proton radius puzzle} stimulated new high precision PRad \cite{Gasparian:2014rna} and MUSE \cite{Cline:2021ehf} experiments to measure the $r_c$ in $e^{\pm}-p$ and $\mu^{\pm}-p$ scattering. In 2019 very unexpected result ($r_c=0.831$) of PRad experiment was published \cite{Xiong:2019umf}, that is even smaller the muonic hydrogen value. As concern the MUSE experiment, where the cross sections of the $e^{\pm}-p$ and $\mu^{\pm}-p$ scattering will be measured simultaneously, first results should appear soon, helping to solve the discrepancies between the Mainz and PRad results \cite{Cline:2021ehf}. The reasons for these discrepancies have been searched in different effects, from new physics and high
order radiative corrections to underestimated systematic effects, arising from the necessary extrapolation of the cross section data to $Q^2=0$. Note that the most recent CODATA evaluation takes into account the muonic hydrogen experiments, giving an updated vaue of 
$r_c=0.8414(19)$ fm \cite{RevModPhys.93.025010}. 

An evaluation of the deuteron charge radius from
electron-deuteron elastic scattering can be found in Ref.
\cite{Sick:1998cvq}. The extracted value: 
$<r^2>=2.1230 \pm 0.0033(stat)\pm 0.002(syst)$ fm is consistent with previous findings and with a calculation based on the present knowledge of nucleon-nucleon phase
shifts. The corrections coming from the deuteron excited states in the 
intermediate state, the meson exchange currents, the six quark
components are estimated to be less than a percent, in the region of
low $Q^2$ values. 

It is expected that the error associated to elastic electron scattering is larger than 
from atomic spectroscopy.  The value deduced from laser spectroscopy $<r^2>  = 2.12718 (13)_{exp}(89)_{ th} $ can be found in Ref.  \cite{10.21468/SciPostPhysProc.5.021} together with a review of light nuclei radii. It is in essential agreement with the  CODATA recommended value for the deuteron rms squared radius 
is $<r^2> = 2.12799 (74)$ fm  \cite{RevModPhys.93.025010}. 

Any revision of the static (and dynamic) properties of the proton
affects directly the description of light nuclei, in particular
deuteron. At relatively large internal distances (small
$Q^2$-values) the deuteron is considered to be a proton and a
neutron, and small corrections would take into account effects
beyond this simple picture.

One problem related to the extraction of the charge radius from form
factor measurements is that it requires an extrapolation of the form
factor for $Q^2\to 0$ (more precisely, of the form factor derivative) introducing an uncertainty. Systematic
errors, as the choice of the fit function, may seriously
affect the extraction of the radius \cite{Pacetti:2016tqi,Pacetti:2018wwk}. Elastic hadron scattering by
atomic electrons (which can be considered at rest) allows to access
a range of very small transferred momenta, even for relatively large
energies of the colliding particles.

Large interest in inverse kinematics (for the case of the elastic
$p+e$ scattering) was arisen due to possible applications - the
possibility to build beam polarimeters, for high energy polarized
proton beams, in the RHIC energy range \cite{Glavanakov:96}. The calculation
of the spin correlation parameters, for the case of polarized proton
beam and electron target, are sizeable and a polarimeter based on this
reaction can measure the polarization of the proton beam \cite{Glavanakov:96}.
Numerical estimations of other polarization observables were done in Ref.
\cite{Gakh:2011sh}. They showed that polarization effects may be sizable
in the GeV range, and that the polarization transfer coefficients for
$\vec{p}+e\rightarrow \vec{p}+e$ could be used to measure the polarization
of high energy proton beams.

Recently inverse kinematics was used to investigate the nuclear reactions which can
be hardly measured by other methods. In the paper \cite{Reifarth:2013bta} it was proposed to measure the neutron
capture cross sections of unstable isotopes. To do so, the authors suggested a combination of a
radioactive beam facility, an ion storage ring and a high flux reactor which allow a direct measurement
of neutron induced reactions over a wide energy range on isotopes with half lives down to minutes.
A direct measurement, in inverse kinematics, of the $^{17}O(\alpha, \gamma)^{21}Ne$ reaction has been
performed at the DRAGON facility, at the TRIUMF laboratory, Canada \cite{Taggart:2019ump}. At this laboratory,
the $^{22}Ne(p, \gamma)^{23}Na$ reaction has, for the first time, been investigated directly in inverse
kinematics \cite{Williams:2019wxo}. A total of 7 resonances were measured. The authors of the paper \cite{Phuc:2019sur}
analyze the (p, 2p) and (p, pn) reactions data measured, in inverse kinematics, at GSI (Darmstadt, Germany)
for carbon, nitrogen and oxygen isotopes in the incident energy range of 300 -- 450 MeV/n (see \cite{R3B:2019gix} and references therein).

In this work, we calculate the differential cross section and polarization observables for the
elastic deuteron-electron scattering in the one-photon-exchange approximation. 
Numerical estimations are given
for various polarization observables. The possibility to build a polarimeter
based on the elastic deuteron-electron scattering is also considered.

The paper is organized as follows. In Sec. II we give the details
of the order of magnitude and the range which is accessible to the
kinematic variables, as they are very specific for this reaction (Sec. II.A).
The spin structure of the matrix element and the
unpolarized cross section are derived and calculated in Sec. II.B in terms of deuteron form factors, the choice of which is discussed in Sec. II.C. 
Sec. III is devoted to the calculation of the polarization observables for the
reaction $d+e\rightarrow d+e$. Analyzing powers in the $\vec  d\,^T +e \to
d+e $ reaction (when deuteron beam is tensor polarized) are calculated
in Sec. III.A. The tensor polarization coefficients in the $d+e \to \vec
\vec d^T+e $ reaction (when the scattered deuteron is tensor polarized) are given
in Sec. III.B. The polarization transfer coefficients from a polarized target  to the polarized recoil electron are calculated in Sec. III.C.  In sections III.D - III.G we give the expressions for double polarization observables and derive various combinations the coefficients of the polarization correlation and polarization transfer between deuteron and electron, provided they have vector polarization. 
In Sec. III. H  the vector polarization transfer from the initial deuteron to the scattered one is calculated. Sec. IV is devoted to discussion and conclusion.

\section{General formalism}

\begin{figure}
\centering
\includegraphics[width=0.322\textwidth]{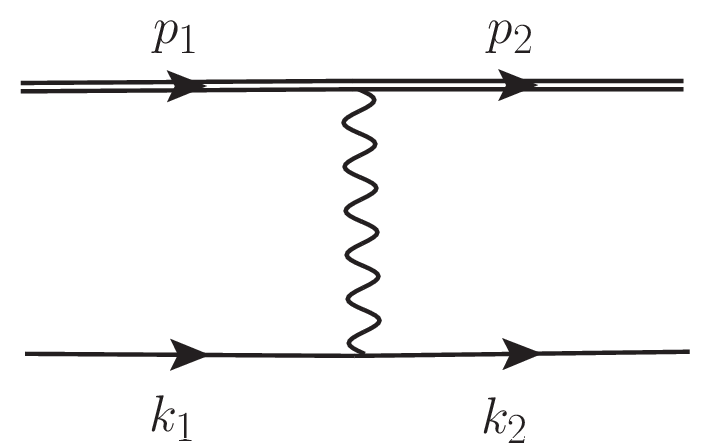}
\hspace{1 cm}
\includegraphics[width=0.322\textwidth]{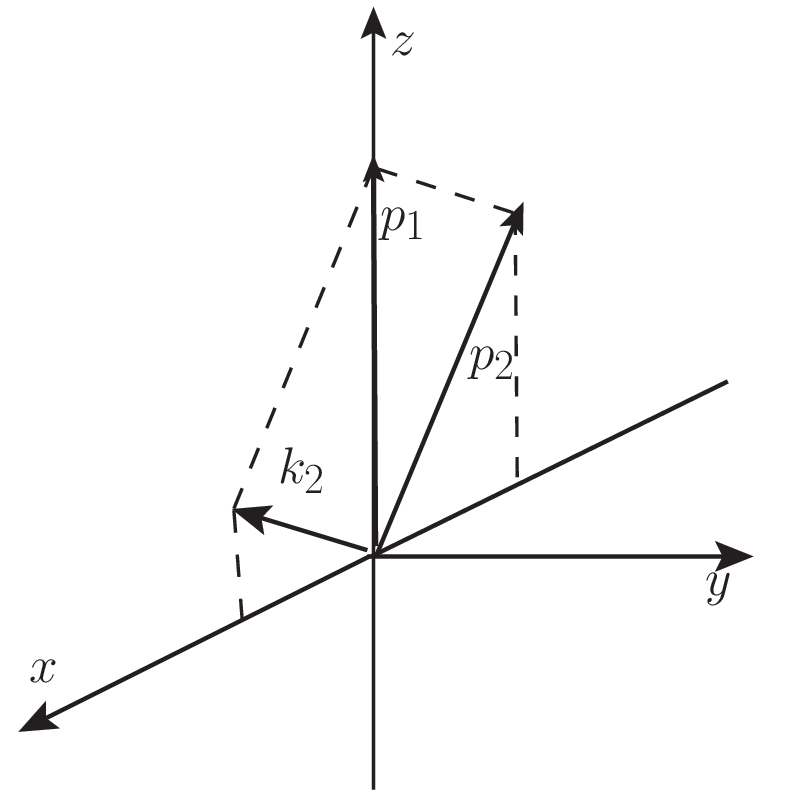}
 \parbox[t]{0.9\textwidth}{\caption{One photon exchange diagram of the process (\ref{eq:1}) and chosen coordinate axes in the rest frame of the initial electron.}\label{fig.1}}
\end{figure}

Let us consider the reaction (Fig. \ref{fig.1})
\begin{equation}\label{eq:1}
d(p_1)+e^-(k_1)\to d(p_2)+e^-(k_2),
\end{equation}
where the particle momenta are indicated in parentheses,  and $k=k_1-k_2=p_2-p_1$ is the four-momentum of the
virtual photon. The reference system is the laboratory (Lab) system, where the electron target is at rest.
A general characteristic of all reactions of elastic and inelastic hadron scattering by atomic electrons (which can be
considered at rest) is the small value of the four momentum squared, even for relatively large energies of colliding
hadrons. Let us first give details of the order of magnitude and the range which is accessible to the kinematic
variables, as they are very specific for this reaction, and then derive the spin structure of the matrix element and
the unpolarized and polarized observables.

\subsection{Kinematics}

In Lab system the four momentum transfer squared is a linear function of the scattered electron energy $\epsilon_2$
\begin{equation}
\label{eq:Q2}
-k^2\equiv Q^2 = -(k_1-k_2)^2=2m(\epsilon_2 -m),
\end{equation}
where $m$ is the electron mass. The conservation of the four-momentum in the reaction (\ref{eq:1}) leads to the following relation between the energy $\epsilon_2$ and the scattering angle $\theta_e$ of the final electron:
\begin{equation}\label{eq:TETA}
\cos\theta_e =\displaystyle\frac{(E+m)(\epsilon_2-m)} {p\,\sqrt{\epsilon_2^2-m^2}},
\end{equation}
where E is the deuteron beam energy, $p=\sqrt{E^2-M^2}$ is its modulus of the 3-momentum ($M$ is the deuteron mass).
From Eq. \ref{eq:TETA} on can see that always $\cos\theta_e\ge 0$ (because $\epsilon_2 > m$), and  the electron can never scatter backward. The following relations hold: 
\begin{equation}\label{eq:eps}
\epsilon_2=m\frac{(E+m)^2+(E^2-M^2)\cos^2\theta_e}{(E+m)^2-(E^2-M^2)\cos^2\theta_e}, \ \ Q^2= \frac{4 m^2 p^2 \cos^2\theta_e}{(E+m)^2-p^2 \cos^2\theta_e},
\end{equation}
which show that the maximum energy of the final electron is achieved  for forward scattering  $(\cos{\theta_e}=1)$:
\begin{equation}\label{eq:epsM}
\epsilon_{2max}=m\frac{2E(E+m)+m^2-M^2}{M^2+2 m E +m^2}.
\end{equation}

From the expression for the $\epsilon_{2max}$, one can obtain the maximum value of the momentum transfer squared
\begin{equation}\label{eq:Q2M}
Q^2_{max} = \frac{4m^2\,p^2}{M^2 + 2 m E + m^2}.
\end{equation}

One can see from Eqs. \,(\ref{eq:epsM}, \ref{eq:Q2M}) that in the inverse kinematics, the available kinematical regions are reduced to small values of
$\epsilon_2$ and $Q^2,$  (compared with $E$ and $E^2$) which are proportional to $m$ and $m^2,$ respectively. For example, at $E=10$ GeV one has 
$\epsilon_2\leq 25$\,MeV and $Q^2\leq 25$\,MeV$^2.$ The upper limits of these quantities increase approximately as $E^2.$

 As in the proton case, for one deuteron angle there may be two values
of the deuteron energy, (and two corresponding values for the
recoil- electron energy and angle, and for the transfer momentum
squared $k^2$). This is a typical situation when the center of mass
velocity is larger than the velocity of the projectile in the center
of mass (c.m.), where all the angles are allowed for the recoil
electron. From momentum conservation, on can find the following
relation between the energy and the angle of the scattered deuteron
$E_2$ and $\theta_d$:
\begin{equation}\label{eq:EDtetaD}
E_2^{\pm}=\frac {(E+m)(M^2+mE)\pm p^2\cos\theta_d\sqrt{m^2 - M^2\sin^2\theta_d}}{(E+m)^2- p^2 \cos^2\theta_d}.
\end{equation}
The two solutions coincide when
the angle between the initial and final hadron takes its maximum
value, which is determined by the ratio of the electron and
scattered hadron masses, $\sin\theta_{d,max}=m/M$. Nevertheless, at fixed values of $\epsilon_2$  or $Q^2$, the energy of the scattered deuteron is unambiguous
\begin{equation}\label{eq:E2}
E_2=E+m-\epsilon_2 = E-\frac{Q^2}{2 m}.
\end{equation}

Hadrons are
scattered from atomic electrons at very small angles, and the larger
is the hadron mass, the smaller is the available angular range for
the scattered hadron.

\subsection{Unpolarized cross section}

In the one-photon-exchange approximation, the matrix element ${\cal
M}$ of  reaction (1) can be written as:
\begin{equation}\label{eq:ME}
{\cal M}=\frac{e^2}{k^2}j^{\mu}J_{\mu}, \end{equation} where
$j_{\mu}(J_{\mu})$ is the leptonic (hadronic) electromagnetic
current. The leptonic current is
\begin{equation}\label{eq:jL}
j_{\mu}=\bar u(k_2)\gamma_{\mu} u(k_1),
\end{equation}
where $u(k_{1,\,2})$ is the bispinor of the incoming (outgoing) electron.
Following the requirements of Lorentz invariance, current
conservation, parity and time-reversal invariance of the hadronic
electromagnetic interaction, the general form of the electromagnetic
current for the deuteron (which is a spin-one particle) is fully
described by three form factors. The hadronic electromagnetic
current can be written as \cite{ARbook}:
\begin{equation}\label{eq:JD}
J_{\mu}=(p_1+p_2)_{\mu}\left [-G_1(Q^2)U_1\cdot
U_2^*+\frac{1}{M^2}G_3(Q^2) \left (U_1\cdot k U_2^*\cdot k + \frac{Q^2}{2}U_1\cdot U_2^*\right )\right ]+
\end{equation}
$$+G_2(Q^2)\left (U_{1\mu}U_2^*\cdot k -U_{2\mu}^*U_1\cdot k\right), $$
where $U_{1\mu}$ and $U_{2\mu}$ are the polarization four
vectors for the initial and final deuteron states. The functions
$G_i(k^2)$, i=1,\,2,\,3, are the deuteron electromagnetic form factors,
depending only on the virtual photon four momentum squared. Due to
the current hermiticity, these form factors are real functions in
the region of space-like momentum transfer.

These form factors are related to the standard deuteron form
factors: $G_C$ (charge monopole) $G_M$ (magnetic dipole) and  $G_Q$
(charge quadrupole) by the following relations:
\begin{equation}\label{eq:FFi}
G_M(Q^2)=-G_2(Q^2), \ \ G_Q(Q^2)= G_1(Q^2)+G_2(Q^2)+2G_3(Q^2),
\end{equation}
$$G_C(Q^2)=\frac{2}{3}\tau \left [G_2(Q^2)-G_3(Q^2)\right ]+\left
(1+ \frac{2}{3}\tau\right ) G_1(Q^2),~\ \tau=\frac{Q^2}{4M^2}. $$
The standard form factors have the following normalization:
\begin{equation}\label{12}
G_C(0)=1, \ \ G_M(0)=\frac{M}{m_N}\mu_d, \ \ G_Q(0)=M^2{\hat Q}_d,
\end{equation}
where $m_N$ is the nucleon mass, $\mu_d=0.857$,  \cite{Mohr:2000ie}, $\hat
Q_d=0.2857$ fm$^2$ \cite{Ericson:1982ei}) is the deuteron magnetic
(quadrupole) moment.

Note also that
$$G_1(Q^2)=\frac{1}{1+\tau}\Big[\tau G_M(Q^2) + G_C(Q^2)+\frac{\tau}{3}G_Q(Q^2)\Big],$$
$$G_3(Q^2)= \frac{1}{2(1+\tau)}\Big[G_M(Q^2)-G_C(Q^2) +\Big(1+\frac{2\tau}{3}\Big)G_Q(Q^2) \Big].$$

The matrix element squared is:
\begin{equation}\label{eq:ME2}
|{\cal M}|^2=16\pi^2\frac{\alpha^2}{Q^4}L^{\mu\nu}H_{\mu\nu},
\end{equation} where $\alpha=e^2/4\pi =1/137$ is the electromagnetic fine
structure constant. The leptonic $L_{\mu\nu}$ and hadronic
$H_{\mu\nu}$ tensors are defined as
\begin{equation}\label{eq:LTHT}
L^{\mu\nu}=j^{\mu}j^{\nu *},~H_{\mu\nu}=J_{\mu}J_{\nu}^*.
\end{equation} The leptonic tensor, $L_{\mu\nu}^{(0)}$, for
unpolarized initial and final electrons (averaging over the initial
electron spin), has the form:
\begin{equation}\label{eq:LT0}
L_{\mu\nu}^{(0)}=-Q^2g_{\mu\nu}+2(k_{1\mu}k_{2\nu}+k_{1\nu}k_{2\mu}).
\end{equation} The contribution to the lepton tensor corresponding to the+
polarized electron target is
\begin{equation}\label{eq:LTs1}
L_{\mu\nu}^{(p)}(s_1)=2im\epsilon_{\mu\nu\rho\sigma}k^{\rho}s_1^{\sigma}, \epsilon_{0123} = -1,
\end{equation} where $s_{1\sigma}$ is the initial electron polarization
four-vector which satisfies following conditions: $k_1\cdot s_1=0, \, s_1^2 = -1$.

The hadronic tensor $H_{\mu\nu}$ is calculated 
in terms of the deuteron electromagnetic form factors, using the
explicit form of the electromagnetic current (\ref{eq:JD}).  
The spin density matrices of the
initial and final deuterons have the 
the following expressions :
\begin{equation}\label{eq:MPD}
\rho_{\alpha\beta}^{(i)} =-\frac{1}{3} \left(
g_{\alpha\beta}-\frac{1}{M^2} p_{1\alpha} p_{1\beta}\right
)+\frac{i}{2M}<\alpha\beta\eta_1p_1>+{\cal Q}_{\alpha\beta}^{(i)},
\end{equation}
$$\rho_{\alpha\beta}^{(f)} =-\left( g_{\alpha\beta}-\frac{1}{M^2}
p_{2\alpha} p_{2\beta}\right
)+\frac{i}{2M}<\alpha\beta\eta_2p_2>+{\cal Q}_{\alpha\beta}^{(f)}, $$
with the notation
$<\alpha\beta ab >= \epsilon_{\alpha\beta \rho \sigma} a^\rho b^\sigma$. Here $\eta_{1\alpha}\,(\eta_{2\alpha})$ and
${\cal Q}_{\alpha\beta}^{(i)}\,({\cal Q}_{\alpha\beta}^{(f)})$ are the four vectors
and tensors describing the vector and tensor (quadrupole)
polarization of the initial (final) deuteron, respectively. The
four-vector of the vector polarization of the initial (final)
deuteron satisfies the following conditions: $\eta_1^2 = -$1,
$\eta_1\cdot p_1=0$ ($\eta_2^2=-$1, $\eta_2\cdot p_2=0$). The tensor
${\cal Q}_{ \alpha\beta}^{(i)}$ satisfies the conditions ${\cal Q}_{\alpha\beta}^{(i)}\,g^{\alpha\beta}=0,$ $Q_{\alpha\beta}^{(i)}=
{\cal Q}_{\beta\alpha}^{(i)},$ ${\cal Q}_{ \alpha\beta}^{(i)}\,p_{1}^{\alpha}=0$. The
tensor ${\cal Q}_{ \alpha\beta}^{(f)}$ satisfies the same conditions, after
substitution: $i\to f$ and $p_{1}^{\alpha}\to p_{2}^{\alpha}$.

The hadronic tensor $H_{\mu\nu}(0)$, corresponding to the case of
unpolarized initial and final deuterons can be written in the
standard form in terms of two spin-independent structure functions:
\begin{equation}\label{eq:HT0}
H_{\mu\nu}(0)= H_1(Q^2) \tilde g_{\mu\nu}+ \frac{1}{M^2} H_2(Q^2)
P_{\mu}P_{\nu},
\end{equation}
where $\tilde g_{\mu\nu} = g_{\mu\nu}- ({k_{\mu}k_{\nu}})/{k^2}$, $P_{\mu}=(p_1+p_2)_{\mu}$.
Averaging over the spin of the initial deuteron, the structure functions
$H_i(Q^2)$, $i=1,2,$ can be expressed in terms of the
electromagnetic form factors as:
\begin{equation}\label{eq:H1H2}
H_1(Q^2)=-\frac{2}{3}Q^2(1+\tau)G_M^2(Q^2),
\end{equation}
$$H_2(Q^2)=M^2\left [G_C^2(Q^2)+\frac{2}{3}\tau G_M^2(Q^2)+\frac{8}{9} \tau^2 G_Q(Q^2)\right ]. $$

The differential cross section is related to the matrix element
squared (\ref{eq:ME2}) by
\begin{equation}\label{eq:ds}
d\sigma=\displaystyle\frac{(2\pi)^4\overline{\left|{\cal M}\right
|^2}}{4\sqrt{(k_1\cdot p_1)^2-m^2M^2}} \displaystyle\frac{d^3\vec
k_2}{(2\pi)^3 2\epsilon_2} \displaystyle\frac{d^3\vec p_2}{(2\pi)^3
2E_2} \delta^4(k_1+p_1-k_2-p_2),
\end{equation}
$$\overline{\left|{\cal M}\right|^2} = 16\pi^2\frac{\alpha^2}{Q^4}\,L^{\mu\nu}(0)\,H_{\mu\nu}(0),$$
where $\vec{p}_2\,(E_2)$ is
the 3-momentum (energy) of the scattered deuteron. 

From this point on, the formalism differs from elastic electron-deuteron scattering,
because we introduce a reference system where the electron is at
rest. In this system, the differential cross section is written
as
\begin{equation}\label{eq:dse0}
\frac{d\sigma}{d\epsilon_2}=\frac{1}{32\pi}\frac{\overline{\left|{\cal
M}\right |^2}}{m\, p^2}.
\end{equation}
 The average over the spins
of the initial particles has been included in the leptonic and
hadronic tensors. Using Eq. (\ref{eq:Q2}) one can write
\begin{equation}\label{eq:dsQ20}
\frac{d\,\sigma}{d\,Q^2}=\frac{1}{64\pi}\frac{\overline{\left|{\cal
M}\right |^2}}{m^2\, p^2}. \end{equation}

The differential cross
section over the electron solid angle can be written as:
\begin{equation}\label{eq:dsTETA0}
\frac{d\sigma}{d\Omega_e}=\frac{1}{32\pi^2}\,\frac{1}{m\,p}\,\frac{|\vec
k_2|^3}{Q^2}\frac{\overline{\left|{\cal M}\right |^2}}{E+m},
\end{equation}
where $d\,\Omega_e= 2\pi\, d\cos\theta$ (due to the
azimuthal symmetry) and we used the relation
\begin{equation}\label{eq:dedO}
d\epsilon_2=\displaystyle\frac{p}{E+m}\displaystyle\frac{|\vec
k_2|^3}{m(\epsilon_2-m)}\displaystyle\frac{d\Omega_e}{2\pi}.
\end{equation}

The differential cross section for
unpolarized deuteron-electron scattering (\ref{eq:dsQ20}), in the coordinate system
where the electron is at rest, can be written as:
\begin{equation}\label{eq:dsQ201}
\frac{d\sigma}{dQ^2}=\frac{\pi\,\alpha^2}{2\,m^2\,p^2}\frac{\cal D}{Q^4}, \ \ {\cal D } = \frac{1}{2}\,L^{\mu\nu}(0)\,H_{\mu\nu}(0),
 \end{equation}
 with
\begin{equation}\label{eq:D(H)}
{\cal D}=(-Q^2+2m^2)H_1(Q^2)+2\left [-Q^2M^2+ 2mE\left(2mE-Q^2\right
)\right ]\frac{H_2(Q^2)}{M^2}.
\end{equation}

The factor ${\cal D}$ has the following form in terms of the deuteron form factors
\begin{equation}\label{eq:D(G)}
{\cal D}=\frac{4}{3}\tau [4m^2(E^2-M^2)-Q^2(m^2-M^2+2mE-2M^2\tau ]G_M^2(Q^2)+
\end{equation}
$$+2[-M^2\,Q^2+2mE(2mE-Q^2)][G_C^2(Q^2)+\frac{8}{9} \tau^2 G_Q^2(Q^2)]. $$
To perform the numerical estimations one needs to know the behaviour all three form factors ($G_M,\,G_C,\,G_Q$) in the region
of small momentum transfer squared. We restrict ourselves to the maximum deuteron beam energy $E=200$\,GeV, {\it i.e.,}$Q^2_{max}$ does not exceed 0.012\, GeV$^2$.

\subsection{About deuteron electromagnetic form factors }

The deuteron form factors are derived from the  differential
cross section of the electron-deuteron scattering. While the magnetic form factor is uniquely
determined by the cross section of unpolarized particles at backward angles, the separation
of charge and quadrupole form factors requires polarization measurements,
either the tensor analyzing powers $T_{20},\,T_{22},\,T_{21}$ or the recoil deuteron tensor polarization $t_{20},\,t_{22},\,t_{21}$  (the electron beam is unpolarized in both cases) \cite{Gourdin:1963ub,SCHILDKNECHT1964254}. This has prompted the development of both,
polarized deuterium targets and polarimeters for measuring the polarization of recoil hadrons \cite{Ferro-Luzzi:1997sqo}.

At storage rings, polarized internal deuteron gas targets from an atomic beam source can be used  \cite{Dmitriev:1985us,Gilman:1990vg,Ferro-Luzzi:1996znh,PhysRevLett.82.3755,Nikolenko:2003zq,Nikolenko:2003fjj}.
The high intensity of the circulating electron beam allows achieving acceptable luminosities despite the very low thickness of the gas targets.

At facilities with external beams, polarimeters can be used to measure the polarization of recoil deuterons \cite{Schulze:1984ms,Garcon:1993vm,JLABt20:2000uor}. High beam
intensities are a prerequisite because the polarization measurement in this case requires a second scattering, what leads to a loss of a few orders magnitude in count rate.

The variation of the scattered electron angle at given momentum transfer squared in unpolarized scattering  allows to separate $G_M^2$ and a combination
(structure function) of the three
form factor squared $A(Q^2) = (2/3)\tau\,G_M^2+ G_C^2 + (8/9)\tau^2\,G_Q^2$.
The measurement of $T_{20}$ and $T_{21}$ (or $t_{20}$ and $t_{21}$) allows to separate also some combination of $G_M\,G_C,\, G_Q^2$ and $G_M^2$ and the product $G_M\,G_Q$, respectively \cite{Haftel:1980zz}.
The three electromagnetic deuteron form factors
have been experimentally determined up to $Q^2\simeq $ 1.7 GeV$^2$
\cite{JLABt20:2000uor}. The structure function $A(Q^2)$ has been measured up
to $Q^2$=6 GeV$^2$  \cite{JeffersonLabHallA:1998xrv} and $G_M^2(Q^2)$ up to 2.8 GeV$^2$
\cite{Bosted:1989hy}. The existing world data on the differential cross
section \cite{JeffersonLabHallA:1998xrv} and $t_{20}$ \cite{JLABt20:2000uor} for electron deuteron
elastic scattering allow to extract $G_C$ and $G_Q$. This has been done in
Ref. \cite{JLABt20:2000qyq} where the world data were collected and three
different analytical parameterizations were suggested, with a number
of parameters varying from 12 \cite{Kobushkin:1994ed} to 33.

The description of these form factors is a challenge for the deuteron models.
The best representation ( very good $\chi^2$ with very small number of parameters)
is based on a generalization of the nucleon two-component picture from Ref. \cite{Iachello:1972nu,Bijker:2004yu} to the
deuteron case \cite{Tomasi-Gustafsson:2005dyd}. The basic idea of tthe model 
is the presence of two components in the deuteron (proton) 
structure: an intrinsic structure, very compact, characterized by a
dipole or monopole $Q^2$ dependence and a meson cloud, which contains
the light vector meson $\rho$, $\phi$ and $\omega$ contributions (not the $rho$ for the deuteron case, due to its isoscalar nature). A very good description of the world data on deuteron electromagnetic form factors has been obtained, with as few
as six free parameters and few evident physical constraints. The form factors are
parameterized as (considering only the contribution of the isoscalar
vector mesons, $\omega$ and $\phi$):
\begin{equation}\label{eq:Gi}
G_i(Q^2)=N_i g_i(Q^2) F_i(Q^2),~i=C,\,Q,\,M
\end{equation}
with:
$$F_i(Q^2)= 1-\alpha_i-\beta_i+
\alpha_i\displaystyle\frac{m_{\omega}^2}{m_{\omega}^2+Q^2}
+\beta_i\displaystyle\frac{m_{\phi}^2}{m_{\phi}^2+Q^2}. $$ 
where $m_{\omega}$ ($m_{\phi}$) is the mass of the $\omega$
($\phi$)-meson.

The terms $g_i(Q^2)$ are written as functions of two  parameters,
also real, $\gamma_i$ and $\delta_i$, generally different for each
form factor:
\begin{equation}\label{eq:gi}
 g_i(Q^2)=1/\left [1+\gamma_i{Q^2}\right ]^{\delta_i},
\end{equation}
and $N_i$ is the normalization of the $i$-th form factor at $Q^2=0$,
$N_C=G_C(0)=1$, $N_Q=G_Q(0)=M^2{\cal Q}_d=25.83,$
$N_M=G_M(0)=\displaystyle\frac{M}{m}\mu_d=1.714,$ where ${\cal
Q}_d$, and $\mu_d$ are the quadrupole and the magnetic moments of
the deuteron.

\begin{figure}
\centering
\includegraphics[width=0.3\textwidth]{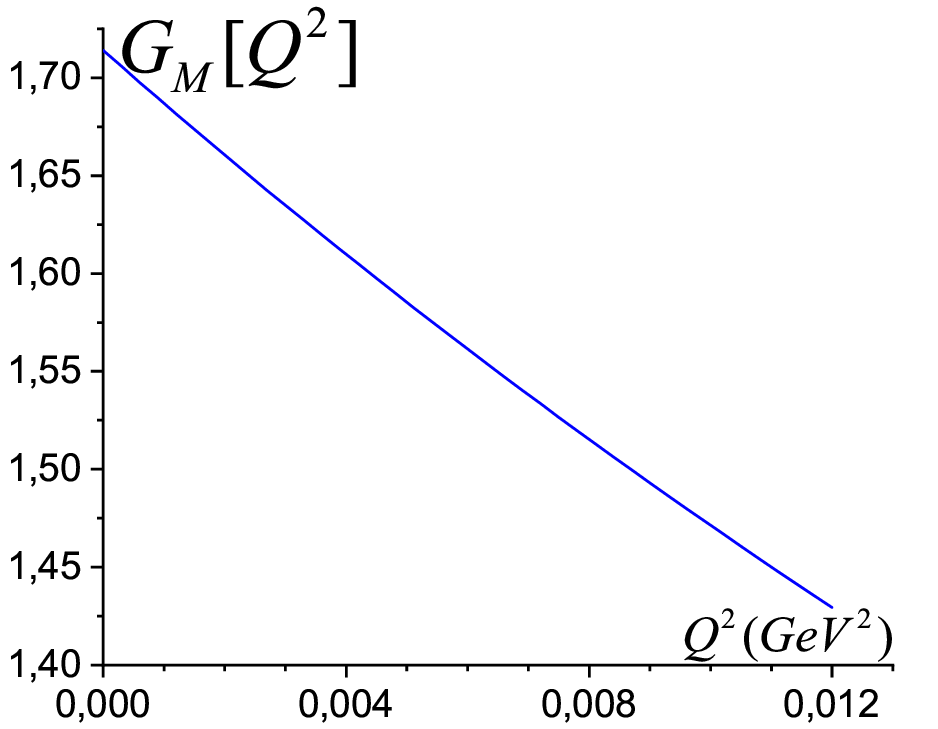}
\includegraphics[width=0.3\textwidth]{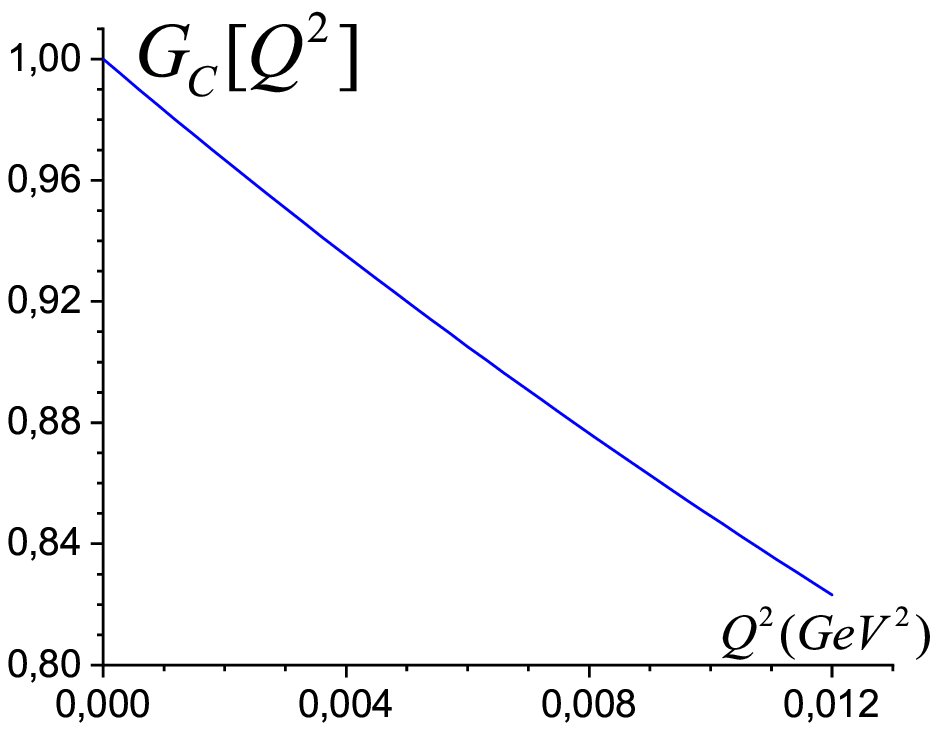}
\includegraphics[width=0.3\textwidth]{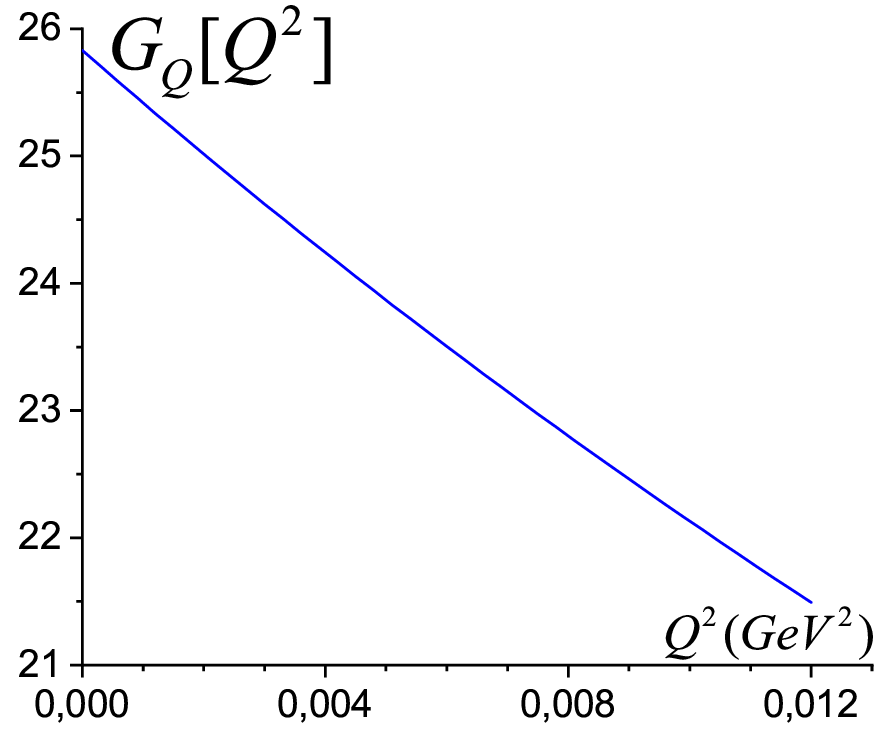}
 \parbox[t]{0.9\textwidth}{\caption{Deuteron electromagnetic form factors defined by Eq.\,(\ref{eq:Gi}) with parameters as in the text.}\label{fig.Gi}}
\end{figure}

The expression  (\ref{eq:Gi}) contains four parameters, $\alpha_i$,
$\beta_i$, $\gamma_i$, $\delta_i$, generally different for different
form factors. We took here the most simple version where  $\delta=1.04$, and
$\gamma=12.1$ are common parameters for the three form factors,  $\alpha
(G_C,\,G_Q,\,G_M)= 5.75$, 4.21 3.77 and $\beta (G_C,\,G_Q,\,G_M)=$ $-$5.11,
$-$3.41, $-$2.86. With the chosen parametrization, the extrapolation to small values of $Q^2$ gives the electromagnetic deuteron form factors shown in Fig.\,\ref{fig.Gi}. One can see that in the region of small $Q^2$ all three form factors are positive and decrease almost linearly with increasing $Q^2$.

\begin{figure}
\centering
\includegraphics[width=0.38\textwidth]{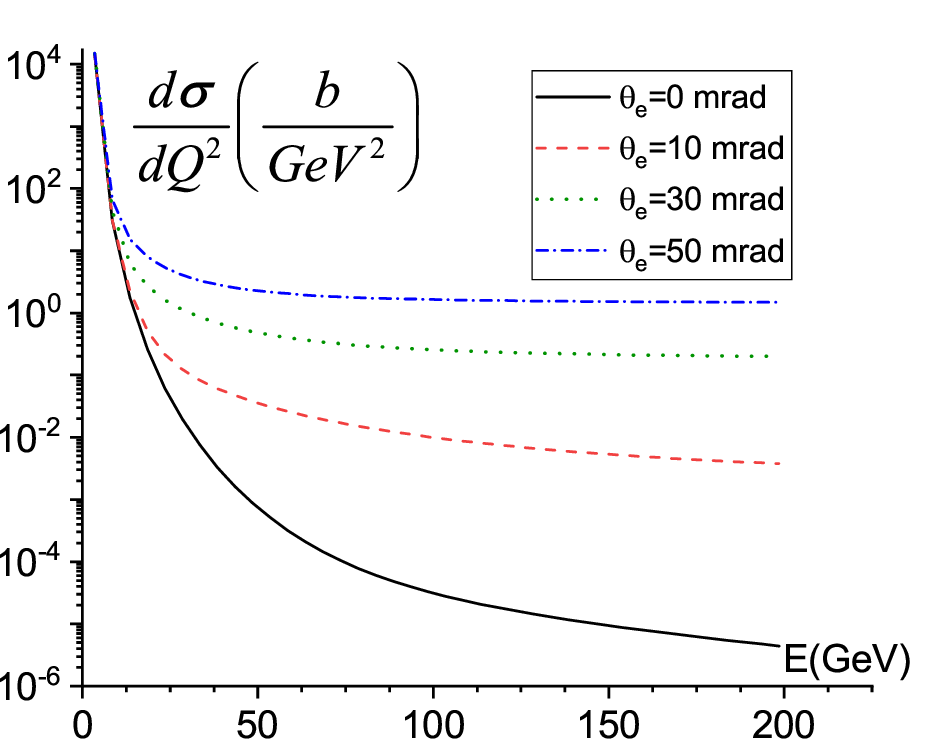}
\includegraphics[width=0.34\textwidth]{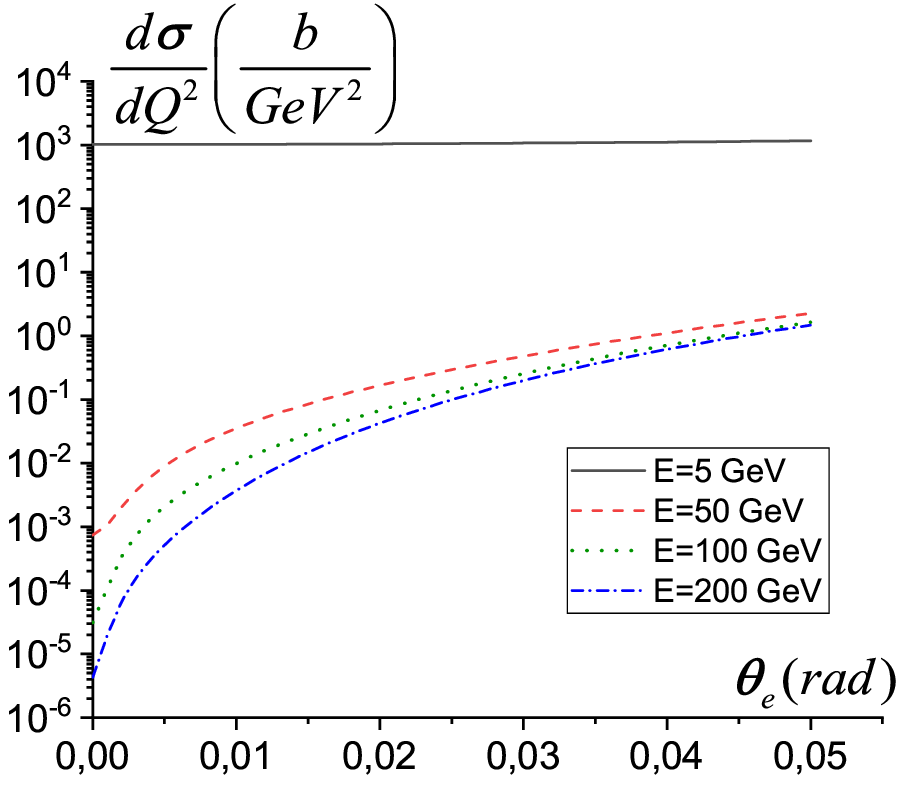}
 \parbox[t]{0.9\textwidth}{\caption{Differential cross section for the unpolarized $d-e$ elastic scattering as defined by Eqs.(\ref{eq:dsQ201}) and (\ref{eq:D(G)}): as a function of incident deuteron beam  energy $E$ for different electron scattering angles (left panel) and as a function of the electron scattering angle at different values of incident  energy $E$ (right panel).}\label{fig.DsBorn}}
\end{figure}

The energy dependence of the differential cross section for
different angles and the angular dependence for different energies are illustrated in Fig.\,\ref{fig.DsBorn}. We restrict ourselves by the values: $E \leq $ 200\,GeV and
$\theta_e \leq$\,50\,mrad.

The unpolarized differential cross
section is divergent at small values of energy, as expected from the
one-photon exchange mechanism. It is monotonically decreasing, not
presenting minima, when the deuteron energy increases and increases when the electron scattering angle increases.

\topmargin=-2.0cm \textheight=24.5cm \textwidth=17.0cm
\oddsidemargin=0.0cm \evensidemargin=0.0cm

\section{Polarization observables}

Several polarization observables can be measured and calculated
for elastic deuteron-electron scattering. Besides the electron
polarization, the initial and final deuterons may have vector and
tensor polarizations. Let us focus here on single and double polarization observables. Among single-spin observables, we consider effects which arise within the one-photon exchange approximation when the amplitude of the process (\ref{eq:1}) is real. 
In this respect, we note that in presenc of the two photon exchange contribution, the scattering amplitude contains an imaginary part. Then, other single-spin effects arise, due to the target electron polarization or due to vector polarization of the deuteron beam. They lead to azimuthal asymmetry of the cross section similar to the one in elastic electron-proton scattering [Ref]. Here we consider the single-spin effects due to the tensor polarization of the initial or final deuteron.

\begin{enumerate}

\item The analyzing powers (asymmetries) due to the tensor polarization
of the deuteron beam, $\vec{d}\,^T+e\to d+e$.

\item The tensor polarization of the scattered deuteron when the other
particles are unpolarized, $ d+ e\to \vec{d}\,^T + e$.

\item The polarization transfer coefficients which describe the polarization
transfer from the polarized electron target to the recoil electron in the
$d+\vec e \to d+\vec e $ reaction.

\item The spin correlation coefficients when the deuteron beam is vectorially
polarized and the initial electron has arbitrary polarization,
$\vec{d}\,^V+\vec e\to d+e$.

\item The polarization transfer coefficients which describe the vector polarization
transfer from a polarized electron target to the scattered deuteron, $ d+\vec e\to \vec{d}\,^V+e$.

\item The spin correlation coefficients when the scattered deuteron has vector
polarization and the final electrons have arbitrary polarization,
$d+e\to \vec{d}\,^V+\vec e$.

\item The polarization transfer coefficients which describe the polarization
transfer from the vector-polarized deuteron beam to the recoil electron in the
 $ \vec{d}\,^V+e\to d+\vec e$ reaction.

 \item The depolarization coefficients which define the dependence of the
scattered deuteron vector polarization on the vector polarization of
the deuteron beam, $\vec{d}\,^V+e\to \vec{d}\,^V+e$.

\end{enumerate}

Let us choose the following orthogonal system: the $z$ axis
is directed along the direction of the deuteron beam momentum $\vec
p$, the momentum of the recoil electron $\vec k_2$ lies in the $xz$
plane ($\theta$ is the angle between the deuteron beam and the
recoil electron momenta), and the $y$ axis is directed along the
vector $\vec p\times \vec k_2$ (see Fig.\,1.). So, the components of the deuteron
beam and recoil electron momenta are the following
$$p_x=p_y=0, \ p_z=p, \ k_{2x}=k_2\sin\theta_e, \ k_{2y}=0, \ k_{2z}=k_2\cos\theta_e, $$
where $p\, (k_2)$ is the magnitude of the deuteron beam (recoil electron) momentum.

To calculate the polarization observables it is necessary to define the polarization 3-vectors of all particles  as well the
components of both deuteron tensor polarizations in their rest frames. In our work we choose the unique coordinate system shown in Fig.\,\ref{fig.1}
(laboratory system) for all these quantities.

The corresponding polarization observables are analytically calculated as functions of $Q^2$ at fixed deuteron beam energy
and their dependence on the kinematical variables is plotted  similarly to the unpolarized cross section in Fig.\,\ref{fig.DsBorn}.

\subsection{The analyzing powers or asymmetries, $A_{ij}$,  due to the tensor polarization
of the deuteron beam, $\vec{d}\,^T+e\to d+e$.}
We consider here the scattering of the tensor polarized deuteron beam on an 
unpolarized electron target. The hadronic tensor corresponding to
a tensor polarized deuteron beam can be written in the
following general form
\begin{equation}\label{eq:HiT}
H_{\mu\nu}({\cal Q}^{(i)})=H_3(Q^2)\bar Q^{(i)}\tilde g_{\mu\nu} +H_4(Q^2)
\frac{\bar Q^{(i)}}{4M^2} P_{\mu}P_{\nu} +H_5(Q^2) (P_{\mu}\tilde
Q^{(i)}_{\nu}+P_{\nu}\tilde Q^{(i)}_{\mu}) +H_6(Q^2) \tilde
{\cal Q}^{(i)}_{\mu\nu},
\end{equation}
where
\begin{eqnarray}
\tilde Q^{(i)}_{\mu}&=&
{\cal Q}^{(i)}_{\mu\nu}k^{\nu}+\frac{k_{\mu}}{Q^2}\bar Q^{(i)},~ \, \tilde
Q^{(i)}_{\mu}k^{\mu}=0, 
\nonumber\\
 \tilde Q^{(i)}_{\mu\nu}&=&{\cal Q}^{(i)}_{\mu\nu}+ \frac{k_{\mu}k_{\nu}}{Q^4}\bar
Q^{(i)}+ \frac{k_{\nu}k^{\alpha}}{Q^2}{\cal Q}^{(i)}_{\mu\alpha}+
\frac{k_{\mu}k^{\alpha}}{Q^2} {\cal Q}^{(i)}_{\nu\alpha},
\label{eq:Qi}
\end{eqnarray}
$$\tilde Q^{(i)}_{\mu\nu}k^{\nu}=0,~\, \bar Q^{(i)}=
{\cal Q}^{(i)}_{\alpha\beta}k^{\alpha}k^{\beta}. $$ 
The structure functions
$H_i(k^2)$ are related to the deuteron electromagnetic form factors by:
\begin{eqnarray}
H_3(Q^2)&=&-G_M^2,\,\ H_4(Q^2)=G_M^2+\frac{4}{1+\tau}G\,G_Q,\ H_5(Q^2)=-\tau(G_M+2G_Q)G_M, \
\nonumber\\
H_6(Q^2)&=&Q^2(1+\tau)G_M^2,\, \  G=\tau\,G_M+ G_C+\frac{\tau}{3}\,G_Q.
\label{eq:HiT(Q2)}
\end{eqnarray}

The contraction of the spin independent leptonic $L^{\mu\nu}(0)$
and spin dependent (due to the tensor polarization of the deuteron
beam) hadronic $ H_{\mu\nu}({\cal Q}^{(i)})$ tensors, in an arbitrary
reference frame, gives:
\begin{equation}\label{eq:L0HT}
C({\cal Q}^{(i)})=L^{\mu\nu}(0)\,H_{\mu\nu}({\cal Q}^{(i)})=a\,k_1^\mu\, k_1^\nu\, {\cal Q}^{(i)}_{\mu\nu}+b\,k_1^\mu\,k^{\nu}\,Q^{(i)}_{\mu\nu}+
c\,k^{\mu}\,k^{\nu}\,{\cal Q}^{(i)}_{\mu\nu},
\end{equation}
where the functions $a, b$ and $c$ have the following form in terms
of the deuteron electromagnetic form factors (in the rest frame of the electron target):

\begin{eqnarray}
a&=&4(1+\tau )Q^2 \,G_M^2, \ b=-16\tau G_M \left [(M^2+m\,E)G_M+2(m\,E-\tau
M^2)\,G_Q\right ],
\nonumber\\
c&=&[Q^2-4m^2+4\frac{m^2}{M^2}E^2]\,G_M^2
+16\tau m\,E\,G_M\,G_Q+ \nonumber\\
&&+\frac{4}{M^2}\frac{G_Q\,G}{1+\tau }\left [-Q^2(M^2+2m\,E)+
4m^2\,E^2\right ].
\label{eq:abc}
\end{eqnarray}

From the condition $p_{1}^{\mu}\,{\cal Q}^{(i)}_{\mu\nu}=0$ one can express the
time components of the quadrupole polarization tensor in terms of
the space components of this tensor. These relations are:
\begin{equation}\label{eq:Qi00}
{\cal Q}^{(i)}_{00}=\frac{p^2}{E^2}{\cal Q}^{(i)}_{zz}, \ \
{\cal Q}^{(i)}_{0x}=\frac{p}{E}{\cal Q}^{(i)}_{xz}, \ \
{\cal Q}^{(i)}_{0y}=\frac{p}{E}{\cal Q}^{(i)}_{yz}, \ \
{\cal Q}^{(i)}_{0z}=\frac{p}{E}{\cal Q}^{(i)}_{zz}.
\end{equation}
The components of the quadrupole polarization tensor ${\cal Q}^{(i)}_{ij}$ defined in the Lab system can be related to the corresponding ones defined in the rest system of the deuteron beam (denoted as $R_{ij}$) by the following relations
$${\cal Q}^{(i)}_{xx}=R_{xx}, \ \  {\cal Q}^{(i)}_{yy}=R_{yy}, \ \
 {\cal Q}^{(i)}_{xz}=\frac{E}{M}\,R_{xz}, \ \  {\cal Q}^{(i)}_{zz}=\frac{E^2}{M^2}\,R_{zz}. $$
The $Q^2$ dependence of the differential cross section of the reaction (\ref{eq:1})
on the polarization characteristics of the deuteron beam, in the case
when beam is tensor polarized, has the following form
\begin{equation}\label{eq:dsA}
\frac{d\sigma}{dQ^2}({\cal Q}^{(i)})= \left (\frac{d\sigma}{dQ^2}\right
)_{un}\left [1+A_{xx}({\cal Q}_{xx}-{\cal Q}_{yy})+ A_{xz}
{\cal Q}_{xz}+ A_{zz} {\cal Q}_{zz}\right ],
\end{equation} 
where $A_{ij}$,
$i,j=x,y,z$ are the analyzing powers (asymmetries) which
characterize the $\vec{d}\,^T-e$ scattering when the deuteron beam is tensor
polarized.

\begin{figure}
\centering
\includegraphics[width=0.3\textwidth]{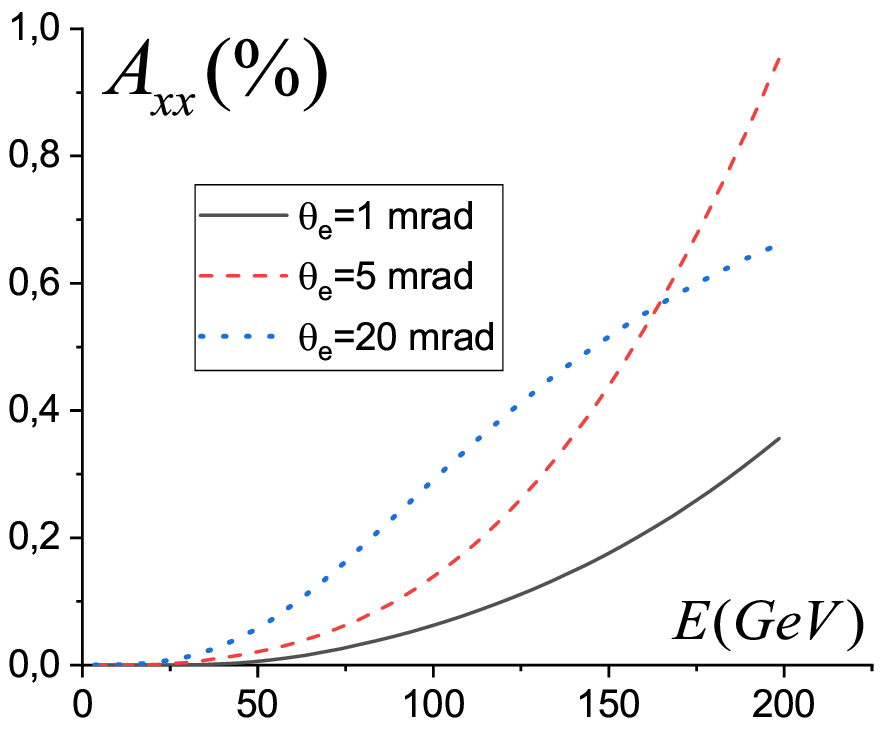}
\includegraphics[width=0.3\textwidth]{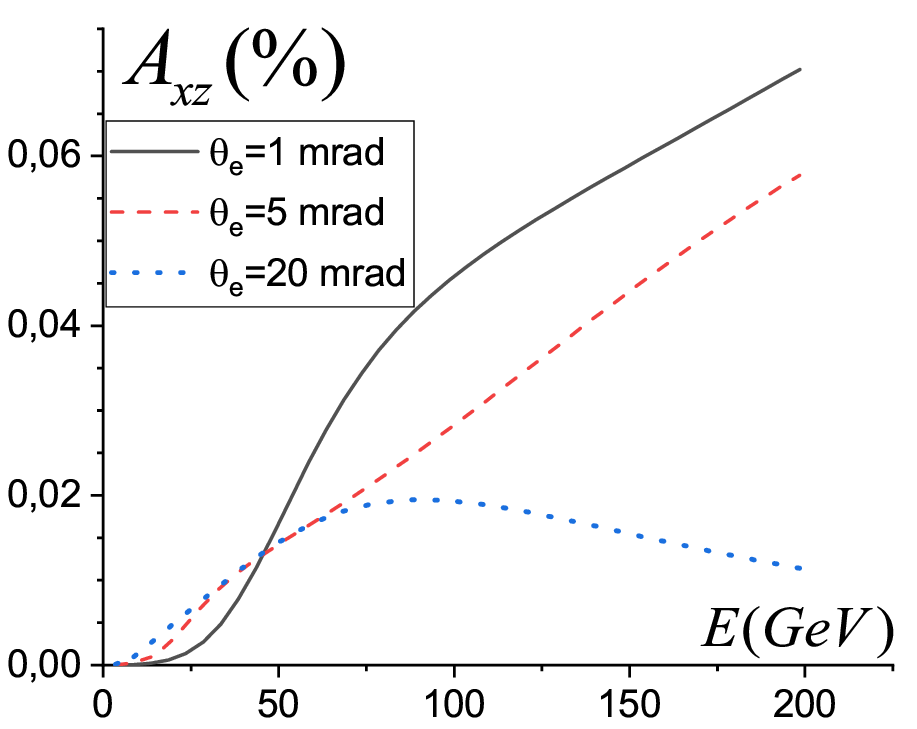}
\includegraphics[width=0.3\textwidth]{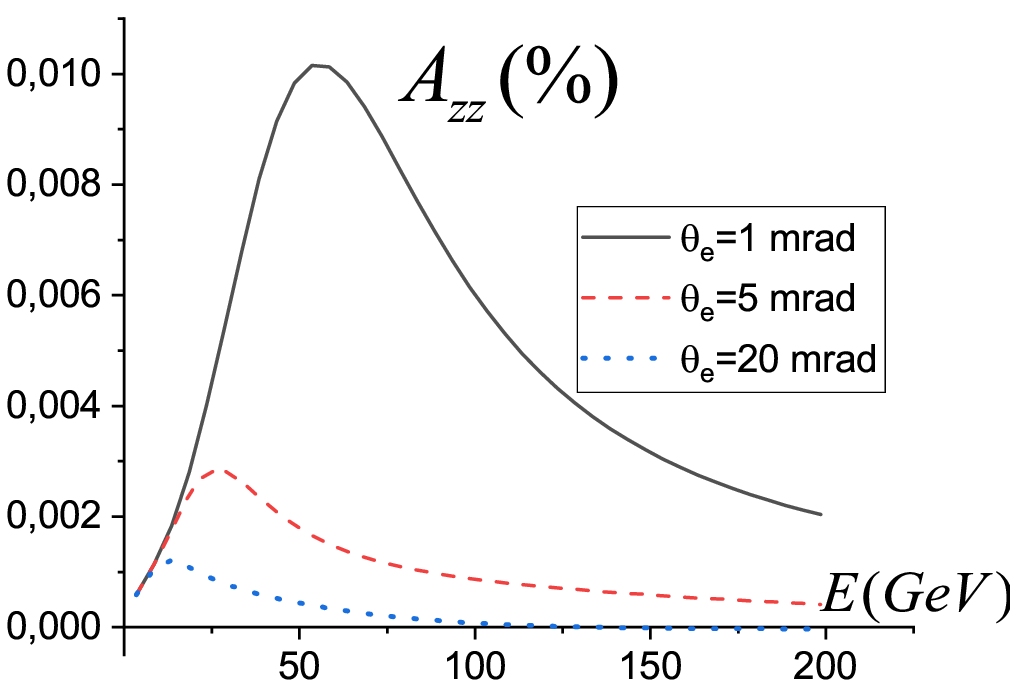}
\vspace{0.5cm}\centering
\includegraphics[width=0.3\textwidth]{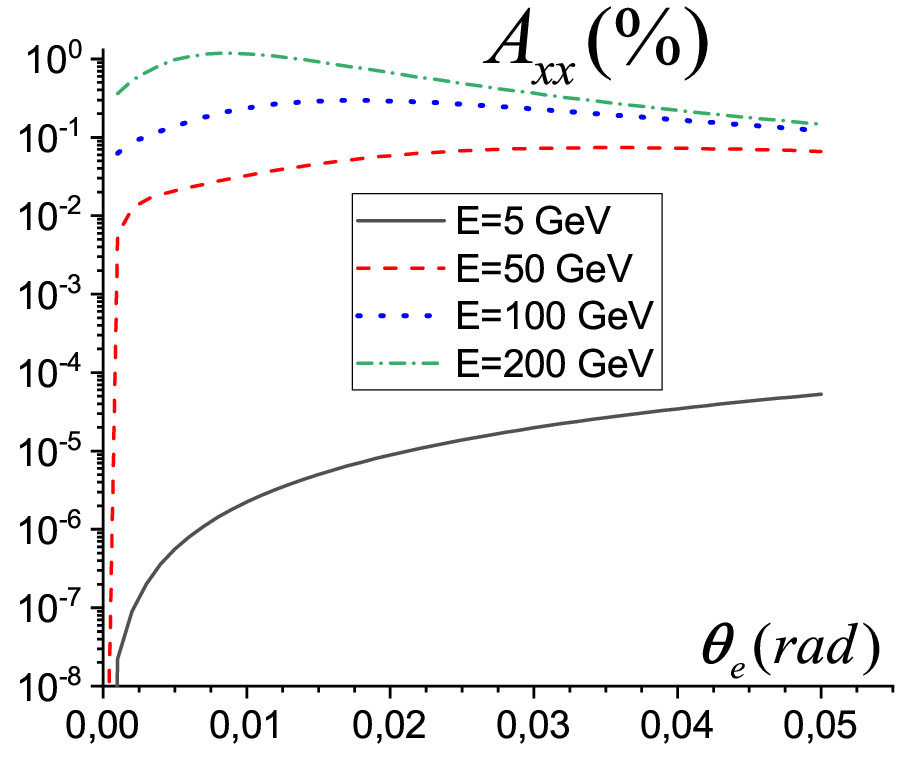}
\includegraphics[width=0.3\textwidth]{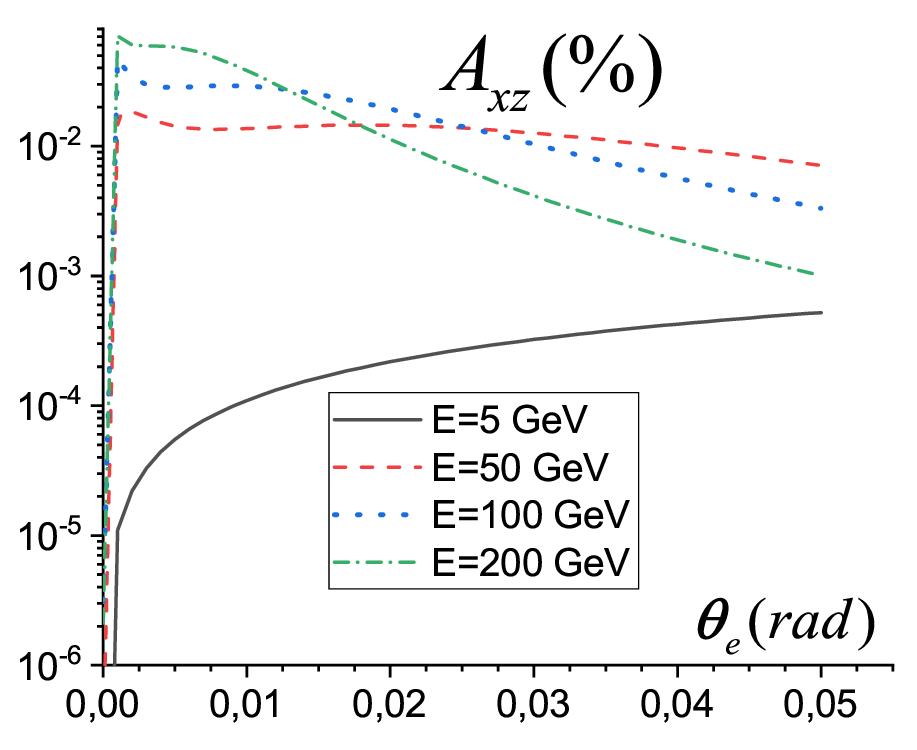}
\includegraphics[width=0.3\textwidth]{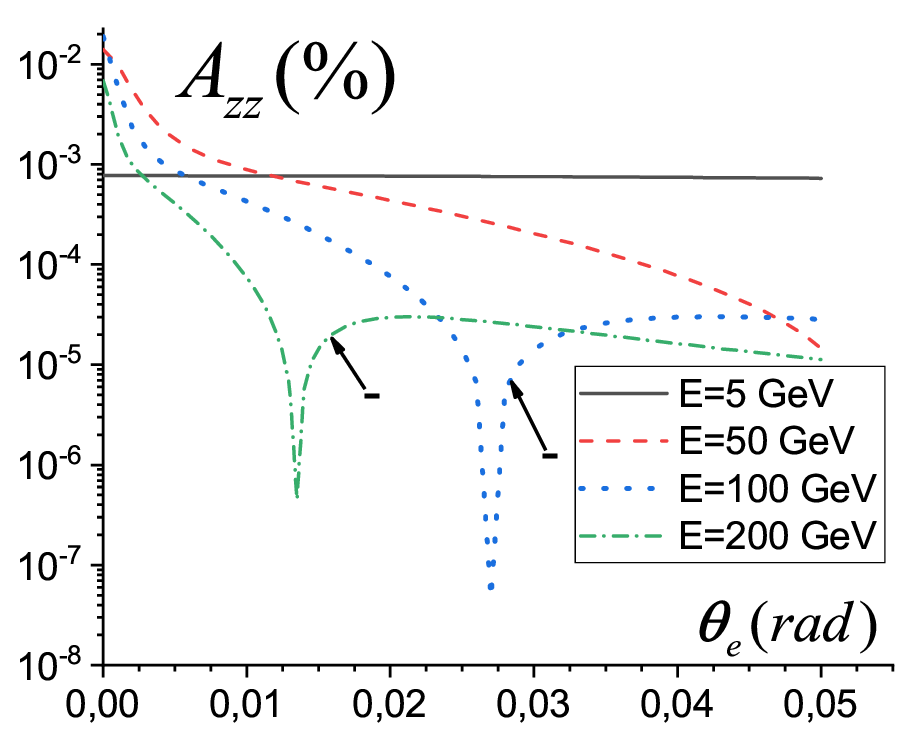}
 \parbox[t]{0.9\textwidth}{\caption{Asymmetries due to tensor polarization of the deuteron beam given by Eqs.(\ref{eq:dsA}, \ref{eq:DA}) as function of the deuteron beam energy $E$ at various values of the electron scattering angle $\theta_e$ (upper row) and as function of $\theta_e$ at different values $E$ (lower row). The arrow with the sign "minus"
 indicates that corresponding asymmetry may become negative and the absolute value is plotted.}\label{fig.Aij}}
\end{figure}

The expressions of these analyzing powers in terms of the deuteron electromagnetic form factors
have the following form
\begin{eqnarray}
{\cal D}A_{xx}&=&\frac{x^2}{M^2}\Big[(m^2p^2+\tau
M^4)G_M^2+mE\,Q^2\,G_M\,G_Q+\frac{(4m^2E^2-M^2Q^2-2mE\,Q^2)}{1+\tau}\,G_Q\,G\Big], \nonumber\\
{\cal D}A_{xz}&=&2\frac{x\tau(M^2+m E)}{m p E}\Big\{M^2Q^2\,G_M^2-2\Big[\frac{m^2p^2(4m E-Q^2)}{M^2+m E}
\nonumber\\
&&-2mE\,Q^2\Big]\,G_M\,G_Q +\frac{4(4m^2E^2-M^2Q^2-2mE\,Q^2)}{1+\tau}\,G_Q\,G\Big\}, \nonumber\\
{\cal D}A_{zz}&=&\frac{Q^2}{E^2}\Big\{\Big [m^2p^2\Big(1+\frac{Q^2}{\,\,\,Q^2_{max}}\Big)+\frac{Q^4}{8} -\frac{3}{4}\,M^2\,x^2\Big]G_M^2+ \nonumber\\
&& \Big[\frac{Q^4}{2}+m E\big[4\tau(Q^2-2m E)-3x^2\big]\Big]G_M\,G_Q\nonumber\\
&& +\frac{Q^2(M^2+2 m E)-4m^2 E^2}{1+\tau}\Big(1-2\tau-3\frac{Q^2}{\,\,\,Q^2_{max}}\Big)G_Q\,G, \Big\}, \nonumber\\
x& =& k_{2x} = - p_{2x} =\left [Q^2\Big(1-\frac{Q^2}{\,\,\,Q^2_{max}}\Big)\right]^{1/2}.
\label{eq:DA}
\end{eqnarray}
The asymmetries due the tensor polarization of the deuteron beam are plotted in Fig.\,\ref{fig.Aij} for the  chosen form factors.



\subsection{Tensor polarization coefficients, $P_{ij}$, in the
$d+e \to \vec{d}\,^T+e $ reaction the tensor polarization of the scattered deuteron is measured }

We consider here the scattering of an unpolarized deuteron beam on an
unpolarized electron target (the polarization of the recoil electron
is not measured). In this case the scattered deuterons may acquire
a tensor polarization. The hadronic tensor which corresponds to
the case of an unpolarized deuteron beam and a tensor polarized
scattered deuteron can be written in the following general form
\begin{equation}\label{eq:HfT}
H_{\mu\nu}({\cal Q}^{(f)})=\bar H_3\bar {\cal Q}^{(f)}\tilde g_{\mu\nu} +\bar H_4
\frac{\bar {\cal Q}^{(f)}}{4M^2} P_{\mu}P_{\nu} -\bar H_5 (P_{\mu}\tilde
{\cal Q}^{(f)}_{\nu}+P_{\nu}\tilde Q^{(f)}_{\mu}) +\bar H_6 \tilde
{\cal Q}^{(f)}_{\mu\nu},
\end{equation}
where the structure functions $\bar H_i$, averaged over the spin of
the initial deuteron, have the following expressions in the terms of the
deuteron electromagnetic form factors: $\bar H_i=H_i/3, i=3,4,5,6$. Note that the tensor structures  in this case can be
obtained from Eq.\,(\ref{eq:HiT}) by the substitution $(p_1 \leftrightarrows - p_2),$, wherein the structure accompanying $\bar{H}_5$ changes sign.

The contraction of the spin independent leptonic $L_{\mu\nu}^{(0)}$
and spin dependent (due to the tensor polarization of the scattered
deuteron) hadronic $ H_{\mu\nu}(Q^{(f)})$ tensors, in an arbitrary
reference frame, gives:
\begin{equation}\label{eq:CfQ}
C({\cal Q}^{(f)})=L^{\mu\nu}{(0)}H_{\mu\nu}({\cal Q}^{(f)})=\bar a\,k_{1}^\mu k_{1}^\nu {\cal Q}^{(f)}_{\mu\nu}+\bar
b\,k_{1}^\mu p_{1}^\nu {\cal Q}^{(f)}_{\mu\nu}+ \bar c\,k^{\mu}k^{\nu}{\cal Q}^{(f)}_{\mu\nu},
\end{equation}
where the coefficients $\bar a, \bar b$ and $\bar c$ are written in terms of the deuteron electromagnetic form factors as:
\begin{eqnarray}
\label{eq:39}
\bar a&=&\frac{2}{3}(1+\tau )Q^2 G_M^2, \nonumber\\
\bar b&=&\frac{8}{3}\tau G_M \left [(M^2+2\tau M^2-k_1\cdot
p_1)G_M+2(\tau M^2-k_1\cdot p_1)G_Q\right ],
\nonumber\\
\bar c&=& \frac{2}{3M^2}\{[(k_1\cdot p_1)^2-Q^2k_1\cdot p_1-m^2M^2+
\tau (1+4\tau )M^4]G_M^2-k^2(2\tau M^2-k_1\cdot p_1)G_MG_Q+ \nonumber\\
&&+(1+\tau )^{-1}[4(k_1\cdot p_1)^2-Q^2(M^2+2k_1\cdot
p_1)]G_Q\,G\} . 
\end{eqnarray}
From the condition $p_{2}^\mu\,{\cal Q}^{(f)}_{\mu\nu}=0$ one can express the time components of
the scattered deuteron quadrupole polarization tensor in terms of
the space components of this tensor. These relations are:
\begin{eqnarray}\label{eq:40}
{\cal Q}^{(f)}_{00}&=&\frac{x^2{\cal Q}^{(f)}_{xx}-2xz{\cal Q}^{(f)}_{xz}+z^2{\cal Q}^{(f)}_{zz}}{E_d^2}, \ \
{\cal Q}^{(f)}_{0z}=\frac{z{\cal Q}^{(f)}_{zz}-y{\cal Q}^{(f)}_{xz}}{E_d},
\\
{\cal Q}^{(f)}_{0x}&=&\frac{z{\cal Q}^{(f)}_{xz}-x{\cal Q}^{(f)}_{xx}}{E_d}, \ \
{\cal Q}^{(f)}_{0y}=\frac{z{\cal Q}^{(f)}_{yz}-x{\cal Q}^{(f)}_{xy}}{E_d}, \ \ E_d= E -\frac{Q^2}{2m}\nonumber
\end{eqnarray}
where $E_d$ is the energy of the scattered deuteron and
$$ z= p_{2z} = p\,\Big[1-\frac{(E+m)\,Q^2}{2m\,p^2}\Big].$$

The components of the quadrupole polarization tensor $Q^{(f)}_{ij}$ which are defined in the Lab system
can be related to the corresponding ones in the rest system of the scattered deuteron (denote them as $V_{ij}$)
by the following relations
\begin{eqnarray}\label{eq:41}
{\cal Q}^{(f)}_{xx}&=&(1+x^2 y)^2V_{xx}-2xyz(1+x^2 y)V_{xz}+(xyz)^2V_{zz}, \ \ {\cal Q}^{(f)}_{yy}=V_{yy}, \\
{\cal Q}^{(f)}_{zz}&=&(xyz)^2V_{xx}-2xyz(1+z^2 y)V_{xz}+(1+yz^2)^2V_{zz}, \nonumber\\
{\cal Q}^{(f)}_{xz}&=&-xyz(1+z^2y)V_{xx}+[1+y(x^2+z^2)+2(xyz)^2]V_{xz}-xyz(1+yz^2)V_{zz}, \nonumber
\end{eqnarray}
where $$y=\frac{1}{M(E_d+M)}.$$

The dependence of the differential cross section of the reaction (1) on
the polarization characteristics of the scattered deuteron (given in the Lab system)in case
when the deuteron beam and electron target are unpolarized has the following form
\begin{equation}\label{eq:42}
\frac{d\sigma}{dQ^2}({\cal Q}^{(f)})= \left (\frac{d\sigma}{dQ^2}\right
)_{un}\left [1+P_{xx}({\cal Q}^{(f)}_{xx}-{\cal Q}^{(f)}_{yy})+ P_{xz}
{\cal Q}^{(f)}_{xz}+ P_{zz} {\cal Q}^{(f)}_{zz}\right ],
\end{equation} where $P_{ij}$,
$i,j=x,y,z$ are the components of the tensor polarization which
describe the $d+e\to \vec{d}\,^T+e$ scattering in case when the scattered
deuteron is tensor polarized. The explicit expressions of these
tensor polarizations, in terms of the deuteron electromagnetic form factors,
can be written as
\begin{eqnarray}
{\cal D} P_{zz}&=&\frac{2}{3\,M^2 d\,E_d}\big[a_1\,G_M^2+ a_2\,G_M\,G_Q+a_3\,G_Q\,G\big], \ \ d= 2 E_d^2-x^2, \nonumber\\
a_1&=& (2z^2-x^2)\,E_d\big\{E^2(m^2 p^2 +\tau M^4) - Q^2(E+m)[m p^2 - \tau M^2(E+m)]\big\}\nonumber\\
&&+p\,d\big[-m z Q^2(M^2+2\tau M^2-mE)+(p E_d-2Ez)(m^2p^2-m E Q^2+\tau M^4+\tau M^2Q^2)\big], \nonumber\\
a_2&=&Q^2\big\{-p^2 d\, m\,E_d^2-2p\,d\,z[\tau M^2(2E+m)-mE(E+m)]+ \nonumber\\
&&+(2z^2-x^2)E\,E_d[2\tau M^2(E+m)-mE(E+2m)]\big\}, \nonumber\\
a_3&=&\frac{4 m^2 E^2 -Q^2(M^2+2m E)}{1+\tau}\big[(2z^2-x^2)E^2\,E_d +p\,d(p\,E_d -2 E\,z)\big]\nonumber\\
{\cal D} P_{xx}&=&\frac{2\,x^2}{3 d\,M^2}[b_1\ G_M^2+
b_2\ G_M\ G_Q+b_3\ G_Q\,G], \label{eq:Pij}\\
b_1&=&mp^2[mE^2-Q^2(E+m)]+\tau M^2[E^2M^2+Q^2(E+m)^2], \nonumber\\
b_2&=&-EQ^2[2(mE-\tau M^2)(E+m)-mE^2], \nonumber\\
b_3&=&\frac{E^2}{1+\tau }[4m^2E^2-Q^2(M^2+2mE)], \nonumber\\
{\cal D} P_{xz}&=&\frac{2 x}{3 d M^2 E_d}\big[c_1 G_M^2+
c_2 G_M G_Q+c_3G_Q\,G\big], \nonumber\\
c_1&=&4 p Q^2(E+m)E_d[2 m p^2-Q^2(E+m)] - 4 z E_d\big[Q^2\tau M^2(E+m)^2 +E^2(m^2 p^2+\tau M^4) \nonumber\\
&&+p^2 Q^2m(E+m)\big] + p d \big[2 E(m^2 p^2 +\tau M^4 +\tau Q^2 M^2)-m Q^2(m E +E^2+p^2-2 \tau M^2)\big],\nonumber\\
c_2&=&-4 z E Q^2 E_d\big[m E^2+2(E+m)(\tau M^2-m E)\big] -p d Q^2\big[2mE(E+m)-2\tau M^2(2E+m)\big],\nonumber\\
c_3&=&\frac{2E}{1+\tau} [(4m^2E^2 - Q^2(M^2+2mE)](p d-2 z E E_d).\nonumber 
\end{eqnarray}

\begin{figure}
\centering
\includegraphics[width=0.3\textwidth]{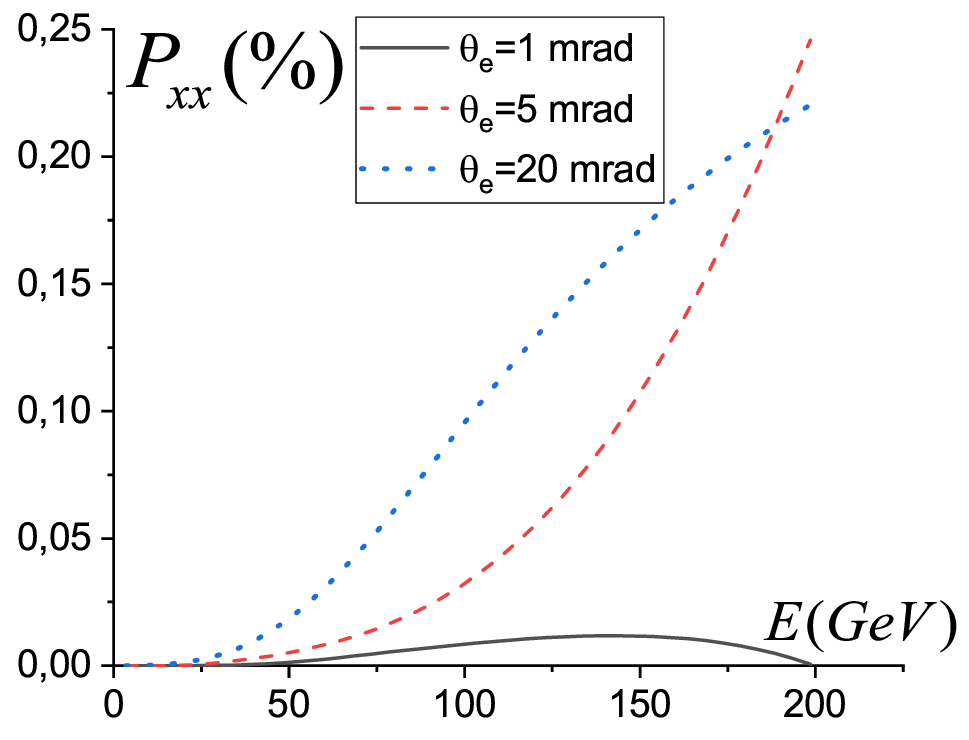}
\includegraphics[width=0.3\textwidth]{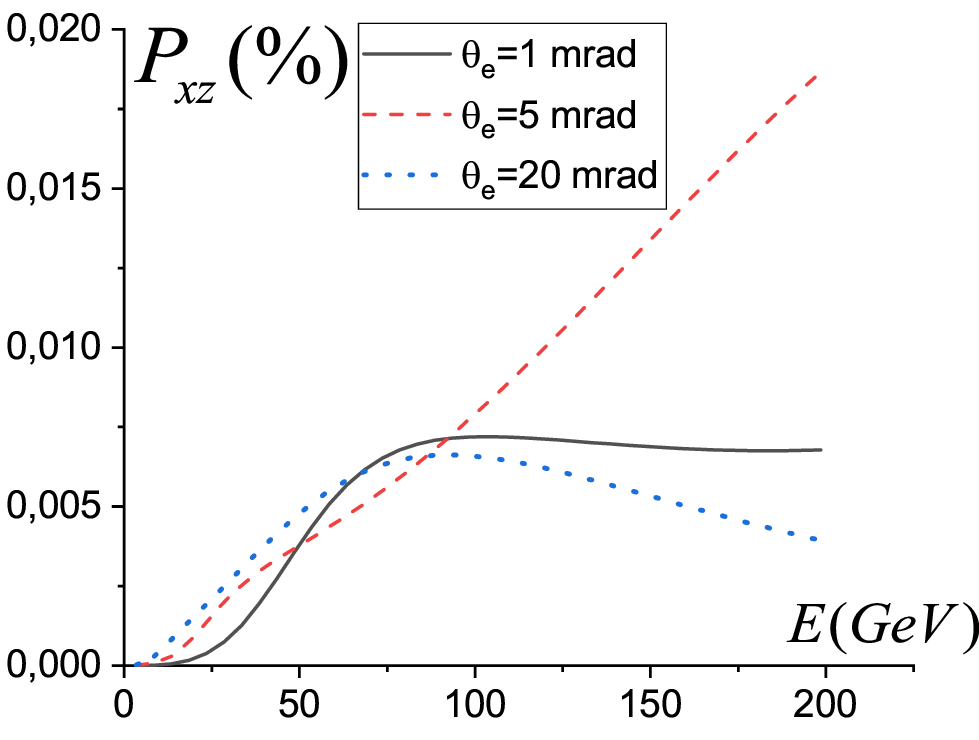}
\includegraphics[width=0.3\textwidth]{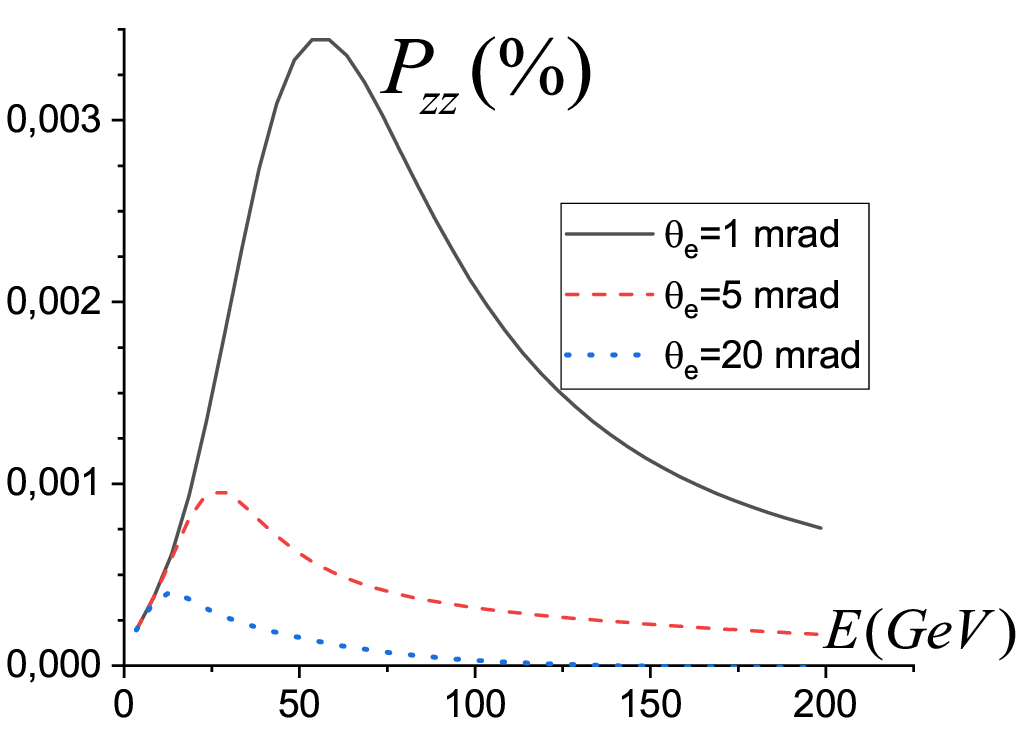}
\vspace{0.5cm}
\includegraphics[width=0.3\textwidth]{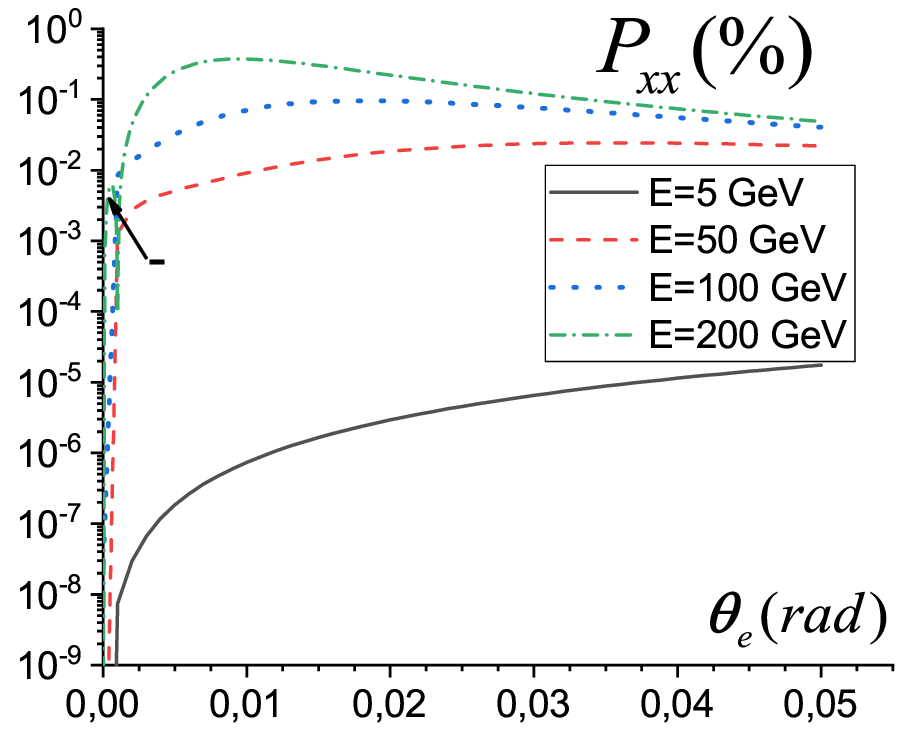}
\includegraphics[width=0.3\textwidth]{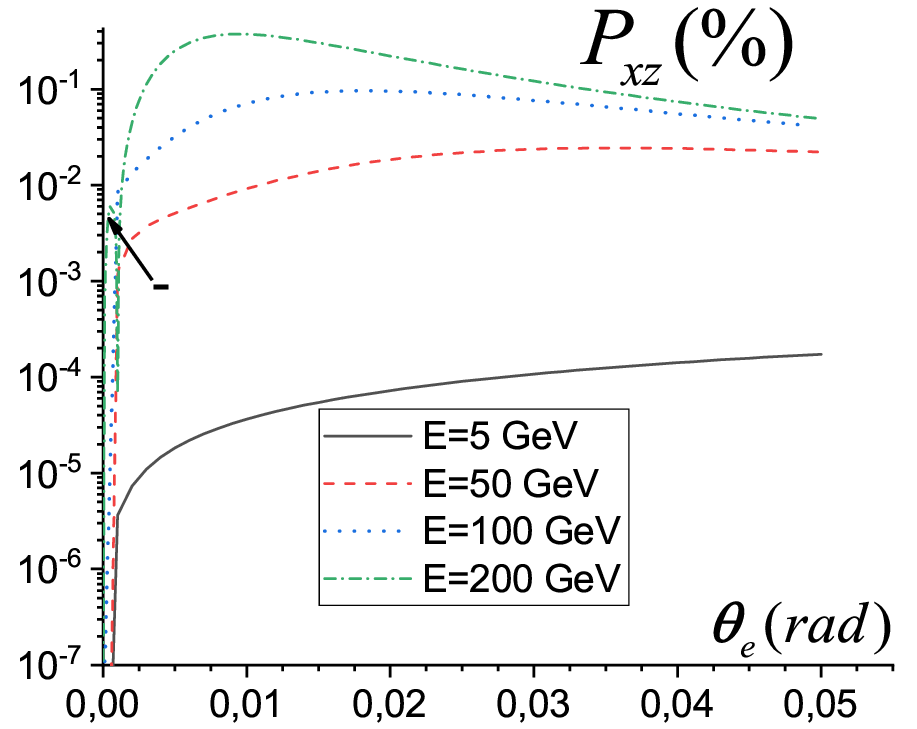}
\includegraphics[width=0.3\textwidth]{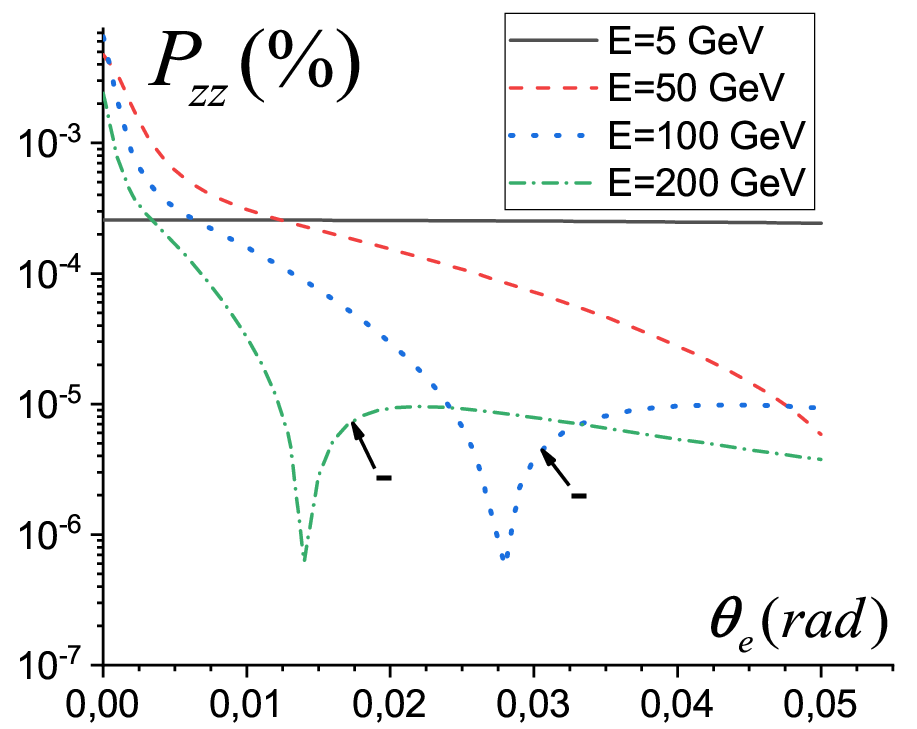}
\parbox[t]{0.9\textwidth}{\caption{Same as Fig.\,\ref{fig.Aij} but for the tensor polarization coefficients $P_{ij}$ given by Eqs.\,(\ref{eq:Pij}),
which describe the tensor polarization of the final deuteron  when the other particles are unpolarized. When the observable has negative values, its modulus is plotted (indicated by arrows). }\label{fig.Pij}}
\end{figure}

The results for the scattered deuteron tensor polarizations are shown in Fig.\,\ref{fig.Pij}.

\subsection{Polarization transfer coefficients from the target to the recoil electron, $t_{ij}$, in the $d+\vec e \to d+\vec e $ reaction}
We consider below the scattering of an unpolarized deuteron beam on a polarized electron target in the case when the polarization of the recoil electron is measured and the polarization of the scattered deuteron is not measured.

The part of the leptonic tensor which corresponds to the case of  polarized target and polarized recoil
electron has the following form
\begin{eqnarray}
L_{\mu\nu}(s_1, s_2)&=&-(k_1\cdot s_2k_2\cdot s_1-\frac{Q^2}{2}s_1\cdot s_2)g_{\mu\nu}-\frac{Q^2}{2}(s_{1\mu}s_{2\nu}+s_{1\nu}s_{2\mu})
-s_1\cdot s_2(k_{1\mu}k_{2\nu}+k_{1\nu}k_{2\mu})+\nonumber\\
&&k_1\cdot s_2(s_{1\mu}k_{2\nu}+s_{1\nu}k_{2\mu})+
k_2\cdot s_1(s_{2\mu}k_{1\nu}+s_{2\nu}k_{1\mu}),  \label{eq:44}
\end{eqnarray}
where $s_{2\mu}$ is the polarization four-vector of the recoil electron which satisfies the following
conditions: $k_2\cdot s_2=0, s_2^2=-1.$

In the Lab system, where the target electron is at rest, the polarization four-vector of the recoil
electron has the following components
\begin{equation}\label{eq:45}
s_2=\left (\frac{\vec k_2\cdot \vec\xi_2}{m}, \vec\xi_2+\frac{\vec k_2(\vec k_2\cdot \vec\xi_2)}{m(m+\epsilon_2)}\right ),
\end{equation}
where $\vec\xi_2$ is the unit vector describing the polarization of the recoil electron in its rest system and $\epsilon_2$ is the recoil electron energy.

The contraction of the spin dependent leptonic tensor $L_{\mu\nu}(s_1, s_2)$
and the spin independent hadronic tensor $ H_{\mu\nu}(0),$ in an arbitrary reference frame, can be written as follows
\begin{eqnarray}
C(s_1, s_2)&=&L^{\mu\nu}(s_1, s_2)H_{\mu\nu}(0)=-2m^2 s_1\cdot s_2 H_1(Q^2)+\nonumber\\
&& +\frac{H_2(Q^2)}{M^2}\Big\{-2s_1\cdot s_2\left [(k_1\cdot P)^2-\frac{P^2 Q^2}{4}\right ]-4M^2(1+\tau )k_1\cdot s_2k_2\cdot s_1-
\nonumber\\
&&
-Q^2 P\cdot s_1P\cdot s_2+ 2P\cdot k_1(k_1\cdot s_2P\cdot s_1+k_2\cdot s_1P\cdot s_2)\Big \}, \label{eq:s1s2}
\end{eqnarray}
where the structure functions $H_{1,2}(Q^2)$ are given by Eq. (19).

The dependence of the differential cross section of the reaction $d+\vec e \to d+\vec e $ on the polarizations
of the initial and recoil electrons has the following form
\begin{equation}\label{eq:DSs1s2}
\frac{d\sigma}{dQ^2}(\vec\xi_1, \vec\xi_2)= \frac{1}{2}\left
(\frac{d\sigma}{dQ^2}\right )_{un}\left [ 1+t_{xx} \xi_{1x}\xi_{2x}+
t_{yy} \xi_{1y}\xi_{2y}+ t_{zz} \xi_{1z} \xi_{2z}+t_{xz} \xi_{1x} \xi_{2z}+
t_{zx}\xi_{1z} \xi_{2x} \right ],
\end{equation}
where $t_{ij}$, $i,j=x,y,z$ are the coefficients of the polarization transfer from the initial electron to the recoil one.

The explicit expressions of the polarization transfer coefficients, as functions of the deuteron form factors, in the Lab system can be written as
\begin{eqnarray}
Dt_{xx}&=&(2m^2+Q^2\sin^2\theta_e)\Big\{H_1(Q^2)+4\frac{H_2(Q^2)}{M^2}[E^2-\lambda (M^2+2mE)]\Big\}+ \nonumber\\
&&+\frac{4H_2(Q^2)|\vec{k}_2|\,\sin^2{\theta_e}}{M^2}\big[4 \lambda m p E \cos{\theta_e} + |\vec{k}_2|(M^2-2E^2)\big],\label{eq:tij} \\
Dt_{yy}&=&2m^2\Big\{H_1(Q^2)+4\frac{H_2(Q^2)}{M^2}[E^2-\lambda (M^2+2mE)]\Big\},
\nonumber\\
Dt_{zz}&=&4|\vec{k}_2|\cos\theta_e \frac{H_2(Q^2)}{M^2}[(M^2-2E^2)|\vec{k}_2|cos\theta_e +4\lambda m p(m+E+E\cos^2\theta_e)]+ \nonumber\\
&&
+2m^2(1+2\lambda \cos^2\theta_e)\Big\{H_1(Q^2)+4\frac{H_2(Q^2)}{M^2}[E^2-\lambda (2E^2-M^2+2mE)]\Big\}, \nonumber\\
Dt_{xz}&=&4|\vec{k}_2|\sin\theta_e \frac{H_2(Q^2)}{M^2}[(M^2-2E^2)|\vec{k}_2|\cos\theta_e +2m p E]+ \nonumber\\
&&+4\lambda \cos\theta_e \sin\theta_e \Big\{m^2H_1(Q^2)+\frac{H_2(Q^2)}{M^2}[4m E(m E+p|\vec{k}_2|\cos\theta_e )-Q^2(M^2+2mE)]\Big\},\nonumber\\
Dt_{zx}&=&4|\vec{k}_2|\sin\theta_e \frac{H_2(Q^2)}{M^2}[(M^2-2E^2)|\vec{k}_2|\cos\theta_e +2m p(2m\lambda +2\lambda E-E)]+ \nonumber\\
&&+4\lambda \cos\theta_e \sin\theta_e \Big\{m^2 H_1(Q^2)+\frac{H_2(Q^2)}{M^2}[4m E(m E+p|\vec{k}_2|\cos\theta_e )-Q^2(E^2+p^2+2m E)]\Big\}, \nonumber
\end{eqnarray}
where $\lambda =Q^2/(4m^2)$. Let us remind that $\sin{\theta_e}$ and $\cos{\theta_e}$ are functions of $Q^2$ and of the deuteron beam energy $E,$ namely
$$|\vec{k}_2|\cos\theta_e=\frac{Q^2}{2 m}\sqrt{1+\frac{4 m^2}{Q^2_{max}}}, \
|\vec{k}_2|\sin\theta_e=\sqrt{Q^2\bigg(1-\frac{Q^2}{Q^2_{max}}\bigg)}, \ |\vec{k}_2|=\sqrt{Q^2(1+\lambda)}.$$
The expression for  $Q^2_{max}$  is given in  Eq. (\ref{eq:Q2M}).

\begin{figure}
\centering
\includegraphics[width=0.32\textwidth]{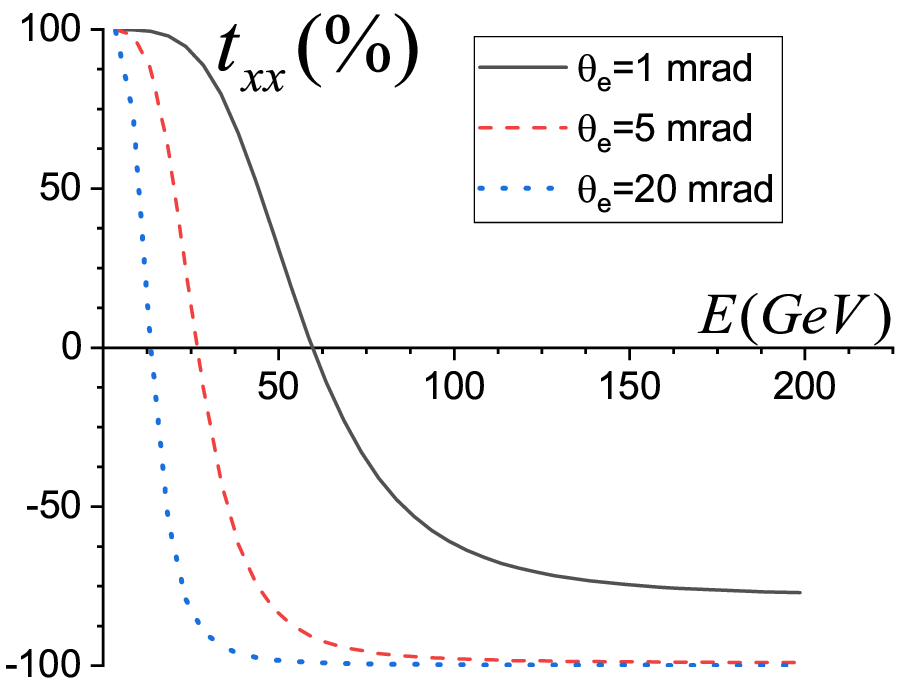}
\includegraphics[width=0.32\textwidth]{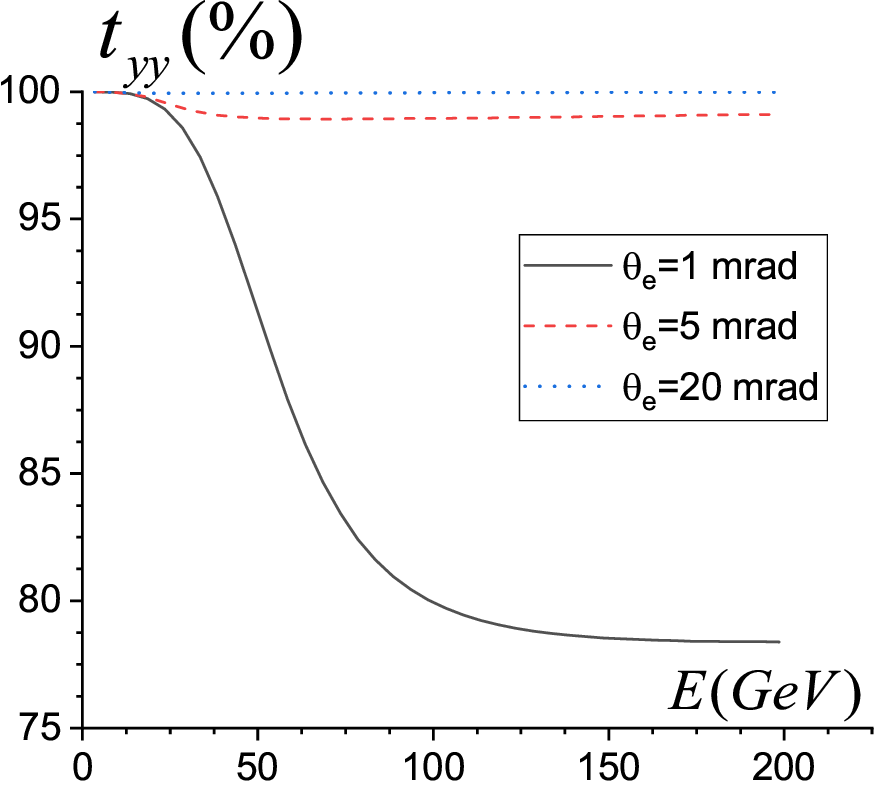}
\includegraphics[width=0.32\textwidth]{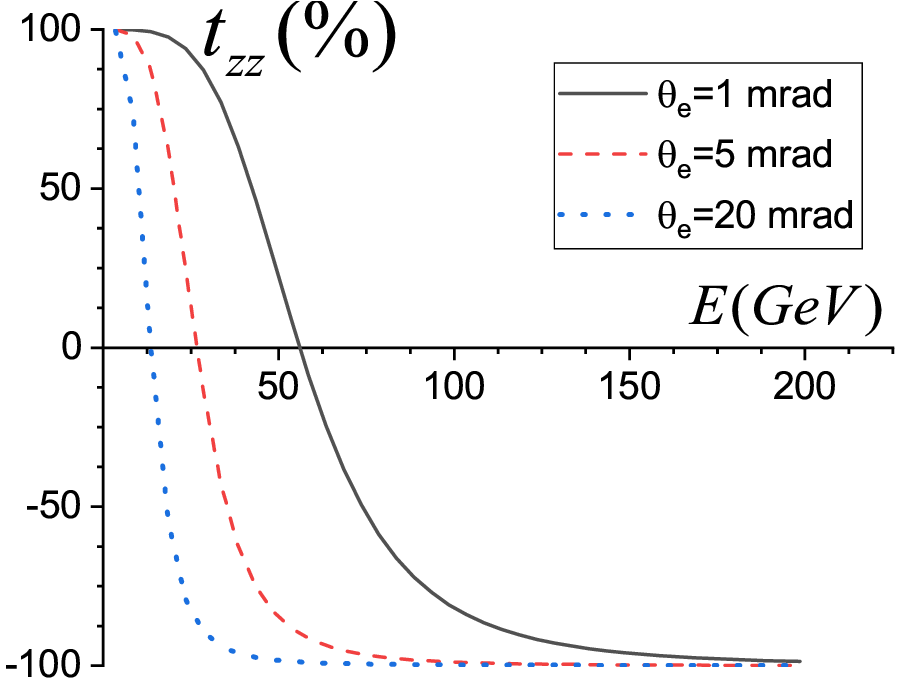}

\vspace{0.2 cm}

\includegraphics[width=0.32\textwidth]{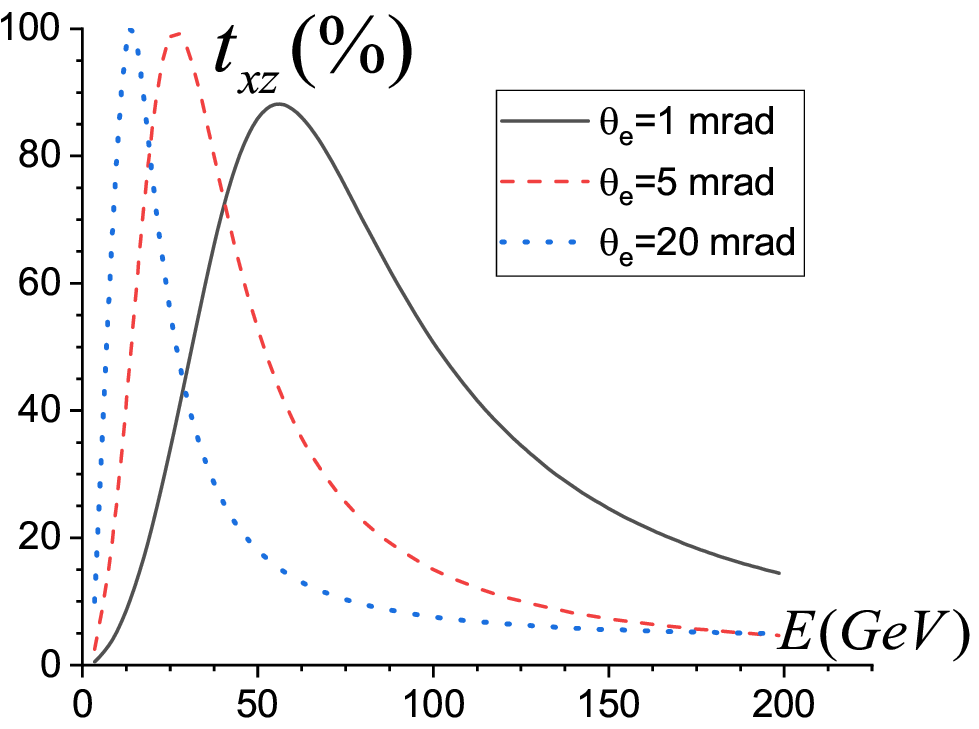}
\includegraphics[width=0.32\textwidth]{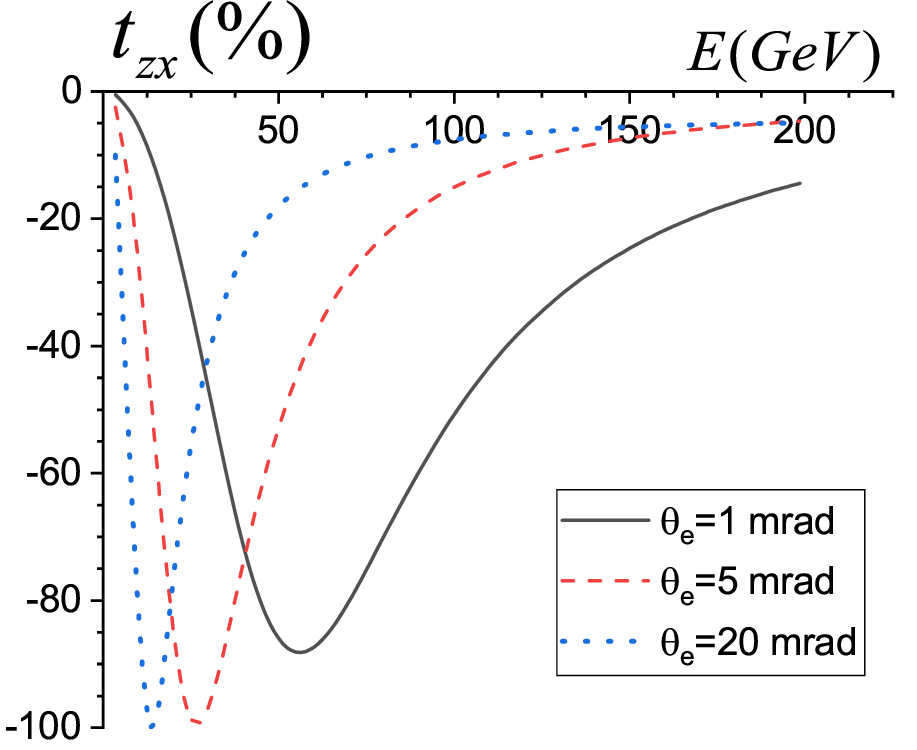}

\vspace{0.5 cm}
\includegraphics[width=0.32\textwidth]{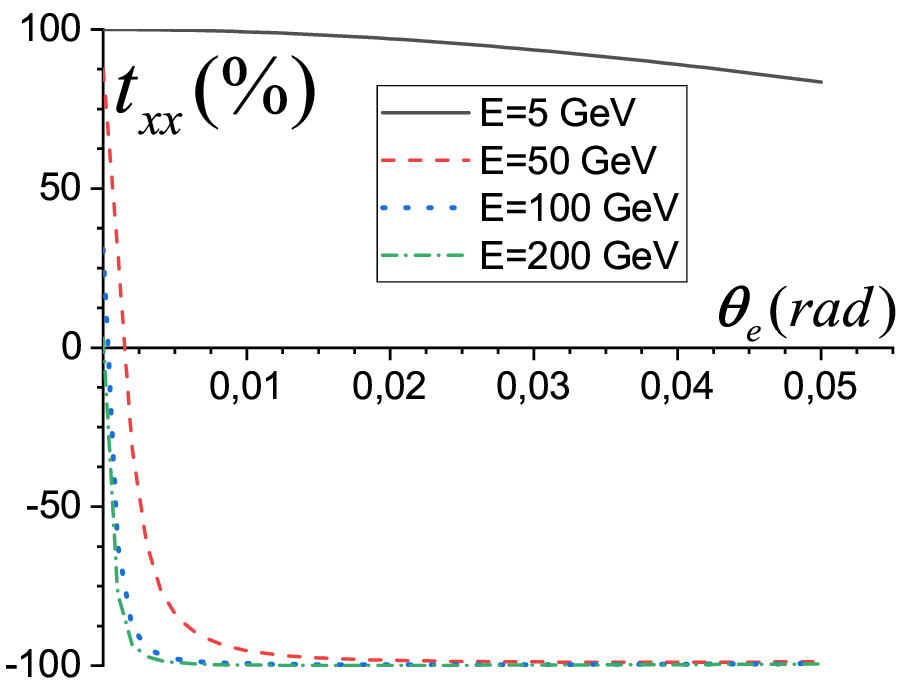}
\includegraphics[width=0.32\textwidth]{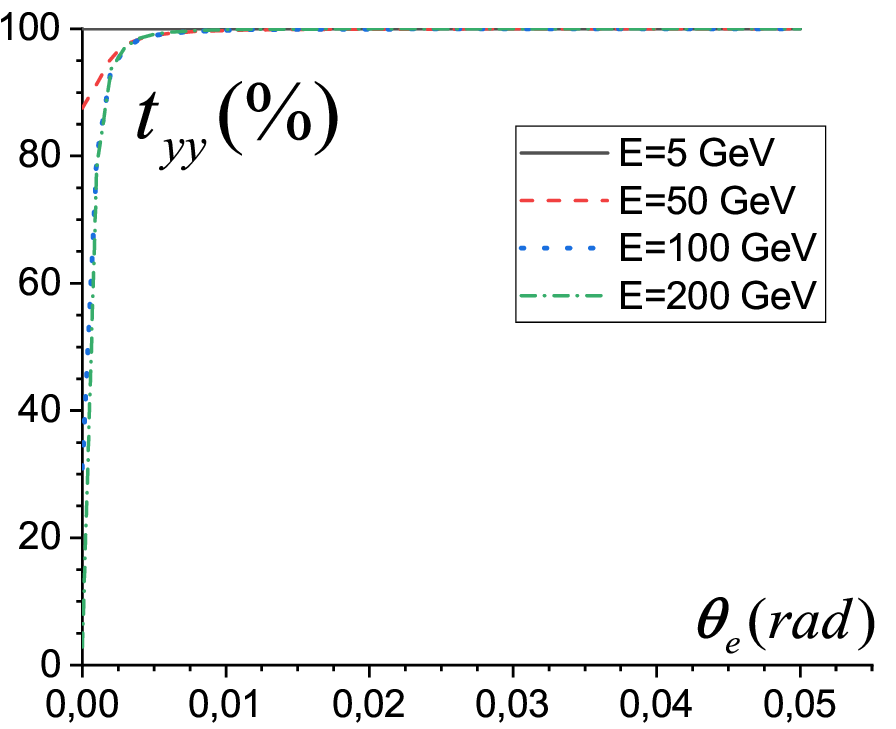}
\includegraphics[width=0.32\textwidth]{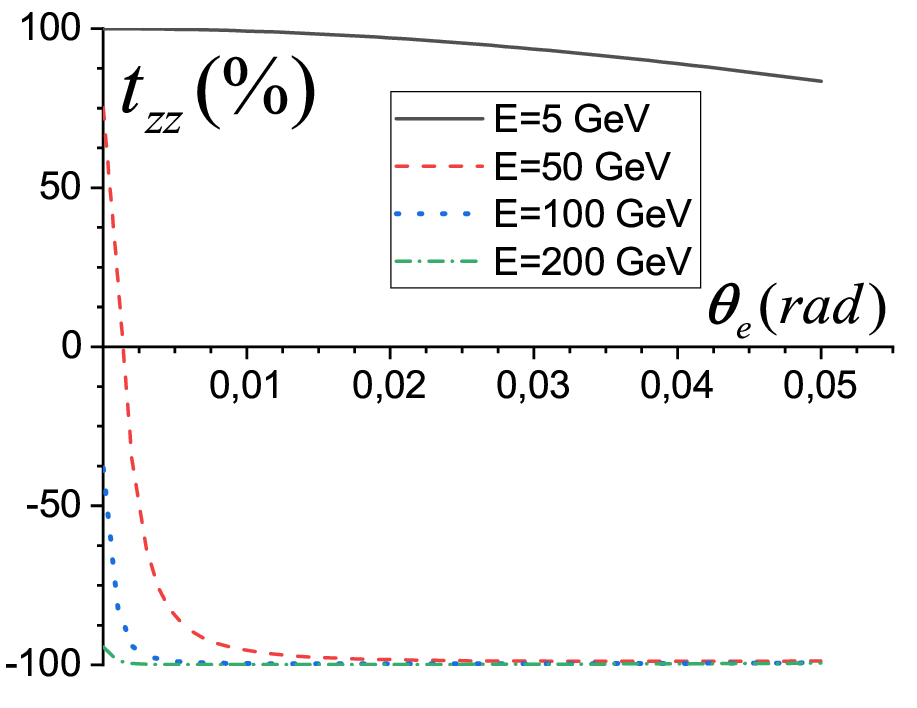}

\vspace{0.2 cm}

\includegraphics[width=0.32\textwidth]{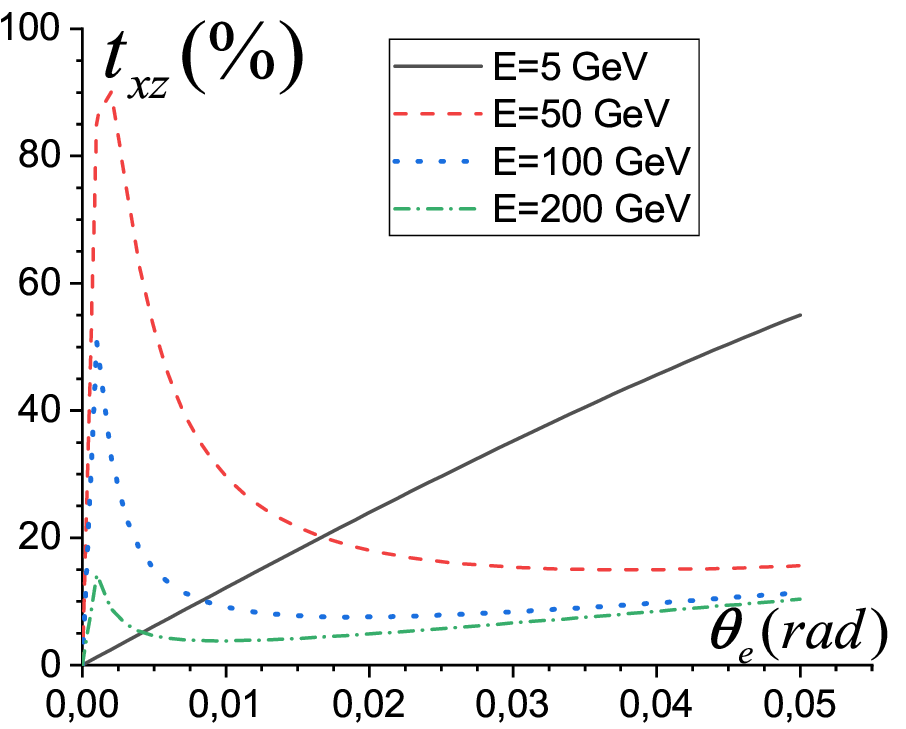}
\includegraphics[width=0.32\textwidth]{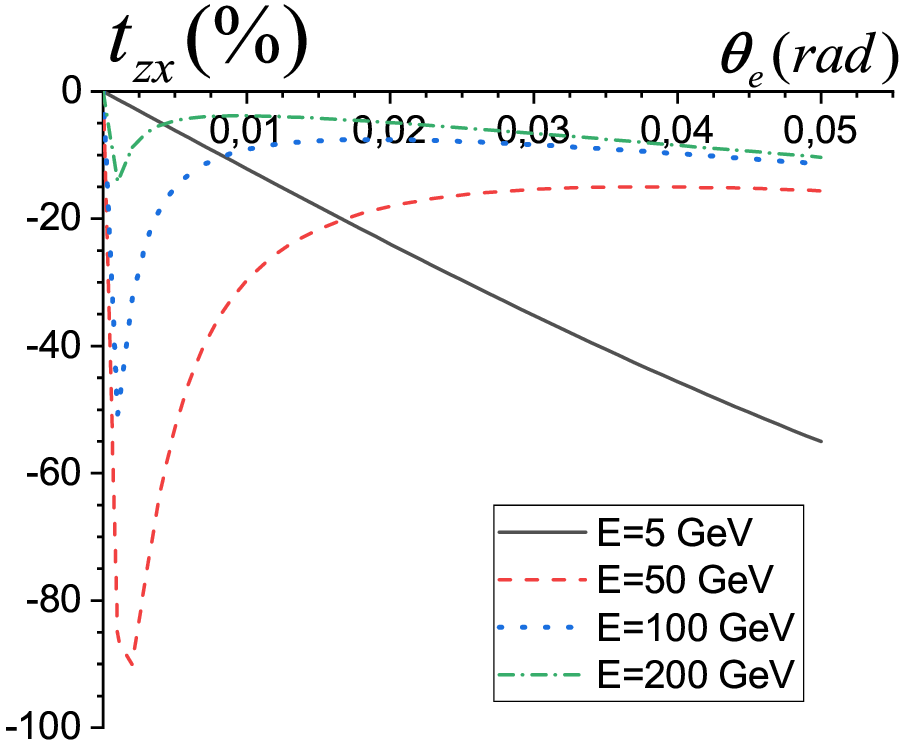}
 \parbox[t]{0.9\textwidth}{\caption{Coefficients of the polarization transfer from the target to the recoil electron 
 in $d+e$ elastic scattering given by Eqs.\,(\ref{eq:tij}).}\label{fig.tij}}
\end{figure}

The electron polarization transfer coefficients $t_{ij}$ are plotted in Fig.\,\ref{fig.tij}.

\subsection{Spin correlation coefficients, $C_{ij}$  due to a polarized electron
target and a vector polarized deuteron beam: $\vec{d}\,^V + \vec{e} \to d + e$}
Let us consider the scattering of a vector polarized deuteron beam (the
polarizations of the final particles are not detected). In this case
a non-zero polarization effects arise only when the electron target
is also polarized. So, the part of the hadronic tensor
$H_{\mu\nu}(\eta_1)$ related to the vector polarized deuteron beam
and unpolarized scattered deuteron can be written as:
\begin{eqnarray}
H_{\mu\nu}(\eta_1) &=&2iM\,G_M\Big[(1+\tau)\widetilde{G}<\mu\nu \eta_1 k> +\frac{\eta_1\cdot k}{4M^2}(G_M-2\widetilde{G})<\mu\nu p_1 k>\Big], \nonumber \\
 \widetilde{G}&=&
G_C+\frac{\tau}{3}G_Q.
\label{eq:Heta1}
\end{eqnarray}


One can see that all correlation coefficients in $\vec{d}\,^V+\vec
e\to d+e$, and polarization transfer coefficients in  the reaction $d+\vec
e\to \vec{d}\,^V +e$, when the deuteron is vector polarized,
are proportional to the deuteron magnetic form factor. It is also
true for the $e+d$ elastic scattering for the corresponding polarization
observables.


The contraction of the spin-dependent leptonic $ L_{\mu\nu}^{(p)}(s_1)$
and hadronic $ H_{\mu\nu}(\eta_1)$ tensors, in an arbitrary
reference frame, gives:
\begin{eqnarray}
C(s_1,\eta_1)&=&L_{\mu\nu}^{(p)}(s_1)H^{\mu\nu}(\eta_1)= 4 m M G_M\{\tau k\cdot \eta_1 (k\cdot s_1 + 2p_1\cdot s_1)G_M+\\
\nonumber\\
&&+2\,\widetilde{G}\,[Q^2 (1+\tau) s_1\cdot \eta_1 +k\cdot \eta_1 (k\cdot s_1 -2\tau p_1\cdot s_1)]\}. 
\label{eq:Ceta1s1}
\end{eqnarray}

In the considered frame, where the target electron is at rest, the
polarization four-vectors of the electron target and of the deuteron
beam have the following components
\begin{equation}\label{eq:s1eta1}
s_1=(0,\vec\xi_1),  \ \ \eta_{1}=\left ( \frac{\vec p\cdot \vec S_1}{M},
\vec S_1+  \frac{\vec p(\vec p\cdot \vec S_1)}{M(E+M)} \right ),
\end{equation}
where $\vec S_1$ is the unit vector describing the vector
polarization of the deuteron beam in its rest system.

Applying the P-invariance of the hadron electromagnetic interaction,
one can write the following expression for the dependence of the
differential cross section on the polarization of the initial
particles:
\begin{equation}\label{eq:Dseta1s1}
\frac{d\sigma}{dQ^2}(\vec\xi_1,\vec S_1)= \left
(\frac{d\sigma}{dQ^2}\right )_{un}\left [1+C_{xx} \xi_{1x} S_{1x}+
C_{yy} \xi_{1y} S_{1y}+ C_{zz} \xi_{1z}S_{1z}+C_{xz} \xi_{1x} S_{1z}+
C_{zx}\xi_{1z} S_{1x}\right ],
\end{equation} where $C_{ij}$,
$i,j=x,y,z$ are the spin correlation coefficients which determine the $\vec{d}\,^V-\vec e$
scattering, when the deuteron beam is vector polarized and electron target is arbitrarily polarized.

The explicit expressions of the spin correlation coefficients, as a
function of the deuteron form factors is:
\begin{eqnarray}
{\cal D} C _{yy}&=&-4m M Q^2 (1+\tau) G_M\,\widetilde{G}, \nonumber\\
{\cal D}C_{xx}&=& 2\tau m M G_M\left [x^2\,G_M -2 Q^2\left (1+\frac{4M^2}{Q^2_{max}}\right)\,\widetilde{G} \right], 
 \nonumber \\
{\cal D} C_{xz}&=&\frac{Q^2}{p} (mE+M^2) \,x \,G_M\,(\tau G_M +2\widetilde{G}),
\nonumber\\
{\cal D} C_{zx}&=&-4m\,M\,p\,x\,G_M\Big[\tau (G_M -2 \widetilde{G})
-\frac{Q^2(E+m)}{4 m p^2}\,(\tau G_M +2\widetilde{G})\Big], \nonumber\\
 {\cal D} C_{zz }&=& -2 Q^2G_M\Big[ 2(mE-\tau M^2)\,\widetilde{G}+ \tau(M^2+mE) G_M- 
\\
&& -\frac{\tau(E+m)M^2}{m p^2}\,(M^2+m E)\,\Big(\tau G_M +2\,\widetilde{G})\Big]. 
\label{eq:Cij}
\end{eqnarray}

\begin{figure}
\centering
\includegraphics[width=0.27\textwidth]{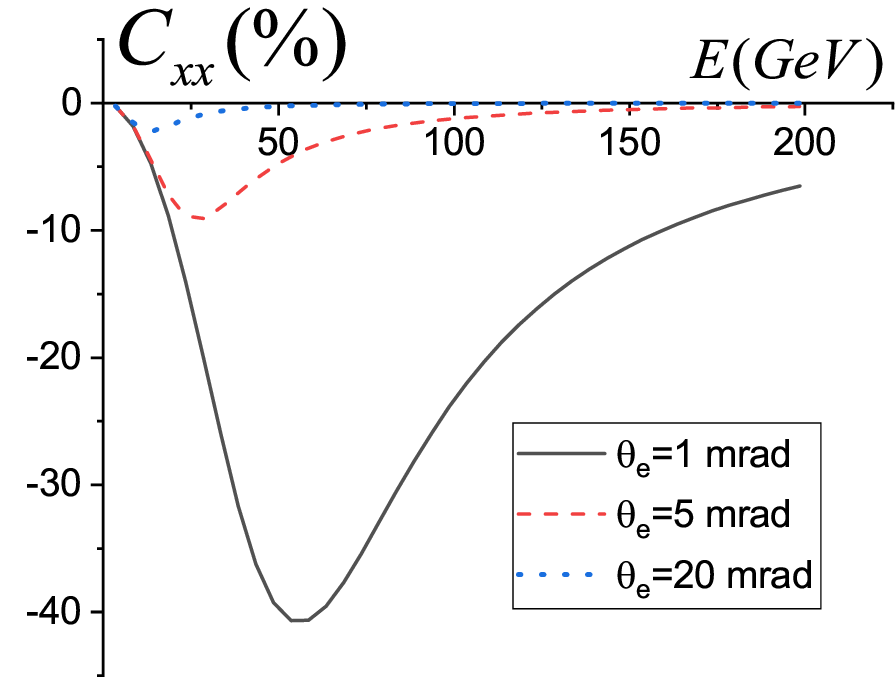}
\includegraphics[width=0.27\textwidth]{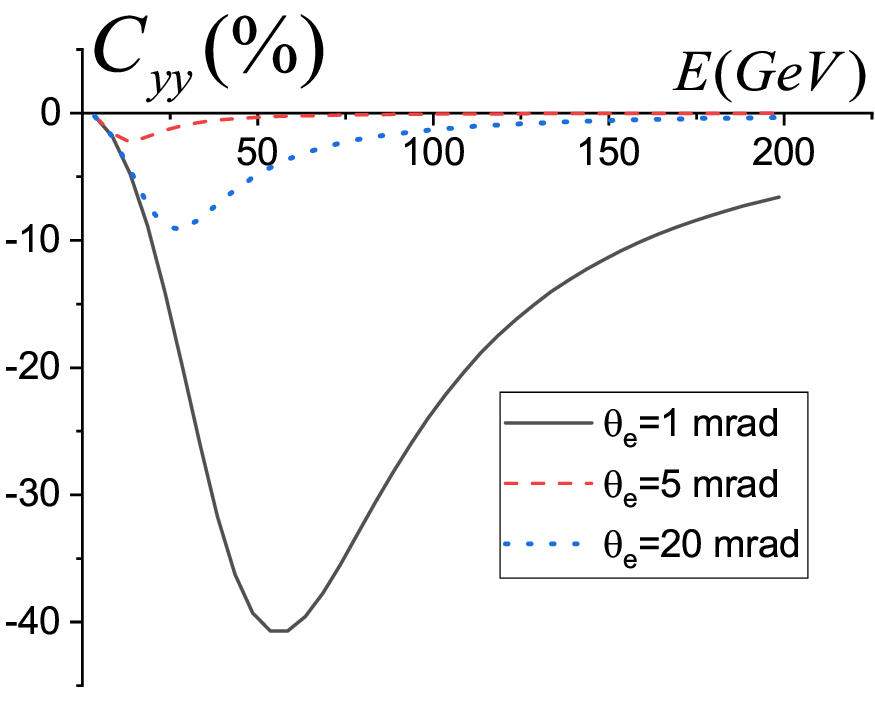}
\includegraphics[width=0.27\textwidth]{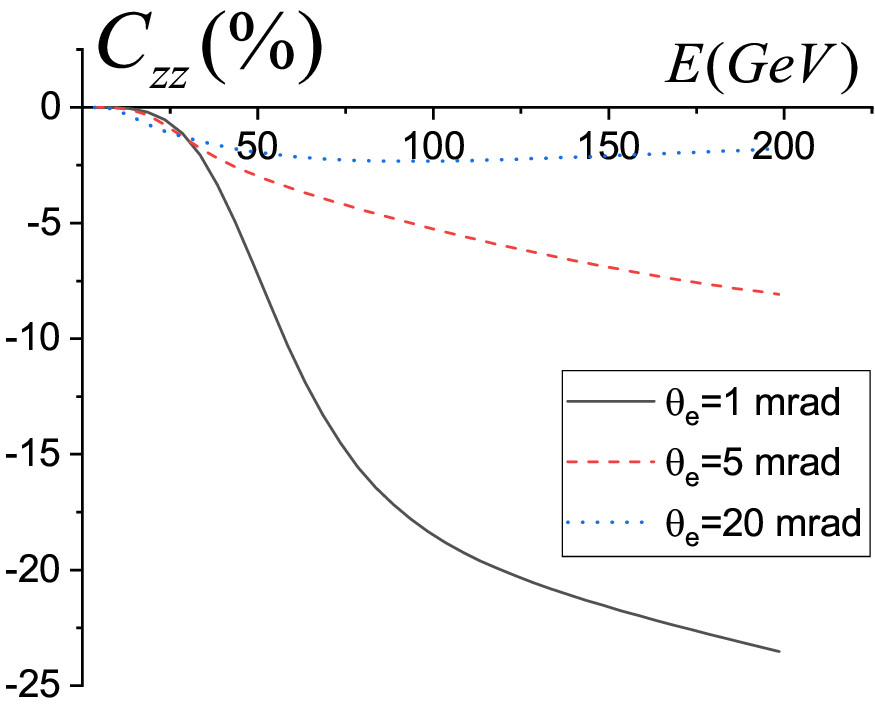}

\vspace{0.2cm}

\includegraphics[width=0.27\textwidth]{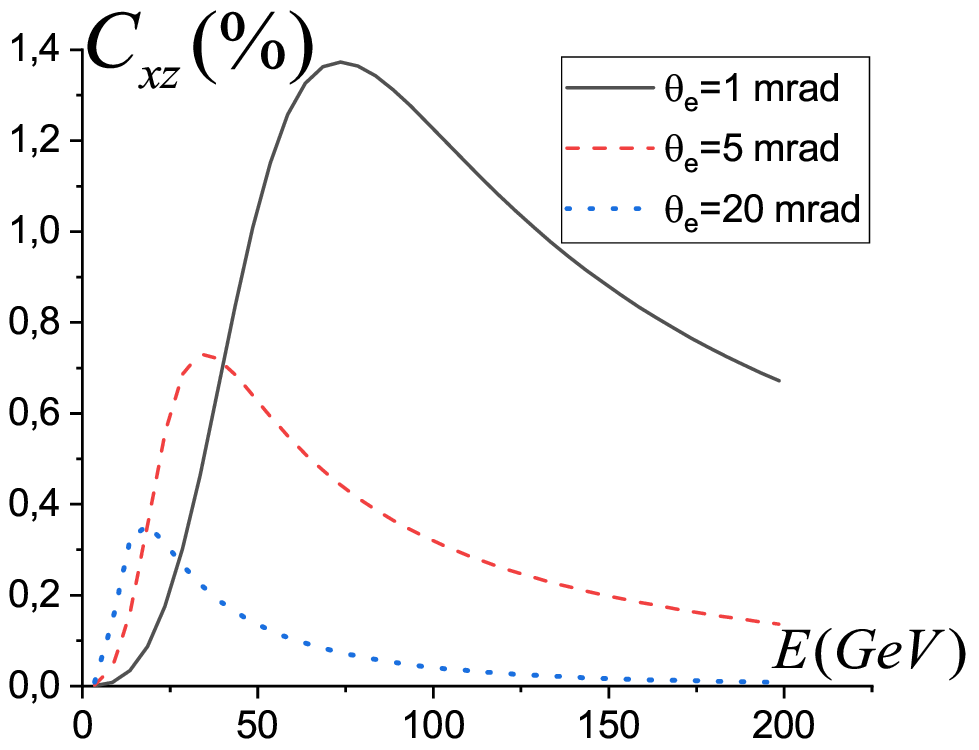}
\includegraphics[width=0.27\textwidth]{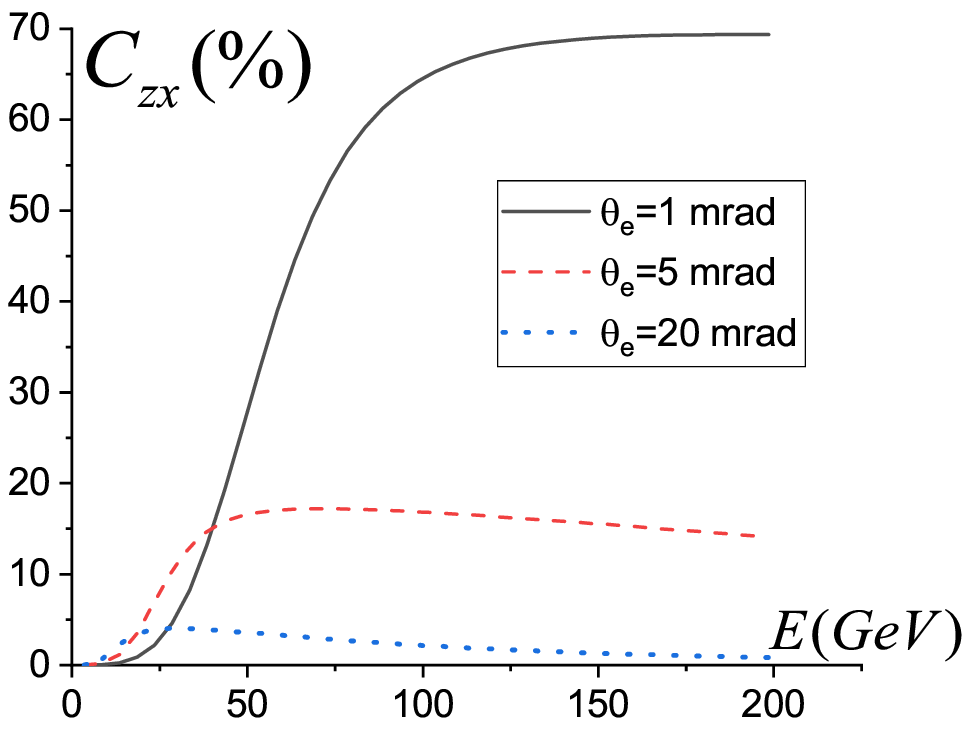}

\vspace{0.5cm}

\includegraphics[width=0.27\textwidth]{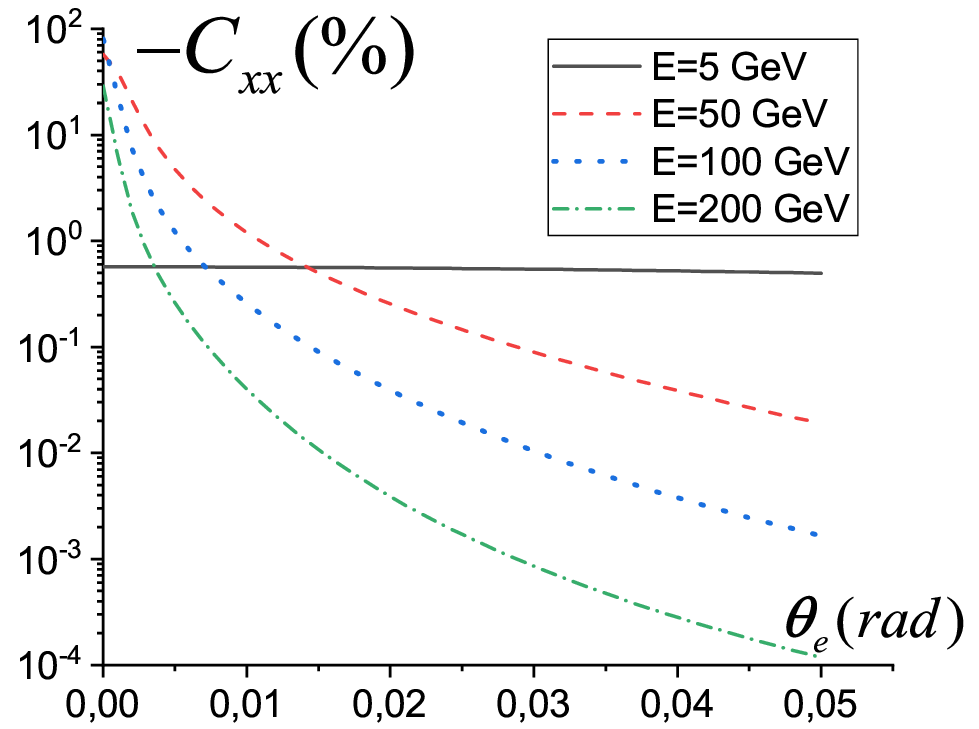}
\includegraphics[width=0.27\textwidth]{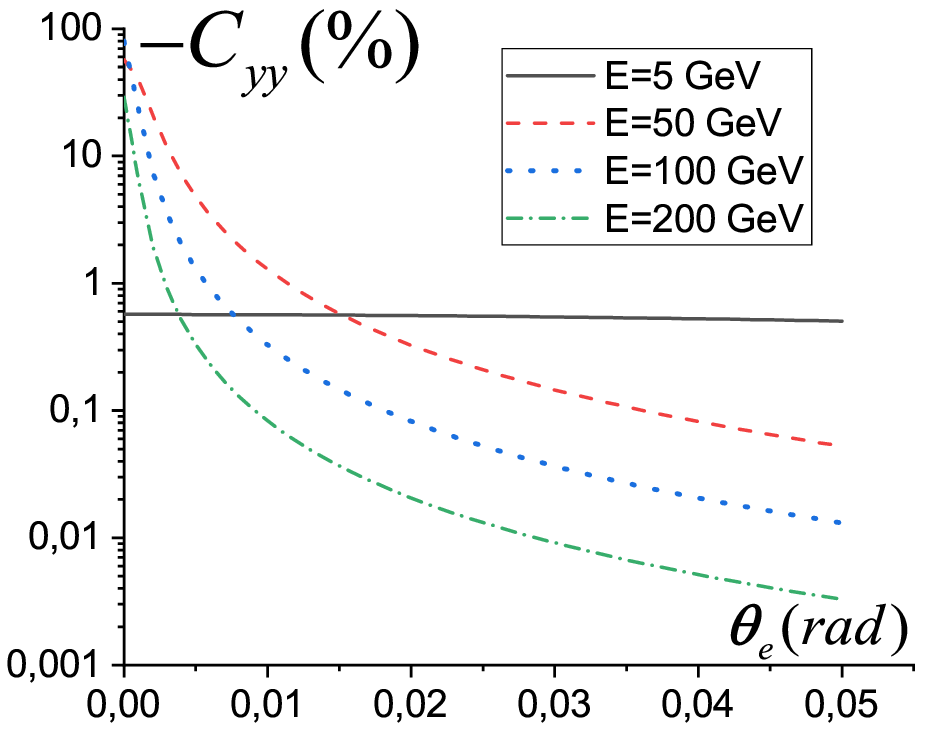}
\includegraphics[width=0.27\textwidth]{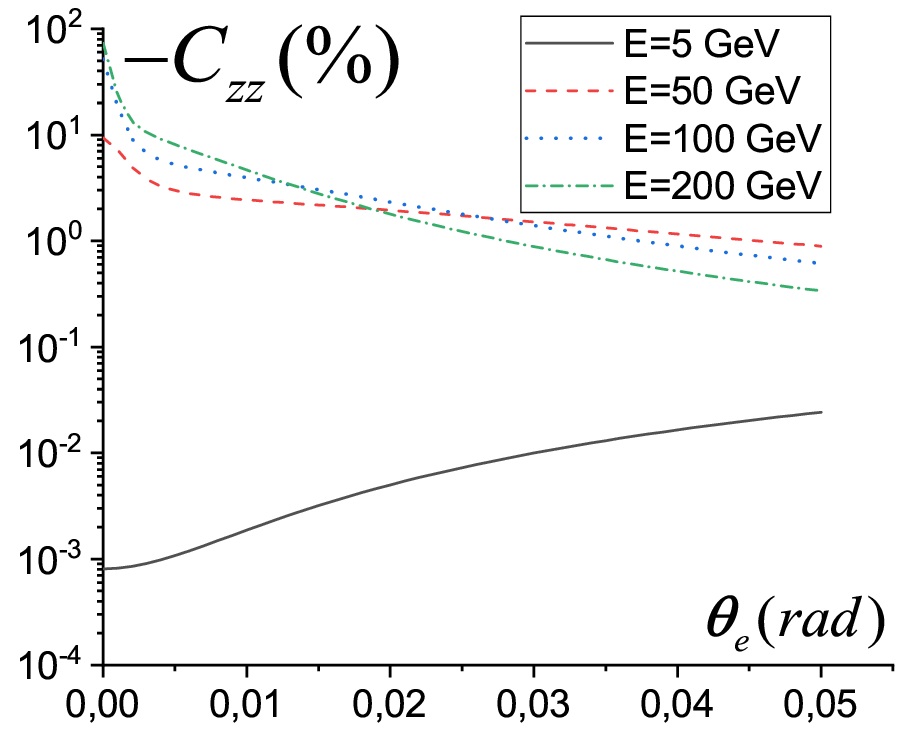}

\vspace{0.2cm}

\includegraphics[width=0.27\textwidth]{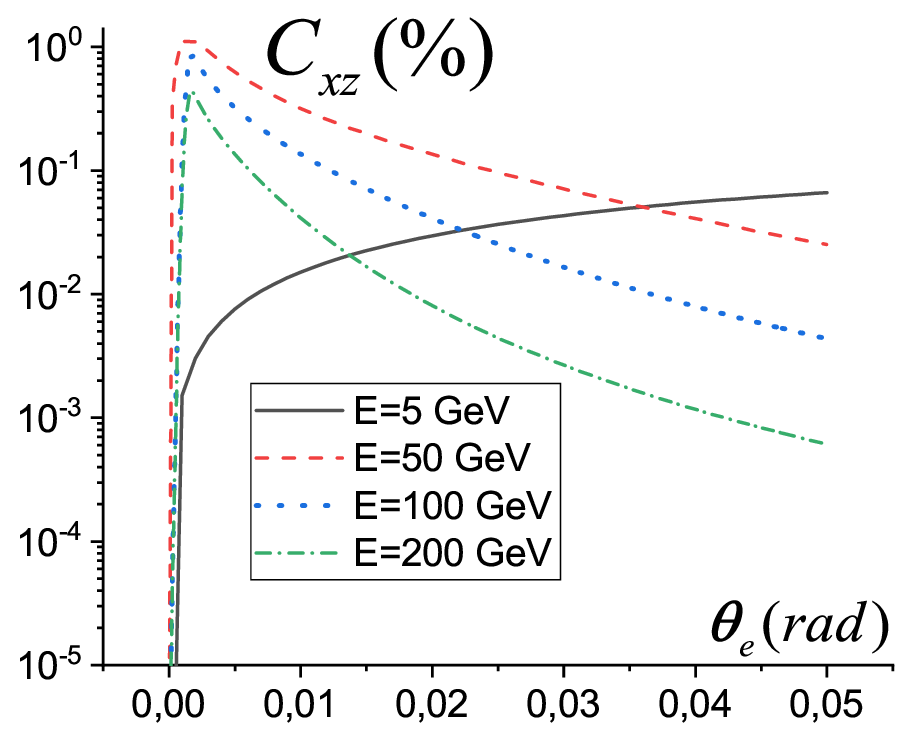}
\includegraphics[width=0.27\textwidth]{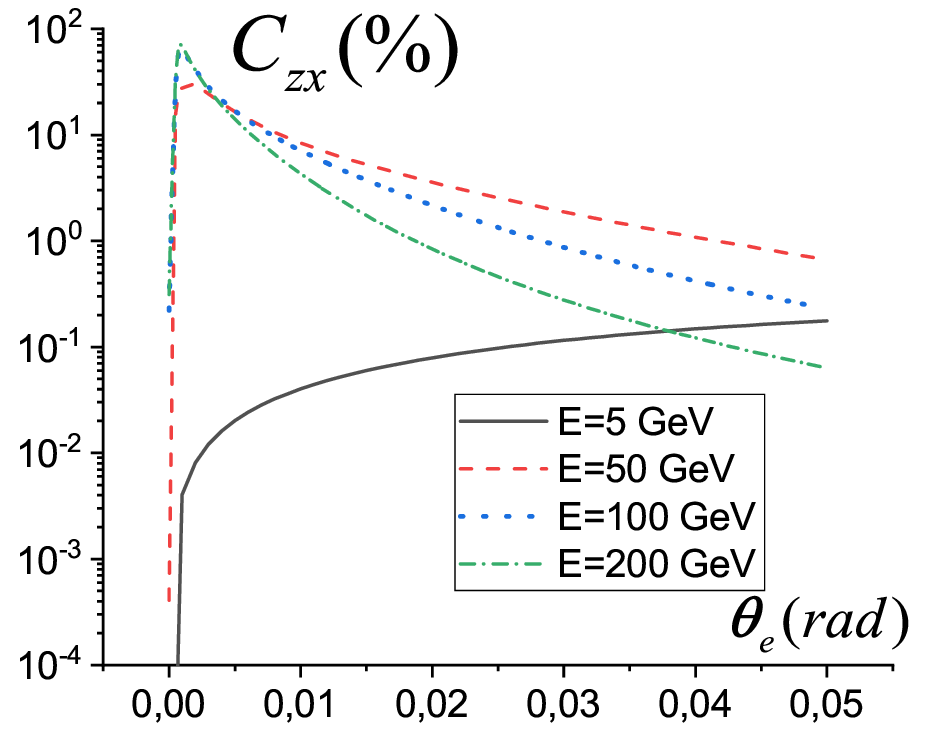}

 \parbox[t]{0.9\textwidth}{\caption{Spin correlation coefficients due to a polarized electron target and a  vector polarized deuteron beam, as given in  Eqs. \,(\ref{eq:Dseta1s1},\,\ref{eq:Cij}).}\label{fig.Cij}}
\end{figure}

The spin correlation coefficient $C_{ij}$ between the vector polarizations of the target electron and the deuteron beam are shown in Fig.\,\ref{fig.Cij}.

\subsection{Polarization transfer coefficients, $T_{ij}$, in the
$d+\vec e \to \vec{d}\,^V+e $ reaction when the target electron has arbitrary polarization and the vector polarization of the scattered deuteron is measured}
Let us consider the case when the initial electron is arbitrary polarized and the scattered deuteron is  vectorially  polarized.
The part of the hadron tensor $H_{\mu\nu}(\eta_2)$ related to the vector polarized scattered deuteron and unpolarized deuteron beam  can be obtained from (\ref{eq:Heta1}) with the substitutions: $\eta_1\to \eta_2$ and $p_1\leftrightarrows - p_2$ and the 
 multiplication by a factor 1/3. The same procedure applies to the calculation of the convolution of the spin depended parts of lepton and hadron tensors. Therefore we have
 \begin{eqnarray}
 \label{eq:Ceta2s1}
C(s_1,\eta_2)&=&L_{\mu\nu}^{(p)}(s_1)H^{\mu\nu}(\eta_2)=
\frac{4}{3} m M G_M\{\tau k\cdot \eta_2 (k\cdot s_1 -
2p_2\cdot s_1)G_M+\nonumber\\
&&
+2\,\widetilde{G} \left [Q^2 (1+
\tau) s_1\cdot \eta_2 +k\cdot \eta_2 (k\cdot s_1 +2\tau p_2\cdot s_1
)\right ]\}. 
\end{eqnarray}
In  Lab system the 4-vector $\eta_2$ is
\begin{equation}\label{eq:eta2}
\eta_{2}=\left ( \frac{\vec{p}_2\cdot \vec S_2}{M},
\vec S_2 +  \frac{\vec{p}_2(\vec{p}_2\cdot \vec S_2)}{M(E_d+M)} \right ),
\end{equation}
where $\vec{S}_2$ is the 3-vector of the scattered deuteron polarization in its rest frame.

The differential cross section can be written as
\begin{equation}\label{eq:Dseta2s1}
\frac{d\sigma}{dQ^2}(\vec\xi_1,\vec S_2)= \left
(\frac{d\sigma}{dQ^2}\right )_{un}\left [1+T_{xx} \xi_{1x} S_{2x}+
T_{yy} \xi_{1y} S_{2y}+ T_{zz} \xi_{1z}S_{2z}+T_{xz} \xi_{1x} S_{2z}+
T_{zx}\xi_{1z} S_{2x}\right ],
\end{equation}
where $T_{ij}$,
$i,j=x,y,z$ are the polarization transfer coefficients which describe the transfer of polarization from the initial electron to the scattered deuteron.

The explicit expressions of the polarization transfer coefficients, in terms of 
the deuteron form factors read:
 \begin{eqnarray}
{\cal D}T_{yy}& =& - \frac{4}{3}Q^2 m M(1+\tau)G_M\,\widetilde{G}, \nonumber\\
{\cal D}T_{xx} &= &\frac{2}{3} m M\,G_M\Big\{x^2 y M \big[2(M+E+2\tau E)\widetilde{G} -\tau (M+E+2\tau M)G_M\big]- 2Q^2(1+\tau)\widetilde{G} \Big\}, \nonumber\\
{\cal D}T_{zx} &= &\frac{xy M^2\,G_M}{3 p}\big\{(E+M+2\tau M)[m(4 p^2-Q^2) - Q^2 E]\tau\,G_M- \nonumber\\
&& -2E\big[4m\tau(p^2-E M) - Q^2(M+E+2\tau E) - Q^2 m (1+2\tau)\big]\big\}, \nonumber\\
{\cal D}T_{xz}& =& \frac{x m }{3}\Big\{\frac{\tau Q^2\,G_M}{m p}\big[m(E+m)-yM(E+M)(m+M)^2\big]G_M  \nonumber\\
&&+4M\big[(1+2\tau)[p-y z M (M+E)] + \frac{y z Q^2}{2}\big]\widetilde{G}\Big\}, \nonumber\\
{\cal D}T_{zz}& =&-\frac{Q^2\,G_M }{3}\Big\{\frac{\tau(z+p)}{p}\big[m(E+m)-yM(E+M)(m+M)^2\big]G_M \nonumber\\
&&+\frac{E}{m}\Big[Q^2\big[y(m+M)^2 + y z (m+M)\frac{E+m}{p}-\frac{(E+m)^2}{p^2}-\frac{m}{M}\big]+4m^2\Big]\widetilde{G}\Big\}.
\label{eq:Tij}
 \end{eqnarray}

\begin{figure}
\centering
\includegraphics[width=0.27\textwidth]{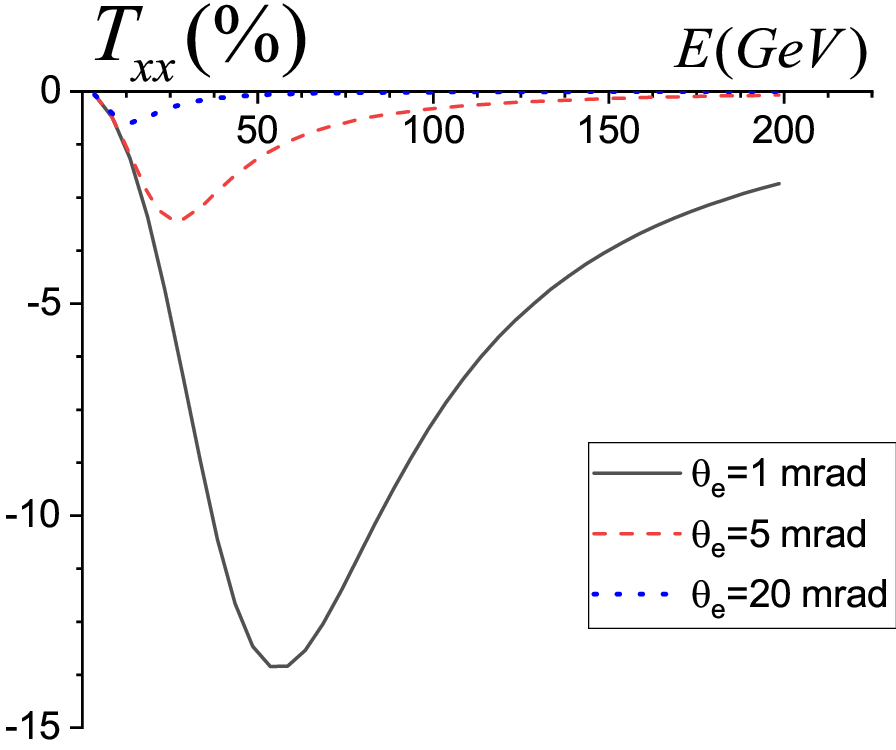}
\includegraphics[width=0.27\textwidth]{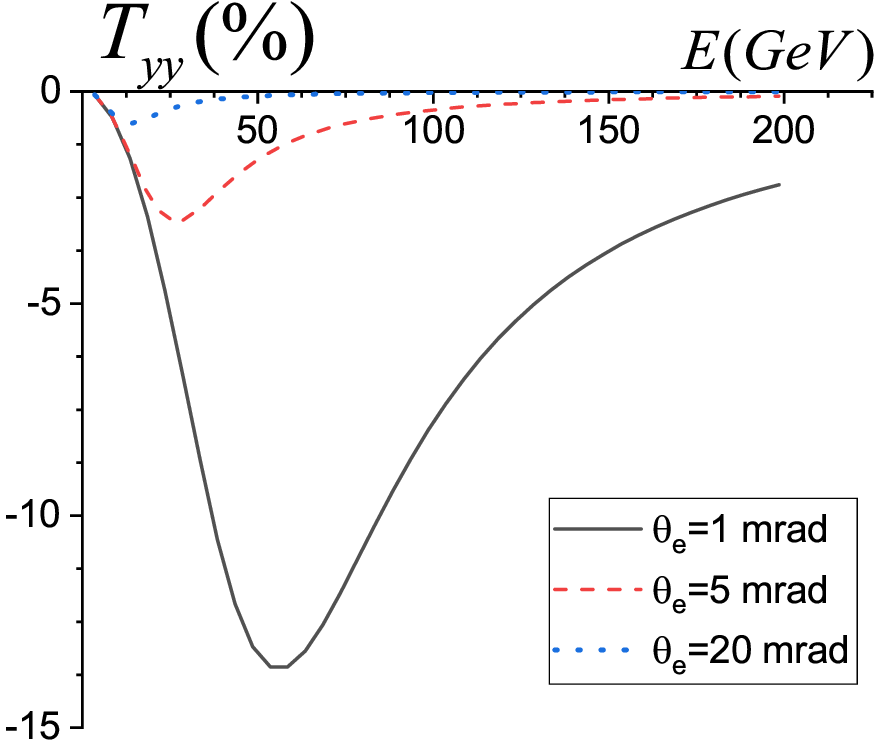}
\includegraphics[width=0.27\textwidth]{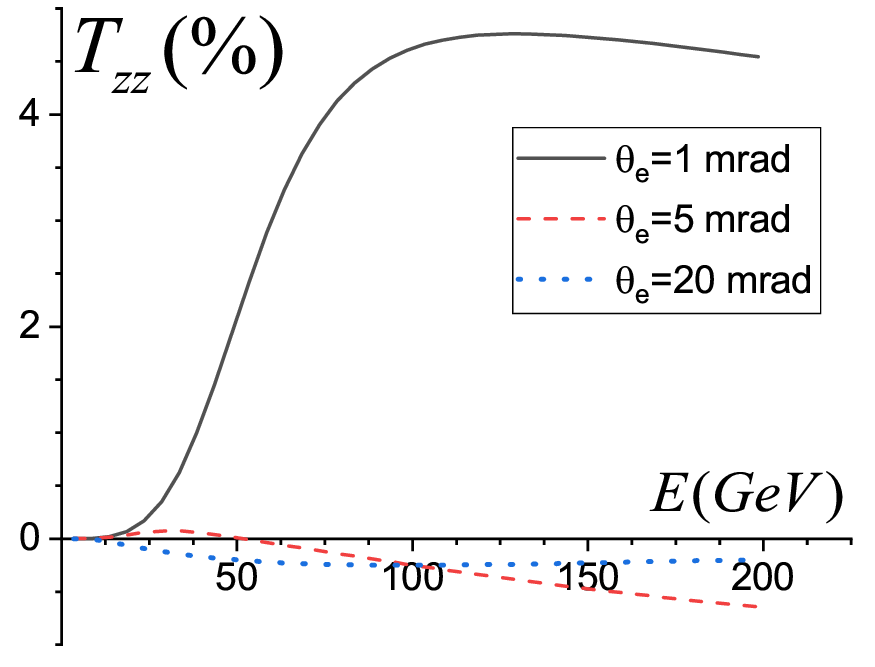}

\vspace{0.2 cm}

\includegraphics[width=0.27\textwidth]{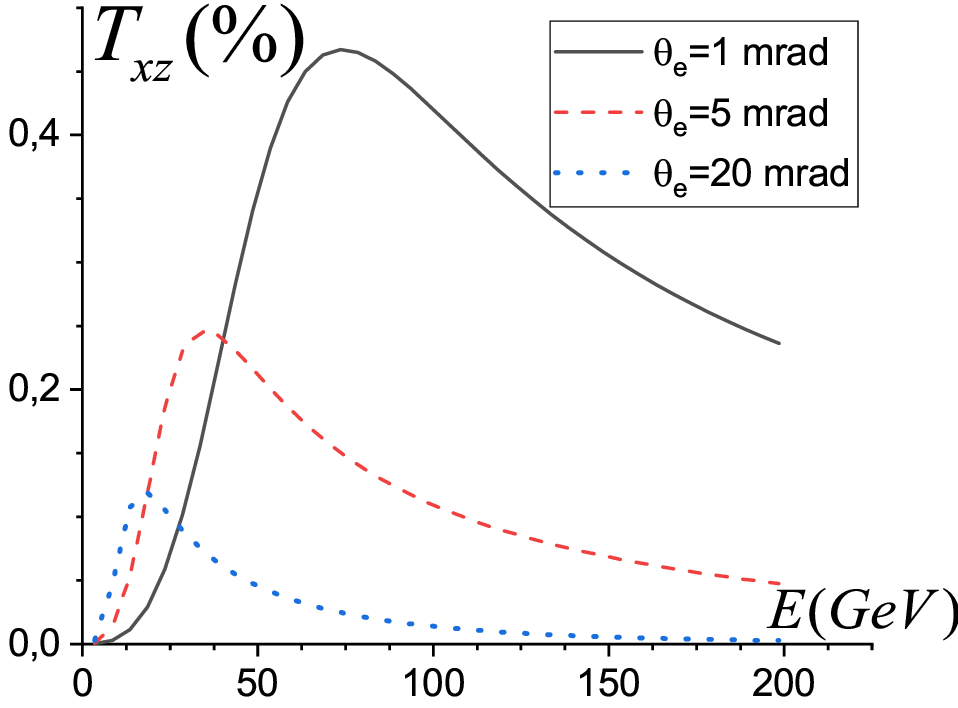}
\includegraphics[width=0.27\textwidth]{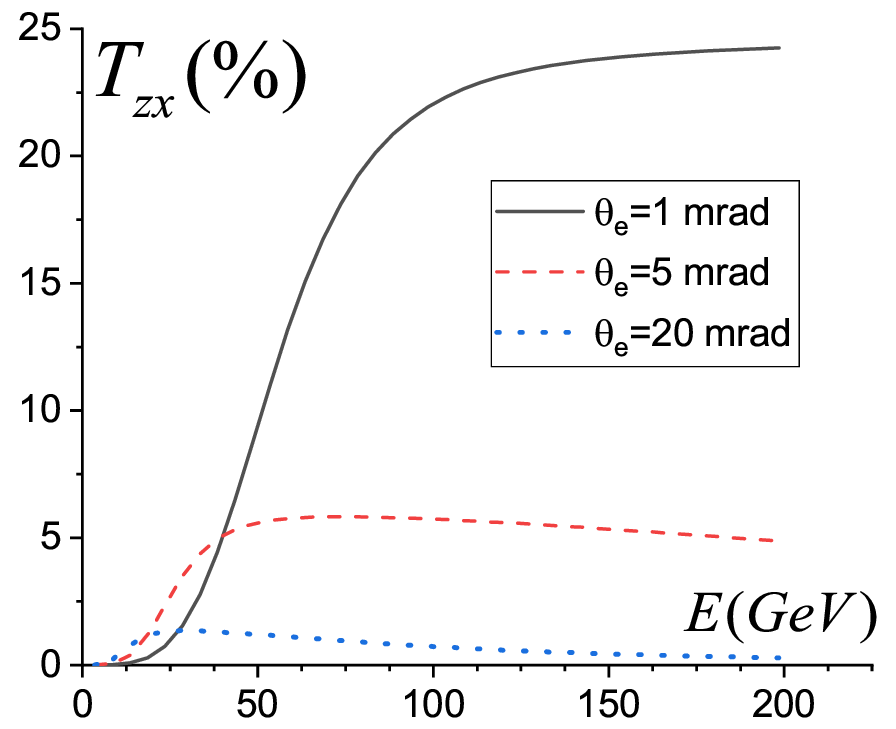}

\vspace{0.5 cm}
\includegraphics[width=0.27\textwidth]{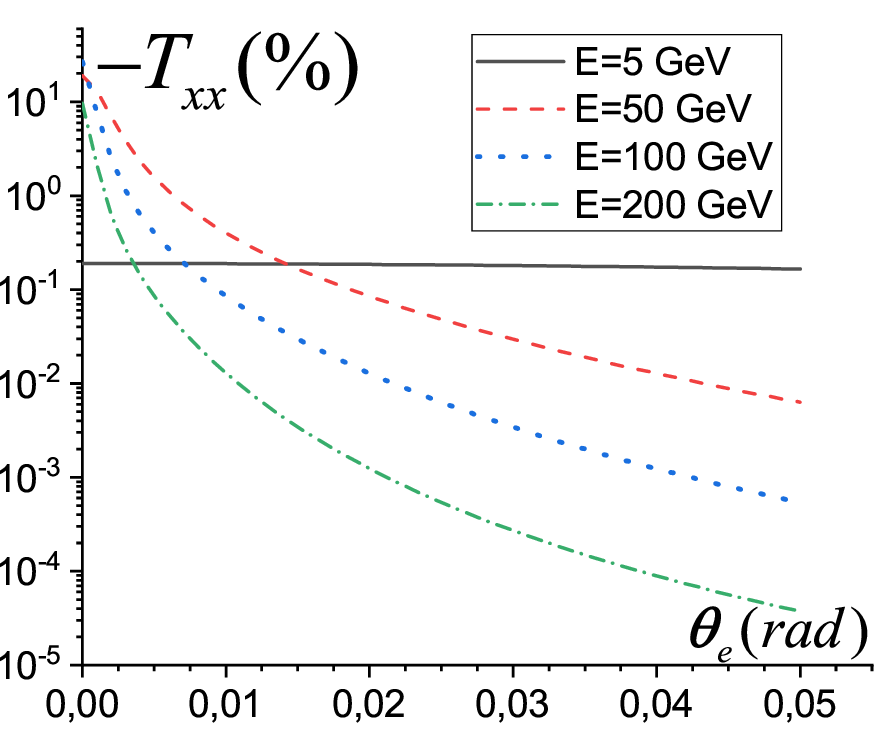}
\includegraphics[width=0.31\textwidth]{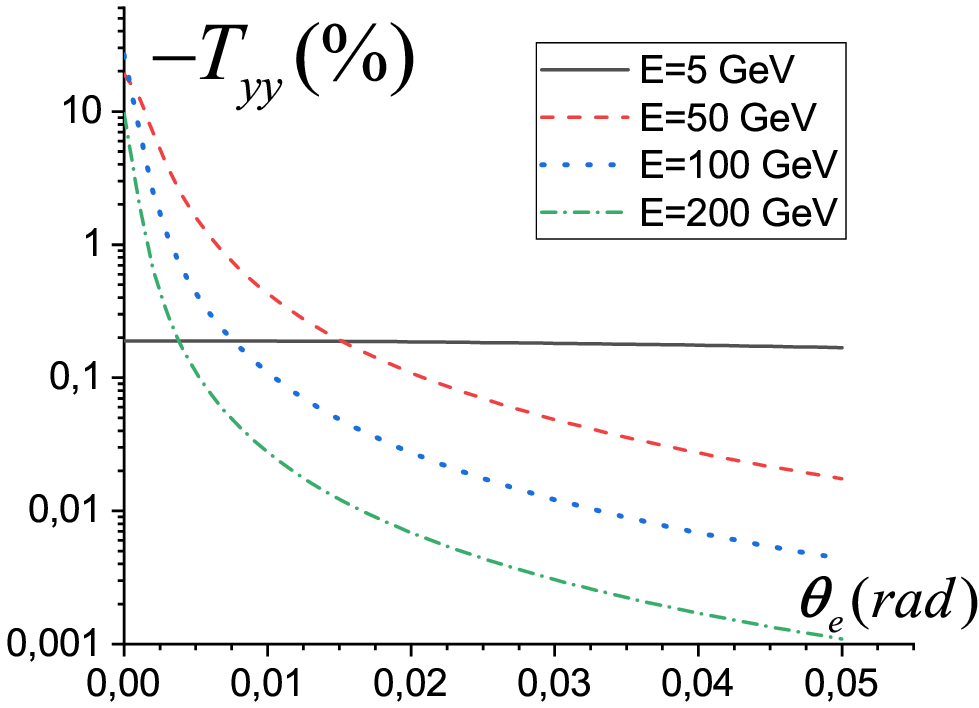}
\includegraphics[width=0.27\textwidth]{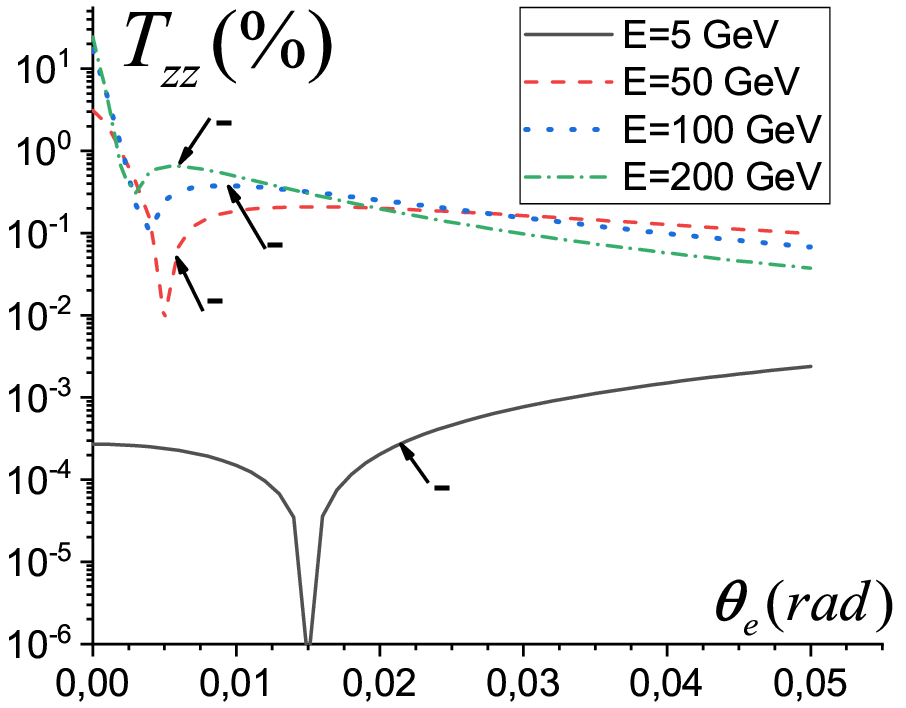}

\vspace{0.2 cm}

\includegraphics[width=0.31\textwidth]{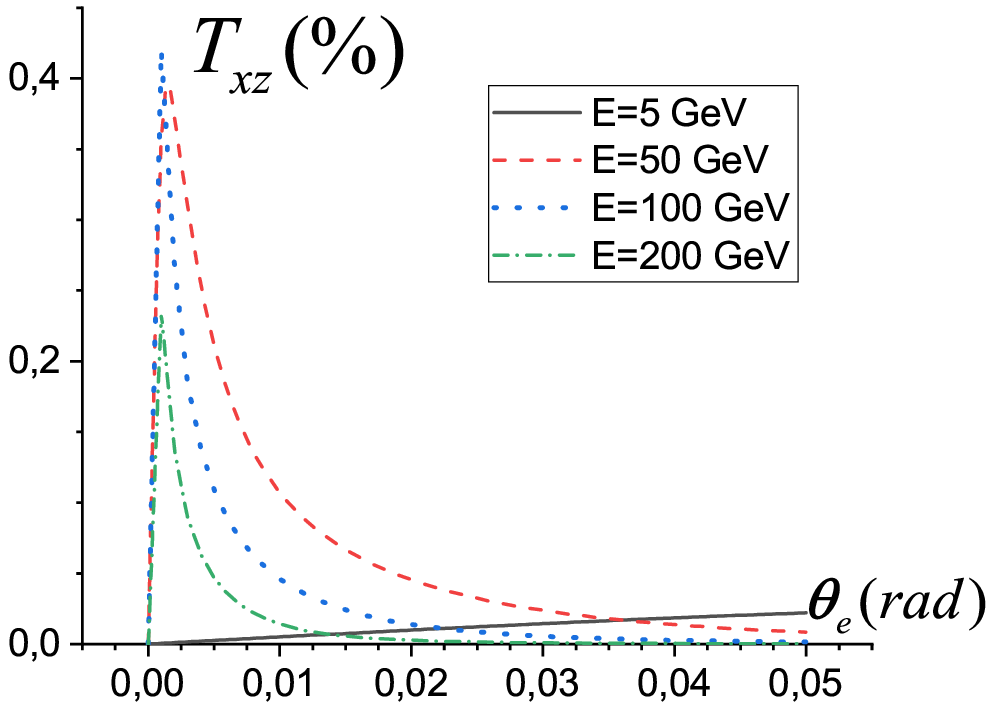}
\includegraphics[width=0.29\textwidth]{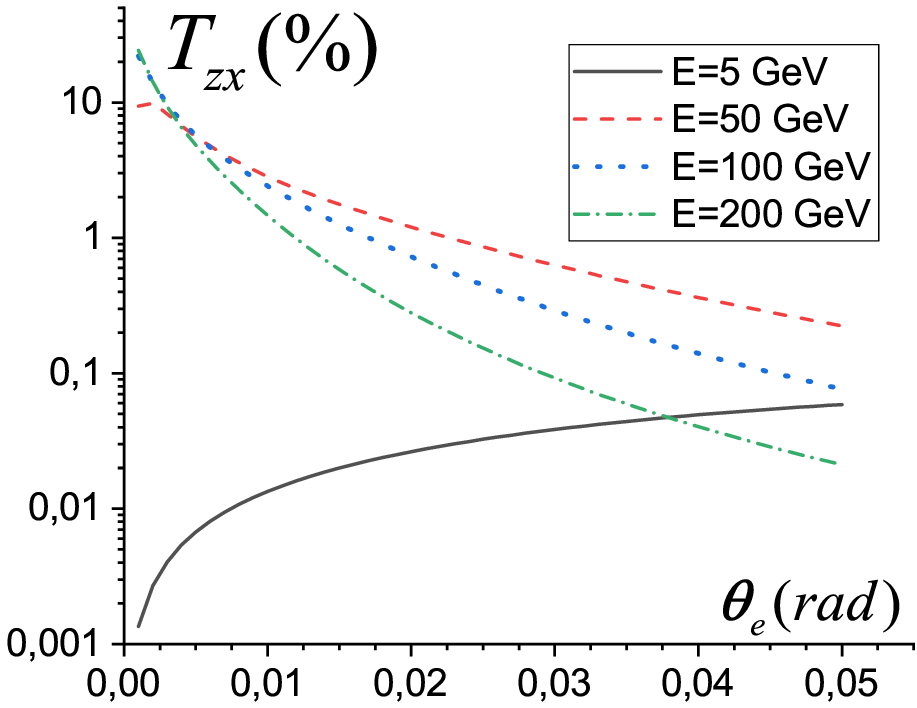}

 \parbox[t]{0.9\textwidth}{\caption{Polarization transfer coefficients from the target electron to the scattered deuteron, given by Eqs.\,(\ref{eq:Tij}), when the target electron has
 arbitrary polarization and the vector polarization of the scattered deuteron is measured.}\label{fig.Tij}}
\end{figure}

The corresponding transfer coefficients are plotted in Fig.\,\ref{fig.Tij}.

\subsection{Polarization coefficients, $c_{ij}$, which describe the correlation between the vector polarization of the scattered deuteron and the arbitrary polarization of the recoil electron in the $ d+e \to \vec d\,^V+\vec{e} $ reaction}

The scattering of an unpolarized deuteron beam by unpolarized electrons is considered here. In this case a correlation exists between
the vector polarization of the scattered deuteron and polarization of the recoil electron. The part of the hadronic tensor $H_{\mu\nu}(\eta_2)$ related to the vector polarized scattered deuteron and unpolarized deuteron beam can be obtained from Eq. (\ref{eq:Heta1}) by the substitutions
 $\eta_1\to \eta_2$ and  $p_1\rightleftarrows - p_2$, so that:
\begin{equation}\label{eq:Heta2}
H_{\mu\nu}(\eta_1) =2iM\,G_M\Big[(1+\tau)\widetilde{G}<\mu\nu \eta_2 k> -\frac{\eta_2\cdot k}{4M^2}(G_M-2\widetilde{G})<\mu\nu p_2 k>\Big].
\end{equation}
The leptonic tensor, $L_{\mu\nu}$, which corresponds to an unpolarized initial electron target and polarized
recoil electron, has the form:
\begin{equation}\label{eq:Ls2}
L_{\mu\nu}=\frac{1}{2}\big[L_{\mu\nu}^{(0)}+L_{\mu\nu}^{(p)}(s_2)\big], \ \ L_{\mu\nu}^{(p)}(s_2)=2im\epsilon_{\mu\nu\rho\sigma}k_{\rho}s_{2\sigma}.
\end{equation}

The contraction of the spin-dependent leptonic $ L_{\mu\nu}^{(p)}(s_2)$
and hadronic $ H_{\mu\nu}(\eta_2)$ tensors, in an arbitrary reference frame, gives:
\begin{equation}\label{eq:Ceta2s2}
C(s_2,\eta_2)=L_{\mu\nu}^{(p)}(s_2)H_{\mu\nu}(\eta_2)= -\frac{4}{3}mM G_M \big \{ \tau k\cdot \eta_2 (2p_2\cdot s_2-k\cdot s_2)G_M+
\end{equation}
$$+2\left (G_C +\frac{\tau}{3}G_Q\right ) \left [k^2(1+\tau) s_2\cdot \eta_2 -k\cdot \eta_2 (k\cdot s_2 +2\tau p_2\cdot s_2
)\right ]\big \}. $$

One can see that all the spin correlation coefficients in $d+e\to\vec d^V+\vec e$ reaction,
when the scattered deuteron has vector polarization and the recoil electron has arbitrary polarization,
are proportional to the deuteron magnetic form factor. It is also true for the $\vec{d}^V +e\to d + \vec{e}$ reaction as well for the $ed$ scattering for the
corresponding polarization observables.


The differential cross section can be written as:
\begin{equation}\label{eq:DSeta2s2}
\frac{d\sigma}{dQ^2}(\vec{S}_2, \vec\xi_2)= \frac{1}{2}\left
(\frac{d\sigma}{dQ^2}\right )_{un}\left [1+c_{xx} S_{2x}\xi_{2x}+
c_{yy} S_{2y}\xi_{2y}+ c_{zz} S_{2z} \xi_{2z}+c_{xz} S_{2x} \xi_{2z}+
c_{zx}S_{2z} \xi_{2x}\right ].
\end{equation}
The explicit expressions  of the corresponding correlation coefficients $c_{ij}$ are
\begin{eqnarray}
{\cal D} c_{yy} &=& -\frac{4}{3}Q^2 m M (1+\tau)\,G_M\,\widetilde{G},\label{eq:cij}\\
{\cal D} c_{xx}& =& \frac{M}{6 m (1+\lambda)}\big(B_{xx}\,G_M^2 + 4\tilde{B}_{xx}\,G_M\widetilde{G}\big),\nonumber \\
B_{xx}&=&yx^2M\tau[Q^2-4m(2E+m)][E+(1+2\tau)M], \nonumber \\
\tilde{B}_{xx} &= &2yx^2 M [(1+2\tau)E+M](m^2-\tau M^2)+ 2y\tau m[E(E-M)-2(1+\tau)M^2] -
\nonumber \\
&&-2(1+\lambda)(1+\tau)Q^2m^2,\nonumber \\
{\cal D} c_{xz}&=&\frac{4xy\tau M^2}{3m^2 p (1+\lambda)}\big( B_{xz}\,G_M^2 -2\tilde{B}_{xz}\,G_M\widetilde{G}\big), \nonumber  \\
B_{xz}&=&[E+(1+2\tau)M]\big\{ m[m^2 p^2-\tau M^2(mE +M^2)]-\tau M^2(E+m)[m(E+m)-\tau M^2]\big\},\nonumber\\
\tilde{B}_{xz}&=& \tau E M^4[E(1+2\tau)+M] + m^3[E^2(E-M)-2E(1+\tau)M^2]+\nonumber\\
&&+m^2 M^2[4\tau(1+\tau)M^2-(1-2\tau)E M -E^2(1+4\tau)] +
\nonumber\\
&&+ \tau m M^2[2M^3+2EM^2(2+3\tau)-E^2(E-M)],\nonumber\\
{\cal D} c_{zx}&=&\frac{4xy\tau M^3}{3m^2 p (1+\lambda)}\big(B_{zx}\tau G_M^2 - 2\tilde{B}_{zx}\,\widetilde{G}\big),\nonumber\\
B_{zx}& = &[\tau M^2-m(2E +m)]\{M^2[E(1+2\tau)+2m(1+\tau)+M] -mE(E-m)\},\nonumber\\
\tilde{B}_{zx} &=& [E+(1+2\tau)M][\tau M^4 -mE(m^2-3\tau M^2)] +
\nonumber\\
&&
+m^2\{(1+2\tau)M[(1+2\tau)M^2-E(2E-M)]-2E^2\},\nonumber\\
{\cal D} c_{zz}&=&\frac{8y\tau M^3}{3m^3 p^2 (1+\lambda)}\big(B_{zz}\tau\,G_M^2 -2[E+(1+2\tau)M]G_M\,\widetilde{G}\big),\nonumber\\
B_{zz}&=& \{M^2[E(1+2\tau)+2m(1+\tau)+M]-mE(E-M)\}\times\nonumber\\
&&\{m^3p^2-\tau M^2[(E+m)(mE+m^2-\tau M^2)+m(mE+M^2)]\},\nonumber\\
\tilde{B}_{zz}&=& m^4E(p^2-\tau M^2)+\tau m^2 M^2[M^2(5\tau E +2(1+\tau)m)-E^2(E+4m)]+\nonumber\\
&&+\tau M^4[EM^2+2m(E^2+M^2)].\nonumber
\end{eqnarray}

\begin{figure}
\centering
\includegraphics[width=0.27\textwidth]{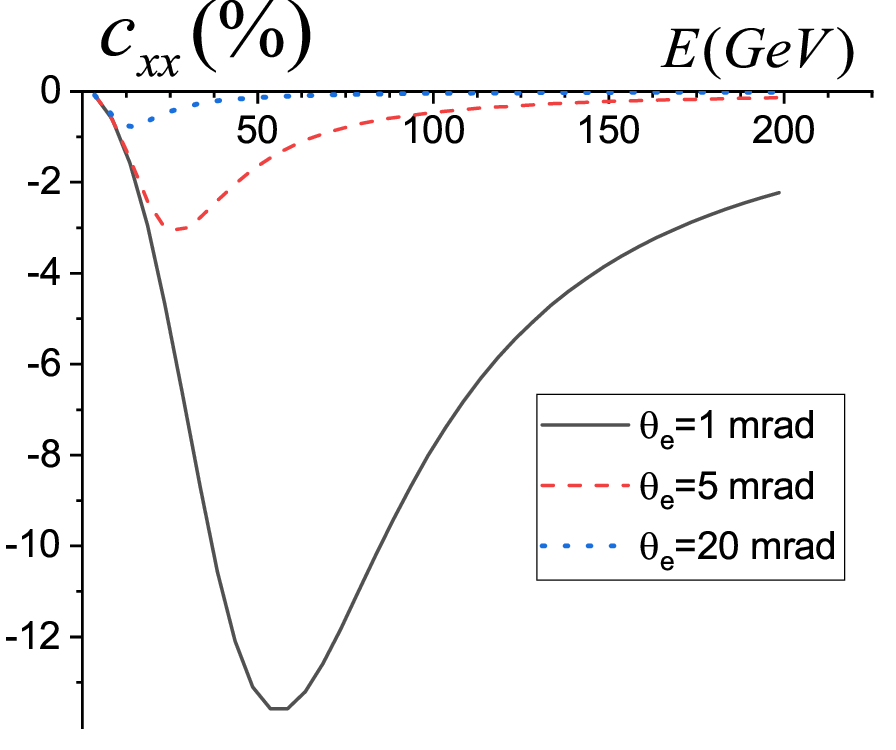}
\includegraphics[width=0.27\textwidth]{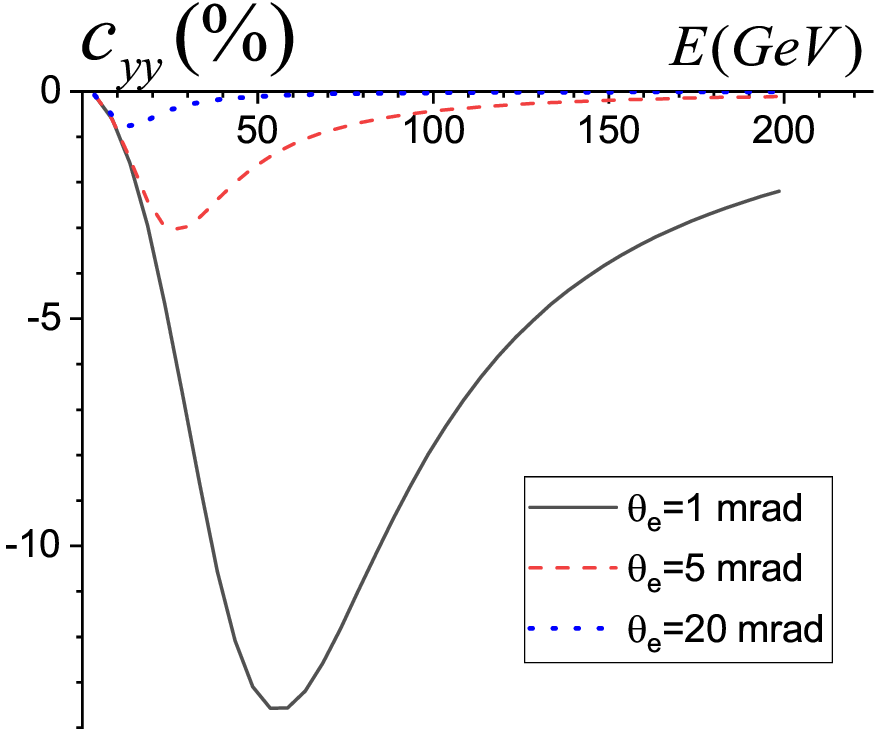}
\includegraphics[width=0.27\textwidth]{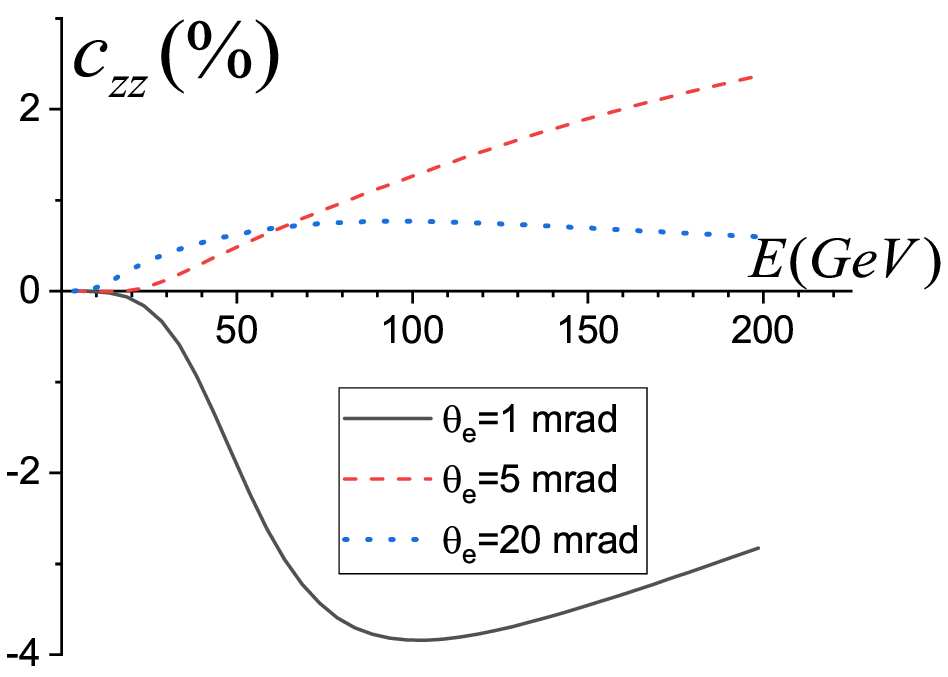}

\vspace{0.2 cm}

\includegraphics[width=0.27\textwidth]{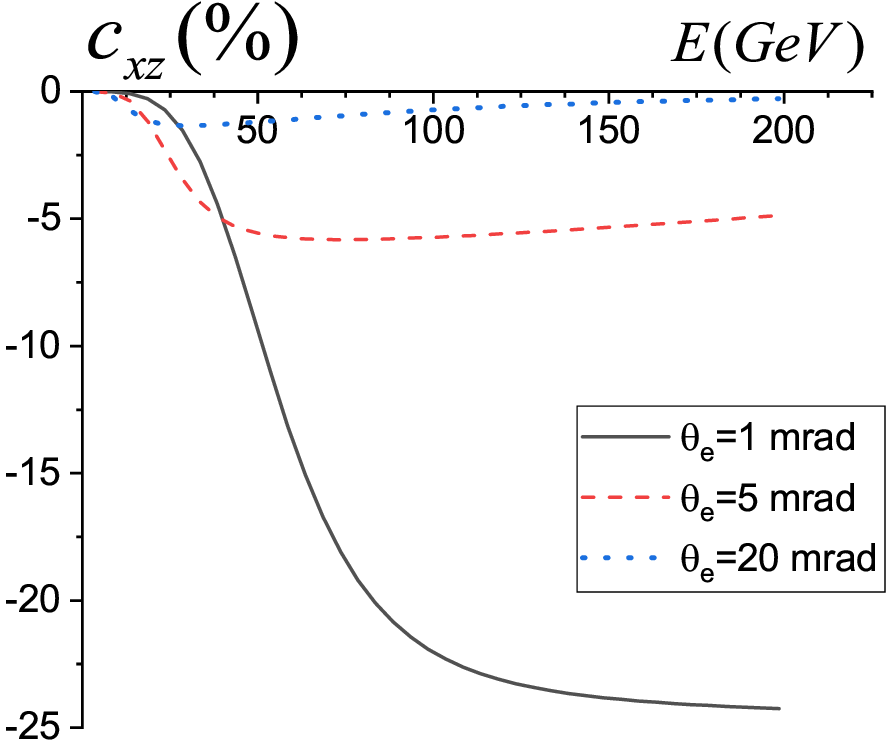}
\includegraphics[width=0.27\textwidth]{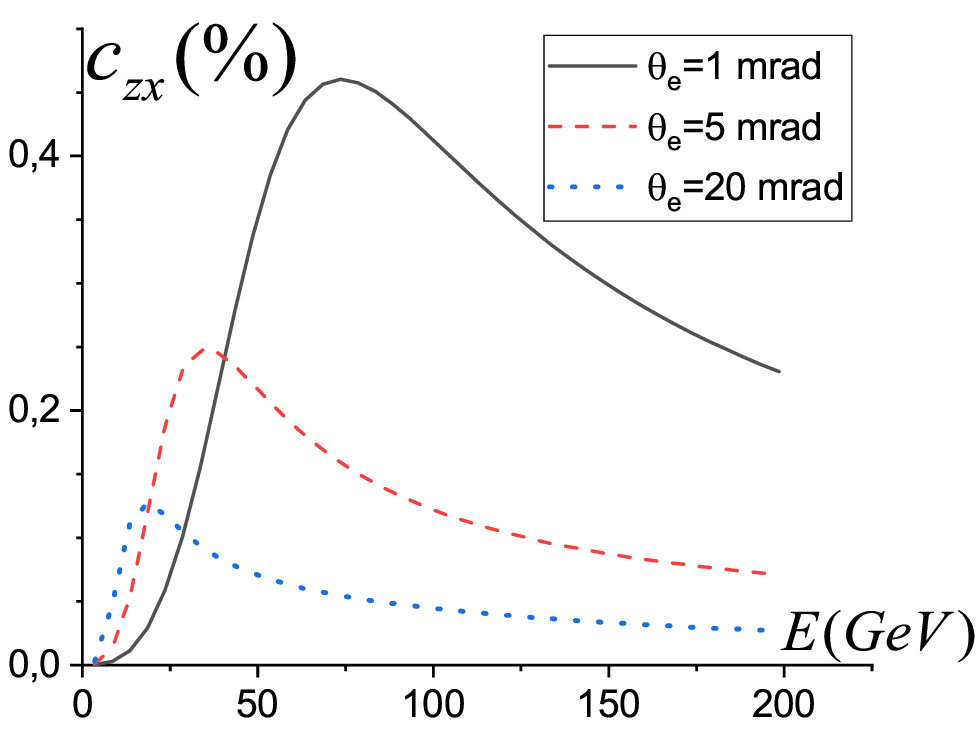}

\vspace{0.5 cm}
\includegraphics[width=0.27\textwidth]{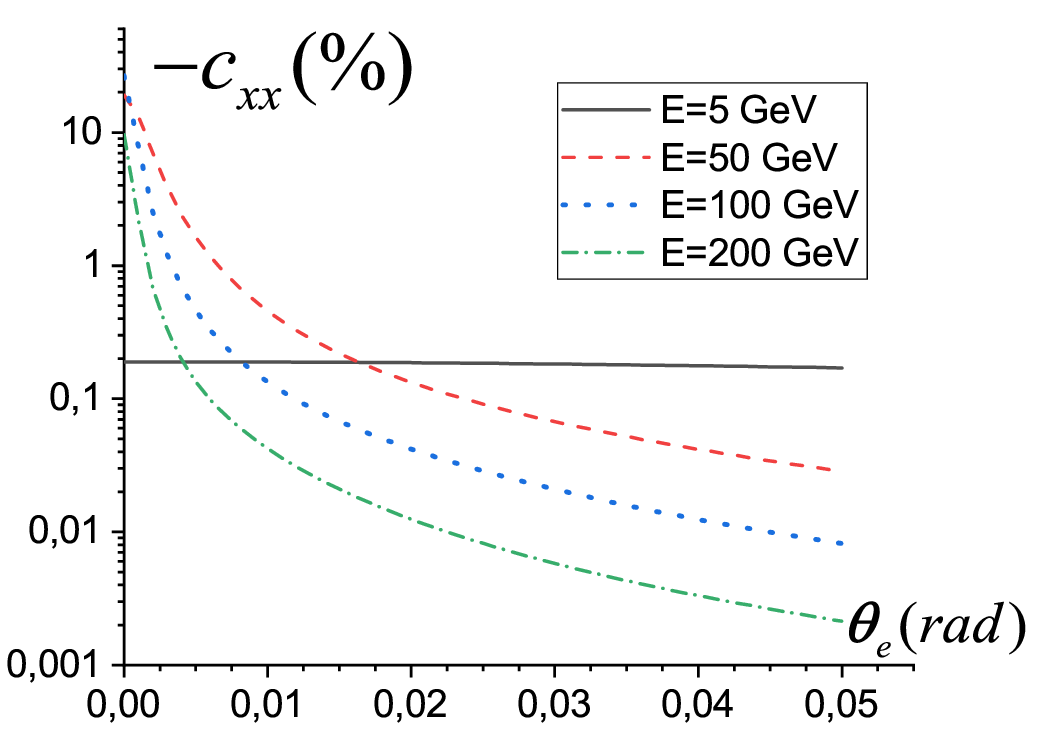}
\includegraphics[width=0.31\textwidth]{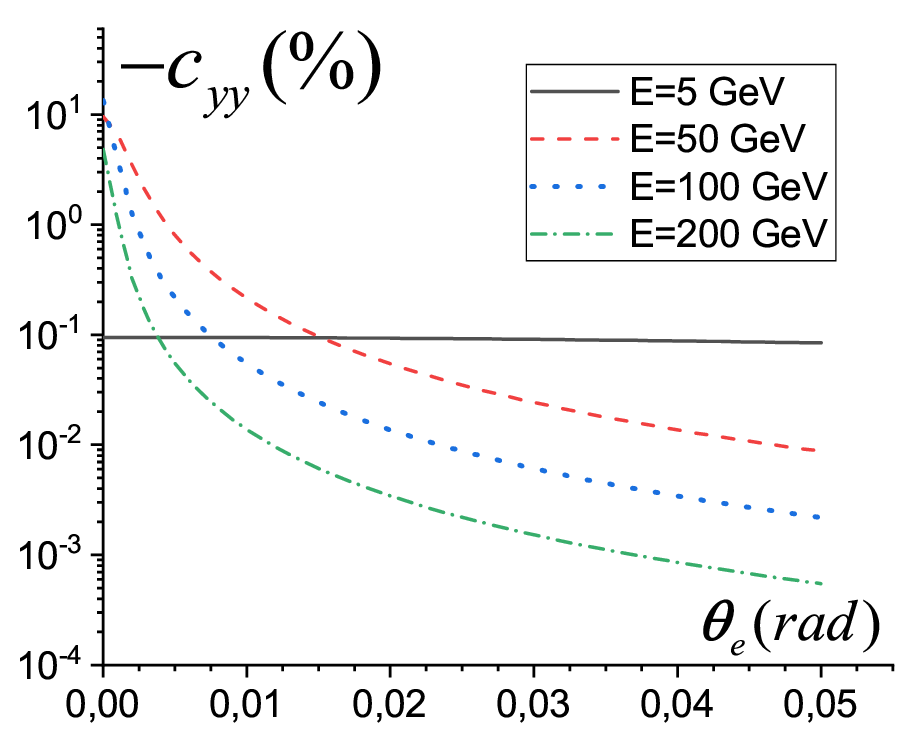}
\includegraphics[width=0.27\textwidth]{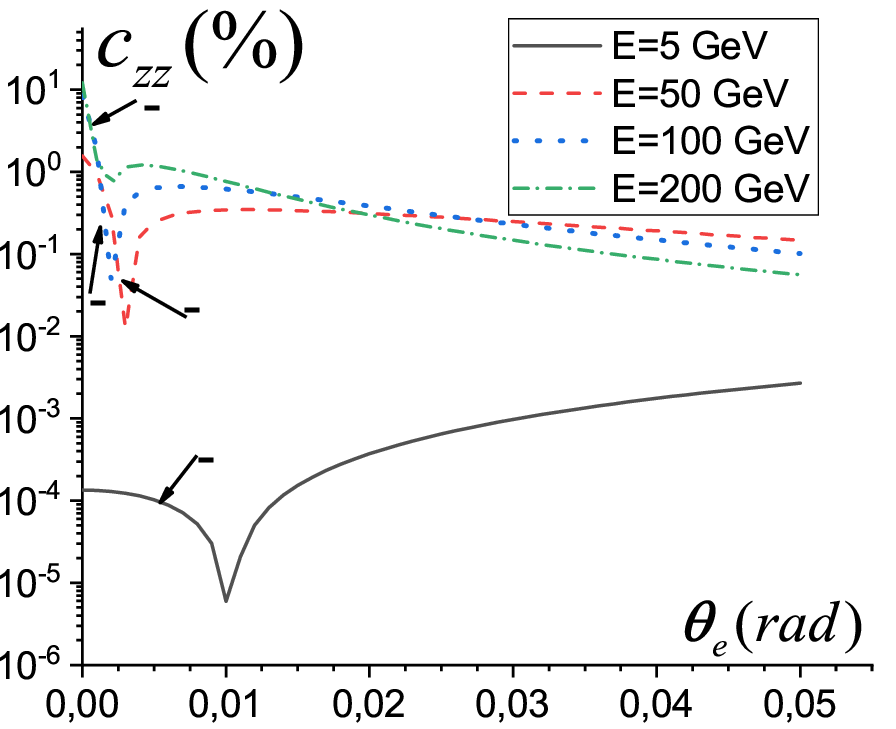}

\vspace{0.2 cm}

\includegraphics[width=0.27\textwidth]{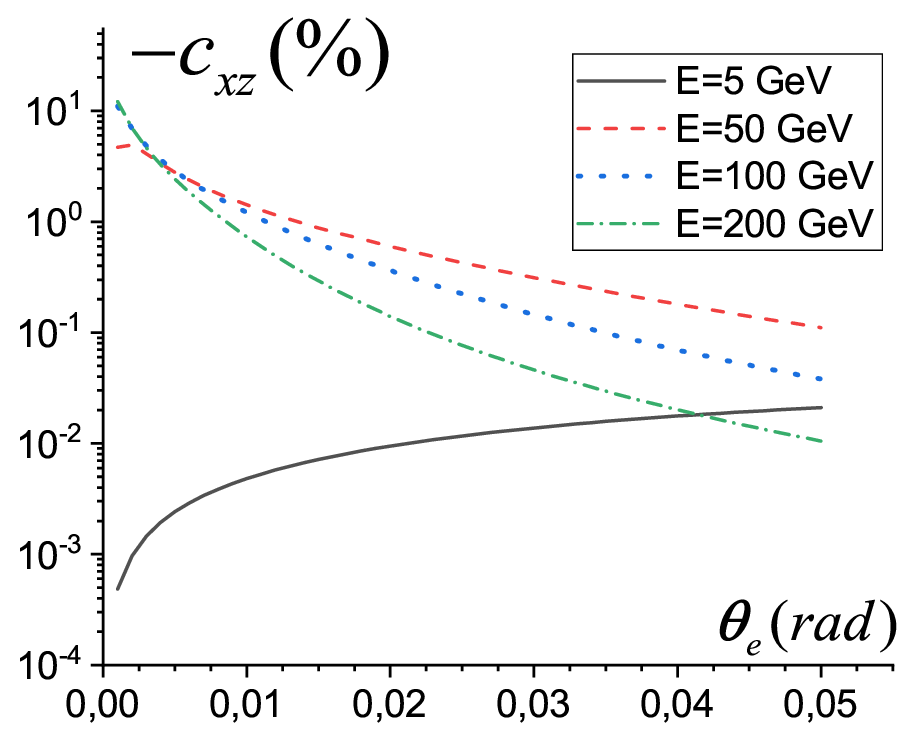}
\includegraphics[width=0.31\textwidth]{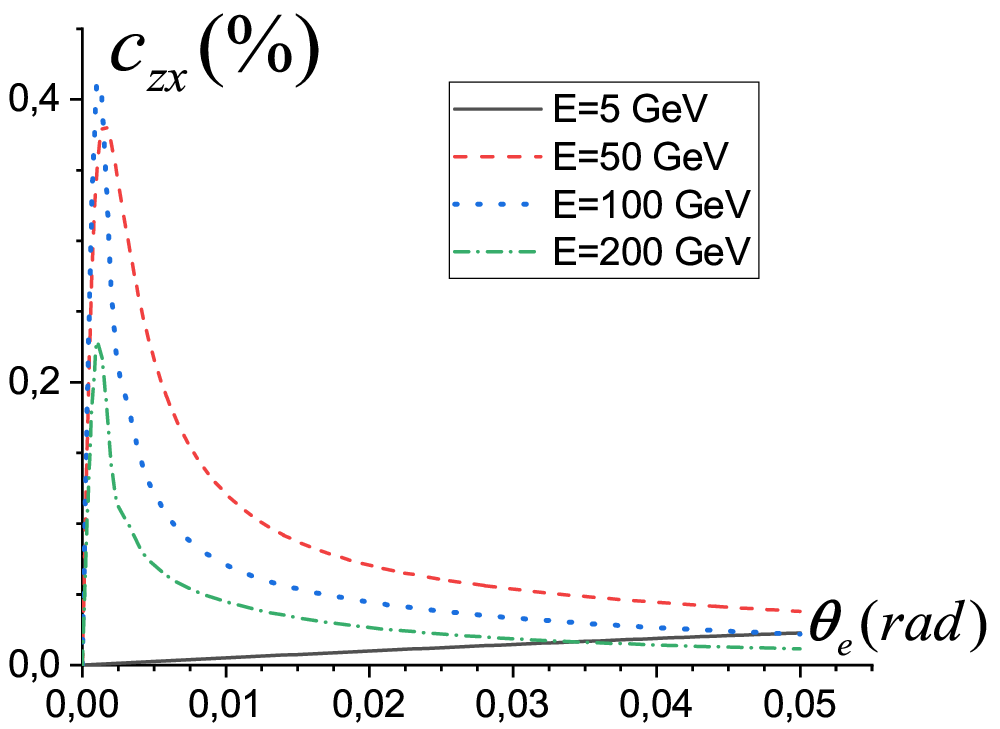}

\parbox[t]{0.9\textwidth}{\caption{Correlation coefficients due the vector polarizations  of the scattered deuteron and the recoil electron
 in $d+e$ scattering, as  given by Eqs.\,(\ref{eq:cij}).}\label{fig.cij}}
\end{figure}

The correlation coefficients $c_{ij}$ between the vector polarizations of the scattered deuteron and the recoil electron are shown in Fig.\,\ref{fig.cij}.


\subsection{Polarization transfer coefficients $\widetilde{T}_{ij}$ which describe arbitrary polarization of the recoil electron when initial deuteron has the vector polarization in the $e +\vec{d}\,^V \to \vec{e} + d $ reaction.}

The vector polarization transfer from the initial vector polarized deuteron to the recoil electron is calculated below. In this case the polarization dependent
part of the hadronic tensor is defined by Eq.\,(\ref{eq:Heta1}) and the corresponding part of the leptonic one by Eq.\,(\ref{eq:Ls2}). The convolution of these polarization dependent terms can be obtained from $C(s_1,\eta_1)$ (see Eq.\,(\ref{eq:Ceta1s1})) by the substitution $s_1\to s_2.$

The corresponding differential cross section can be written in terms of the polarization transfer coefficients $\widetilde{T}_{ij}$ as :
\begin{equation}\label{eq:Dseta1s2}
\frac{d\,\sigma}{d\,Q^2}= \frac{1}{2}\Big(\frac{d\,\sigma}{d\,Q^2}\Big)_{un}\big[1+S_{1x}\xi_{2x}\widetilde{T}_{xx} +S_{1y}\xi_{2y}\widetilde{T}_{yy}
+ S_{1z}\xi_{2z}\widetilde{T}_{zz} + S_{1x}\xi_{2z}\widetilde{T}_{xz} + S_{1z}\xi_{2x}\widetilde{T}_{zx}\big].
\end{equation}
The explicit expressions of the $\widetilde{T}_{ij}$ are
\begin{eqnarray}
{\cal D}\widetilde{T}_{yy} &=& -4m M Q^2(1+\tau)\,G_M\,\widetilde{G},\nonumber\\
{\cal D}\widetilde{T}_{xx} &=& 4m M\Big\{x^2\tau\Big(\frac{E+m}{m(1+\lambda)}-\frac{1}{2}\Big)G_M^2-\Big[x^2\Big(1+
\frac{2(\tau E-m)}{m(1+\lambda)}\Big)+Q^2(1+\tau)\Big]G_M\,\widetilde{G}\Big\},\nonumber\\
{\cal D}\widetilde{T}_{zz}& =& \frac{8\tau M^2}{m^2}\Big\{\Big[mE(\tau M^2-m^2)-m^2 M^2(1-\tau)+,\nonumber\\
&&+\frac{M^2(E+m)^2}{p^2}\Big(2M^2-mE
- \frac{2\tau(\tau M^4-m^3E)}{m^2(1+\lambda)}\Big)\Big]\tau G_M^2 +
\nonumber\\
&&
+ \Big[2 \tau m^2 M^2+mE(\tau M^2-m^2) +,\nonumber\\
&&+\frac{\tau M^2(E+m)^2}{p^2}\Big(\frac{2(m^3 E+\tau M^4)}{m^2(1+\lambda)}-m E - 2\tau M^2\Big)\Big]2\,G_M\,\widetilde{G}\Big\},\nonumber\\
{\cal D}\widetilde{T}_{xz} &=& 4 \tau x M\Big\{\Big[\frac{\tau M^2(E+m)}{p}\Big(\frac{2(E+m)}{m(1+\lambda)}-1\Big)-m p\Big] G_M^2 +,  \label{eq:Ttij}  \\
&&2\Big[mp -\frac{M^2(E+m)}{p}\Big(\frac{2m(\tau E-m)}{m^2(1+\lambda)} +1 \Big)\Big] G_M\,\widetilde{G}\Big\},\nonumber\\
{\cal D}\widetilde{T}_{zx}&=& \frac{4 \tau x M^2}{p}\Big\{\Big[M^2-Em +\frac{2 M^2 E}{m} + \frac{2(E+m)(m^3 E-\tau M^4)}{m^3(1+\lambda)}\Big] \tau G_M^2 +\nonumber\\
&&2\Big[M^2(1-2\tau)-mE-\frac{2\tau M^2 E}{m}+ \frac{2(E+m)(m^3 E + \tau^2 M^4)}{m^3(1+\lambda)}\Big] G_M\,\widetilde{G}\Big\}. \nonumber
\end{eqnarray}

\begin{figure}
\centering
\includegraphics[width=0.3\textwidth]{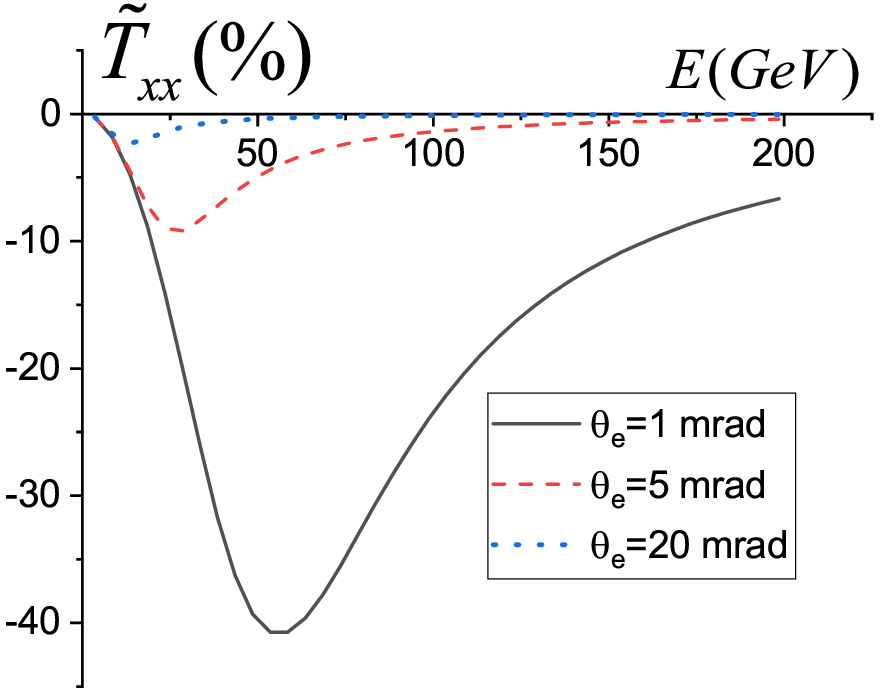}
\includegraphics[width=0.3\textwidth]{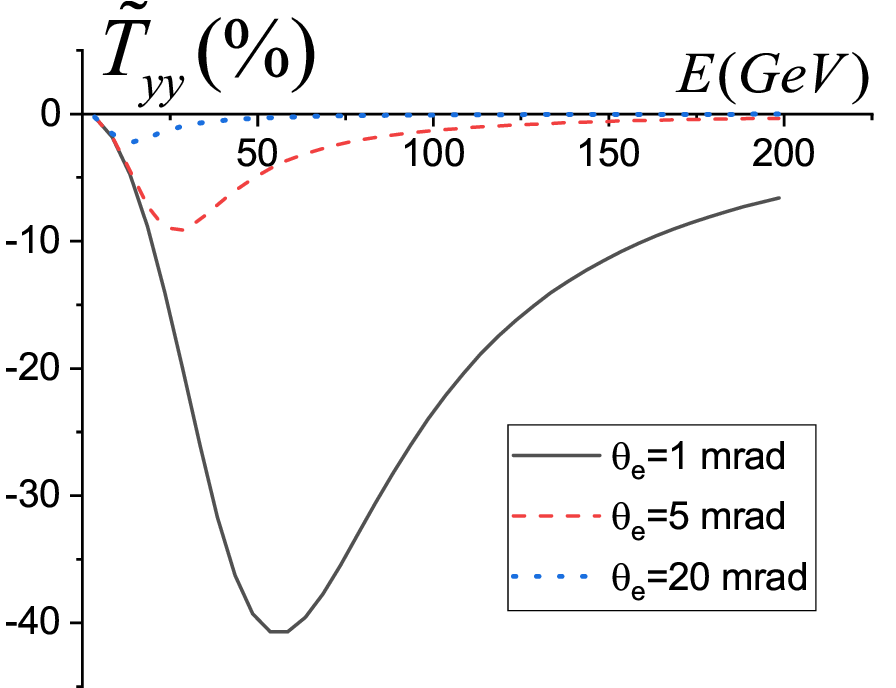}
\includegraphics[width=0.3\textwidth]{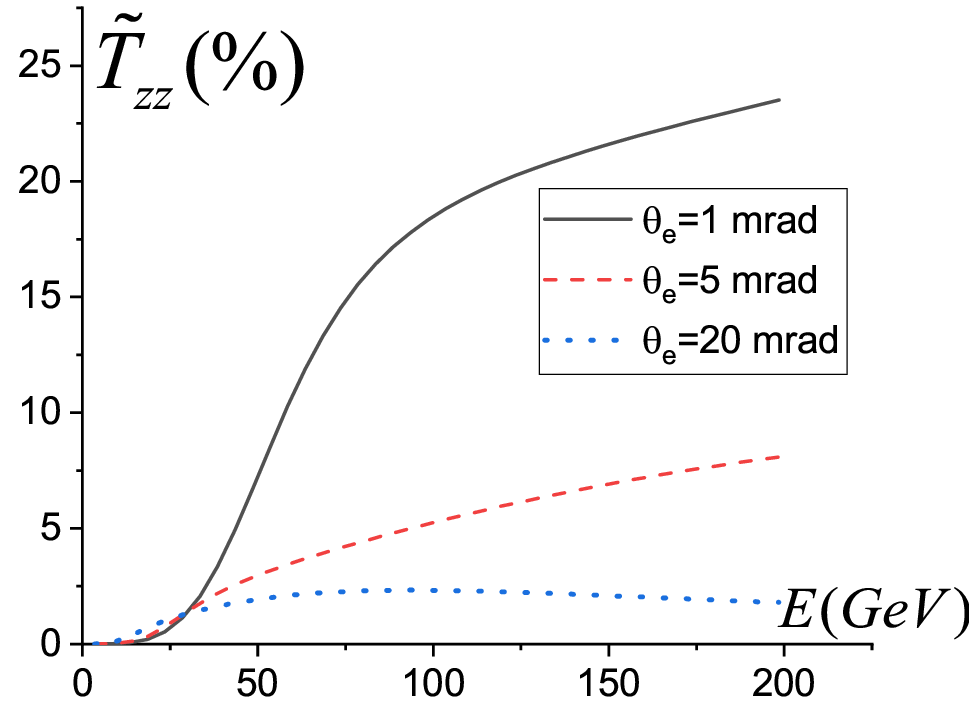}

\vspace{0.2 cm}

\includegraphics[width=0.3\textwidth]{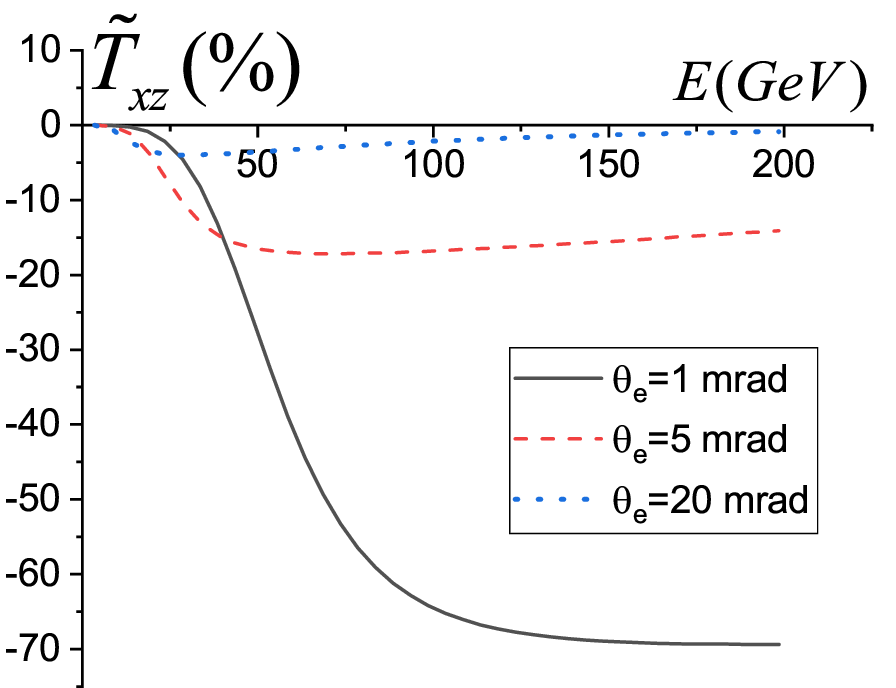}
\includegraphics[width=0.3\textwidth]{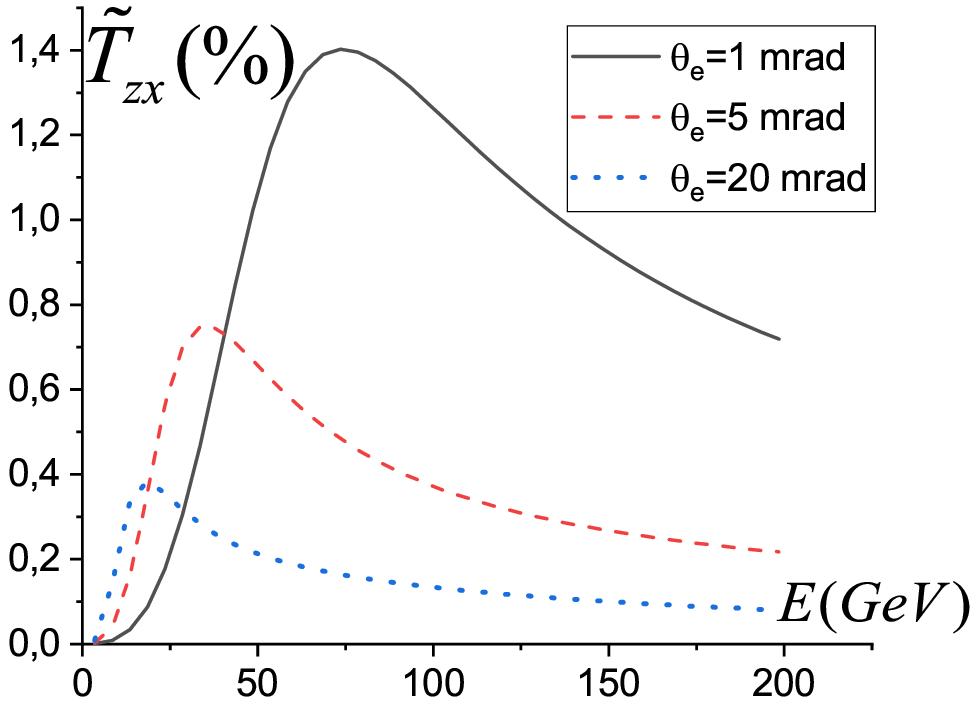}

\vspace{0.5 cm}
\includegraphics[width=0.3\textwidth]{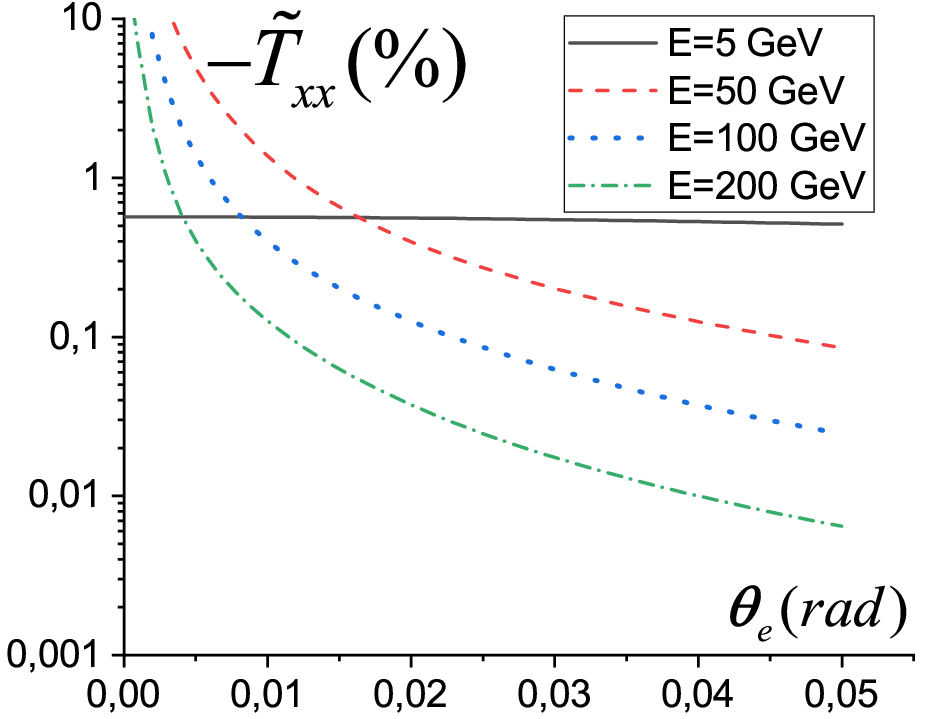}
\includegraphics[width=0.33\textwidth]{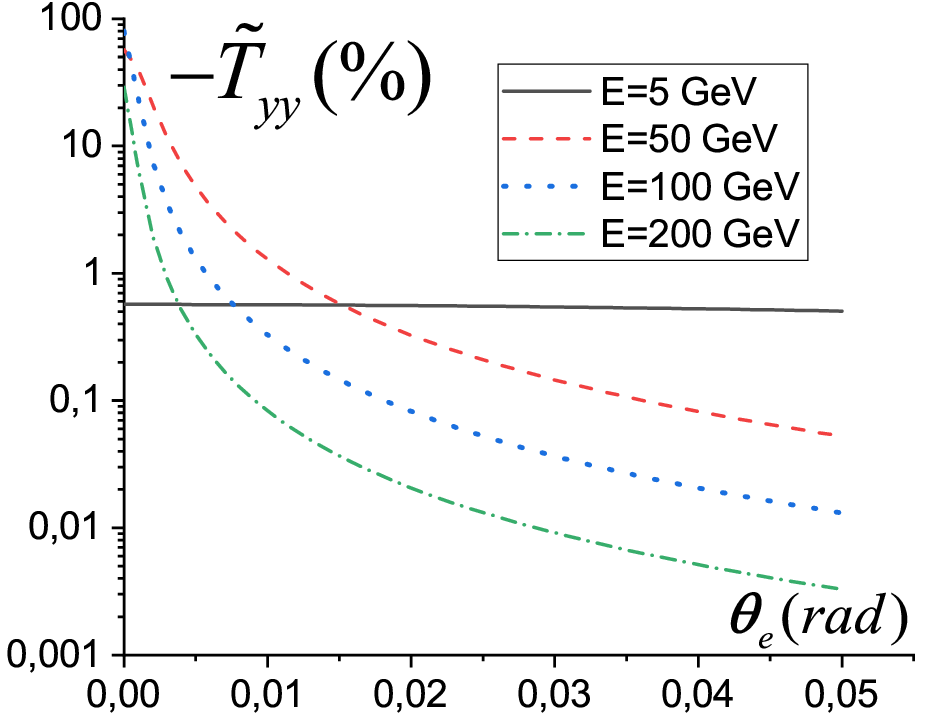}
\includegraphics[width=0.3\textwidth]{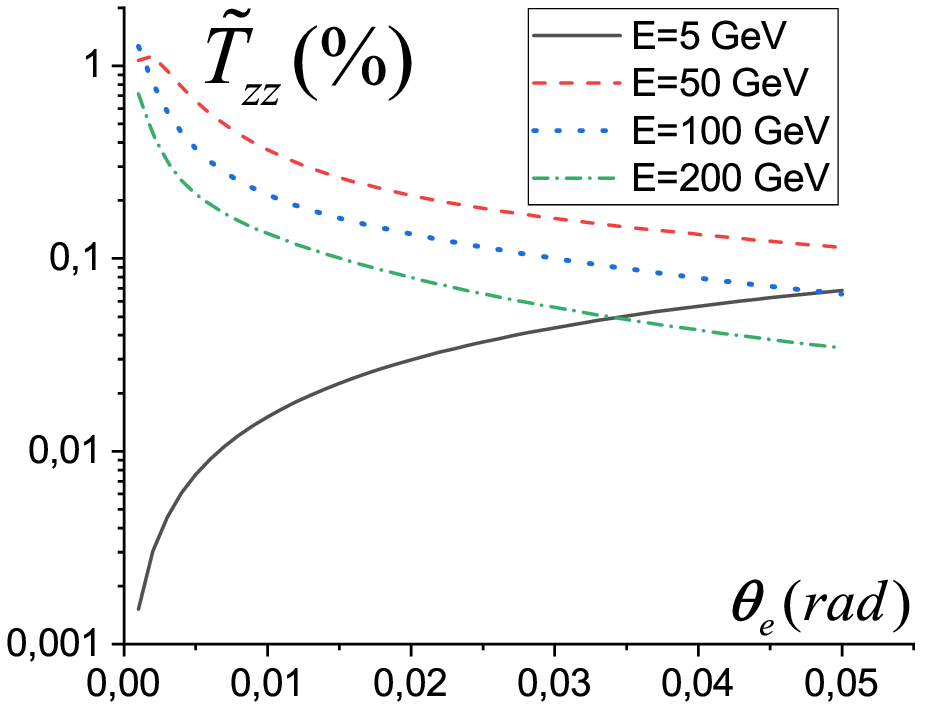}

\vspace{0.2 cm}

\includegraphics[width=0.3\textwidth]{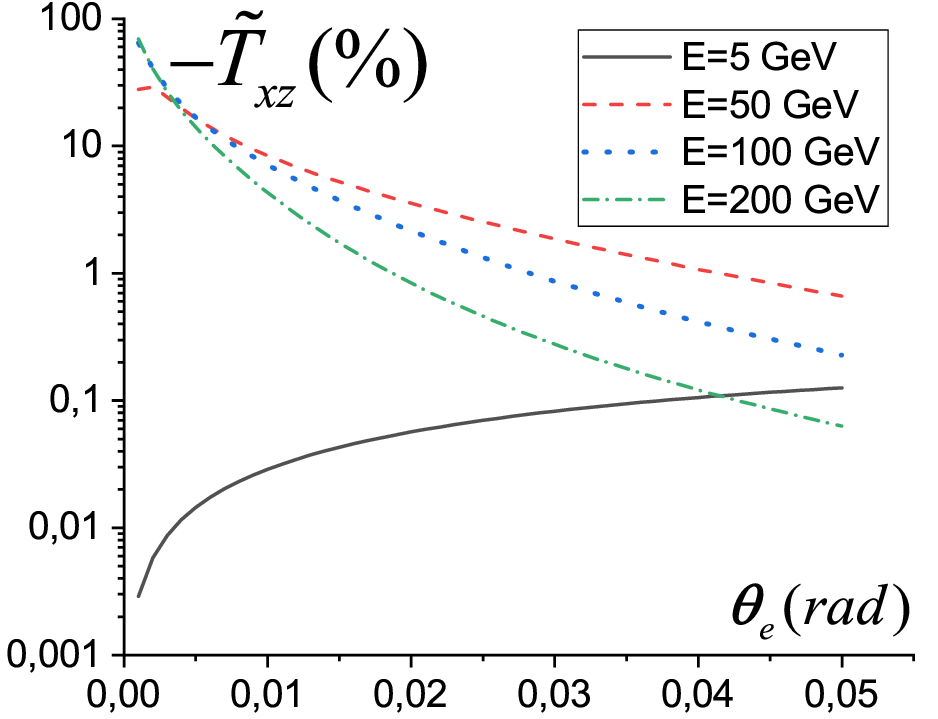}
\includegraphics[width=0.33\textwidth]{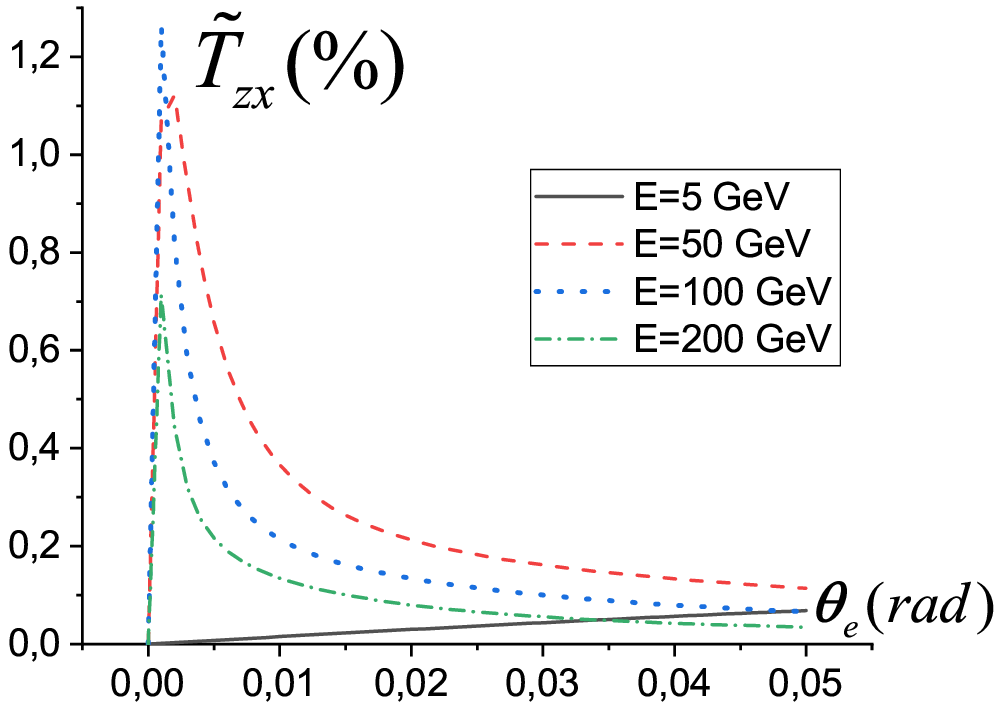}

 \parbox[t]{0.9\textwidth}{\caption{Polarization transfer coefficients which describe arbitrary polarization of the recoil electron when initial deuteron has the vector polarization in the $e +\vec{d}\,^V \to \vec{e} + d $ reaction given by Eqs.\,(\ref{eq:Ttij}).}\label{fig.wTij}}
\end{figure}

The polarization transfer coefficients $\widetilde{T}_{ij}$  are plotted in Fig.\,\ref{fig.wTij}.

\subsection{Polarization transfer coefficients $V_{ij}$ describing  the vector polarization  transfer  from the initial to the scattered deuteron  in the reaction $\vec d\,^V + e \to \vec d\,^V +e. $}

\begin{figure}
\centering
\includegraphics[width=0.3\textwidth]{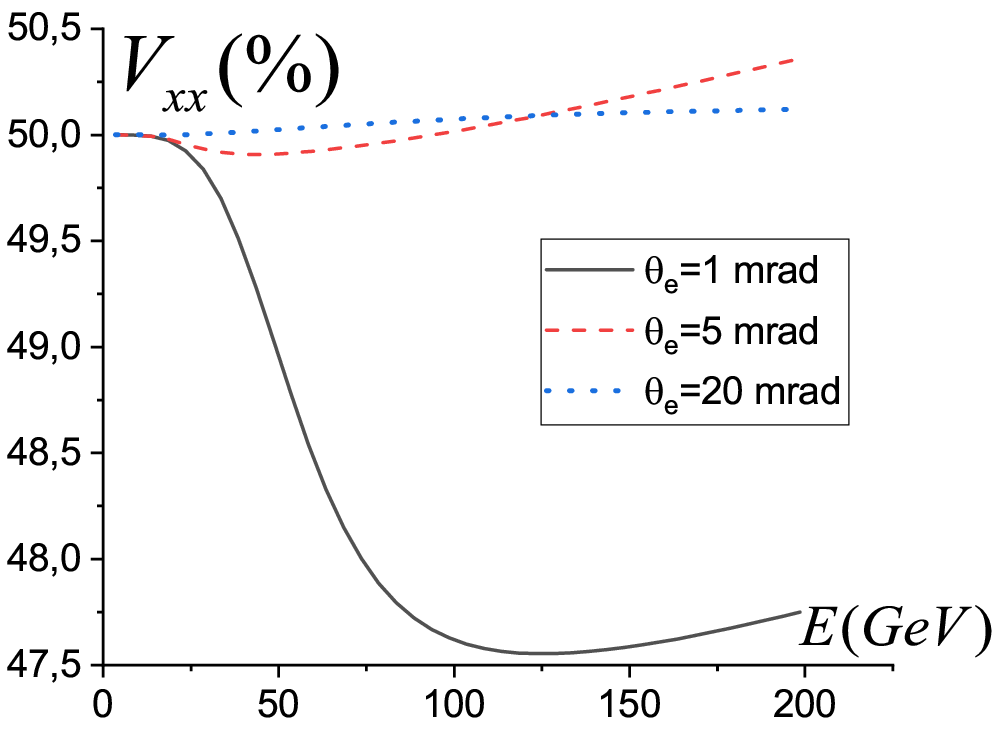}
\includegraphics[width=0.3\textwidth]{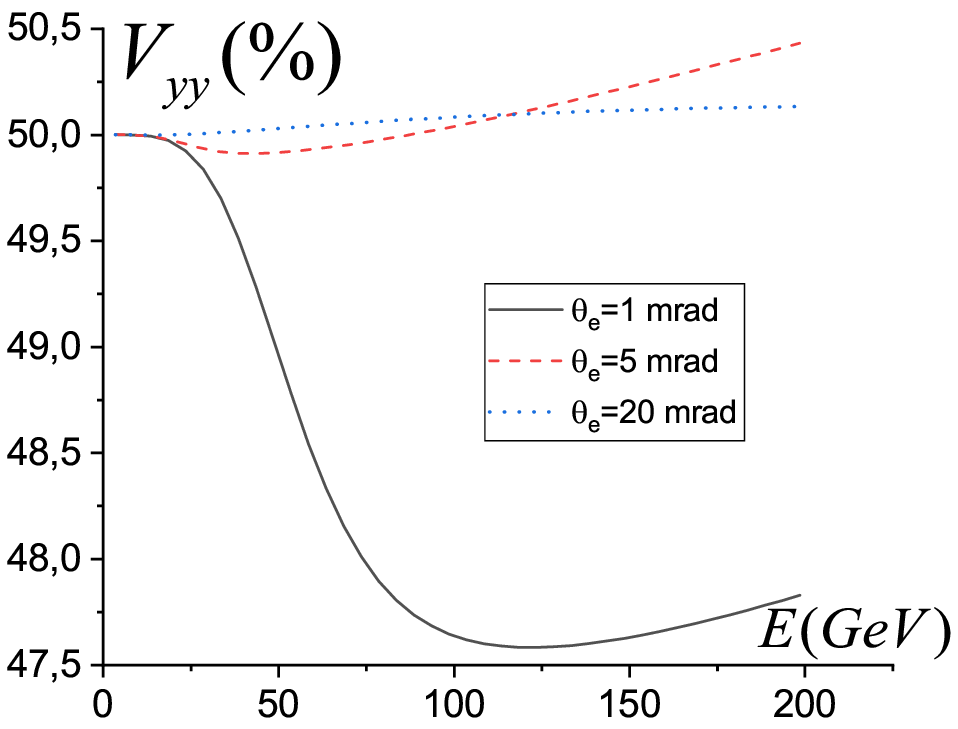}
\includegraphics[width=0.3\textwidth]{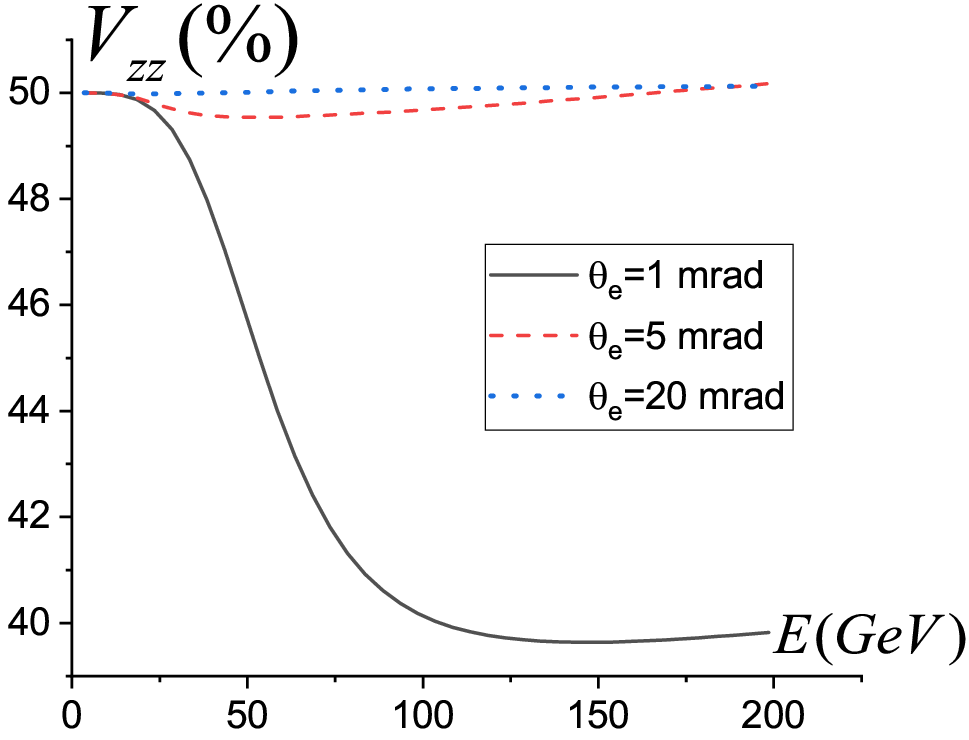}

\vspace{0.2 cm}

\includegraphics[width=0.3\textwidth]{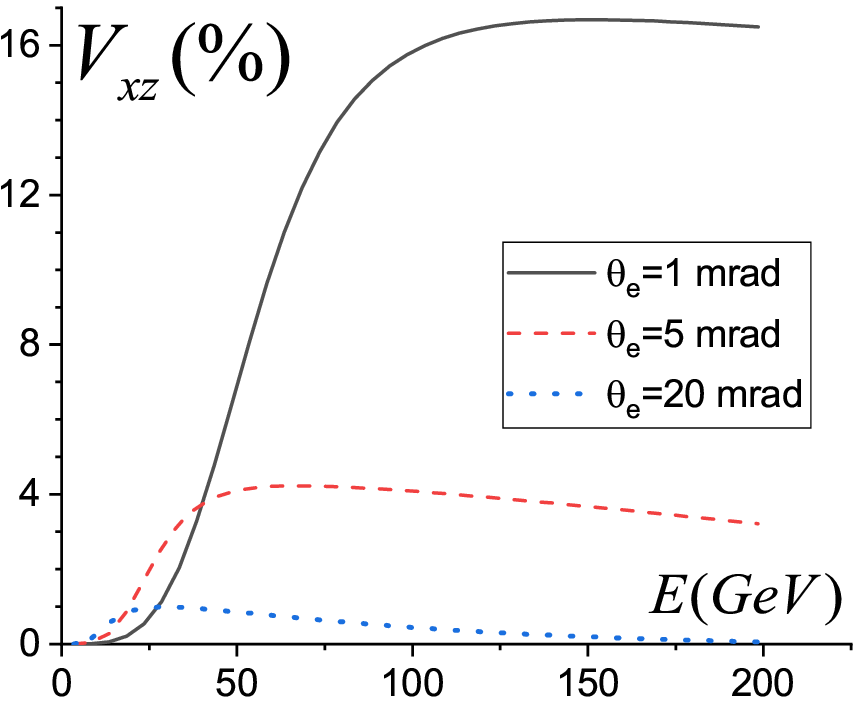}
\includegraphics[width=0.3\textwidth]{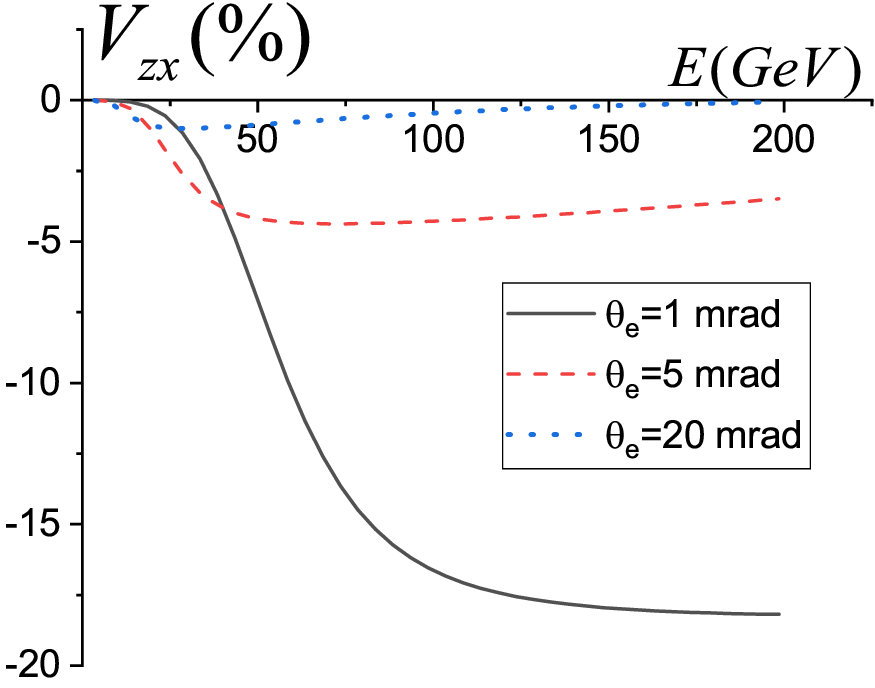}

\vspace{0.5 cm}
\includegraphics[width=0.3\textwidth]{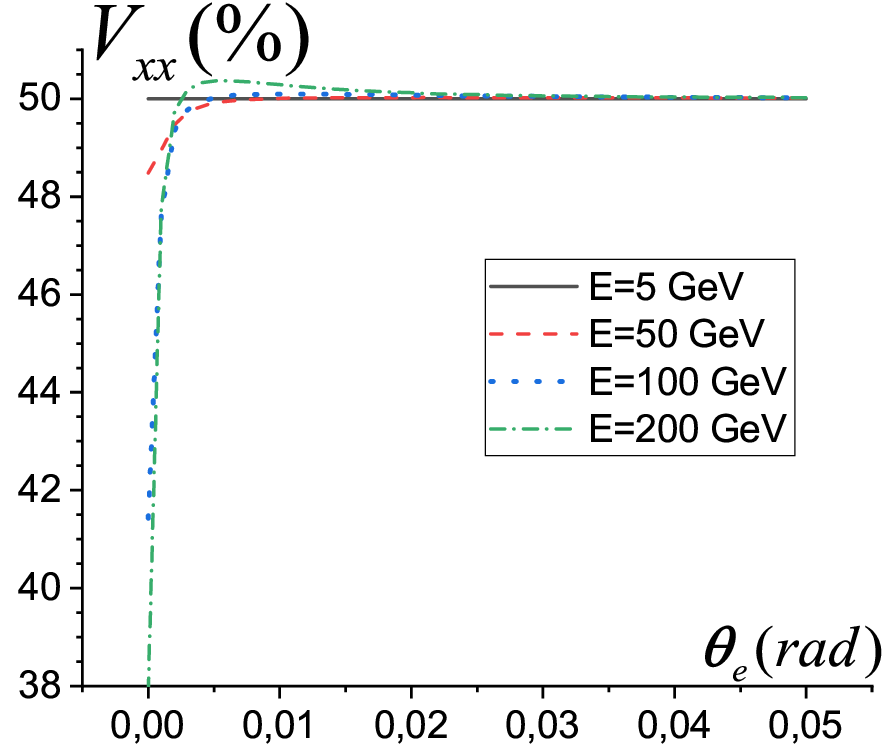}
\includegraphics[width=0.3\textwidth]{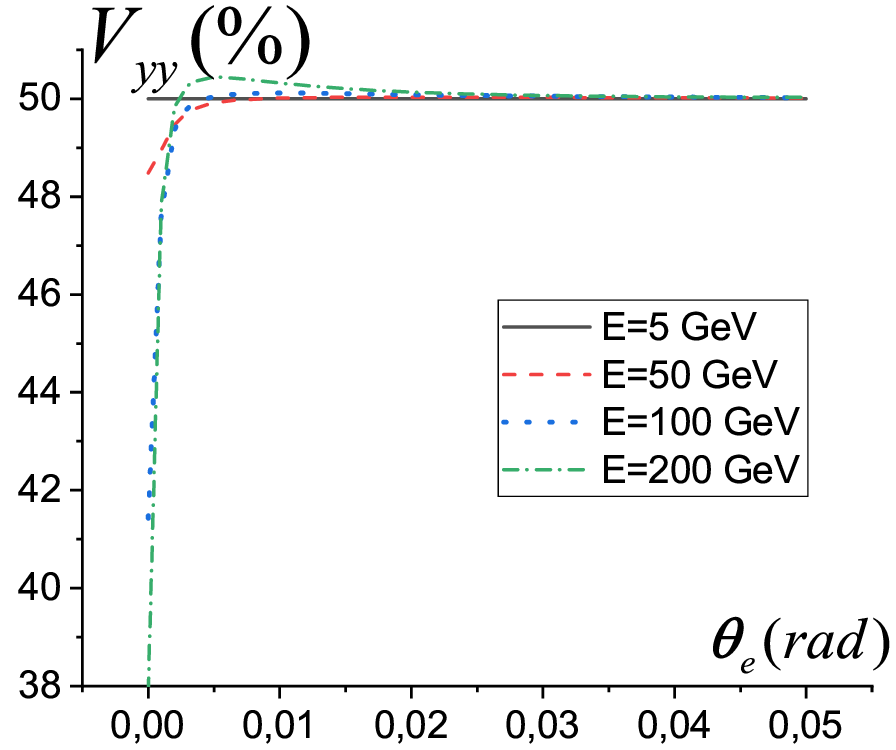}
\includegraphics[width=0.3\textwidth]{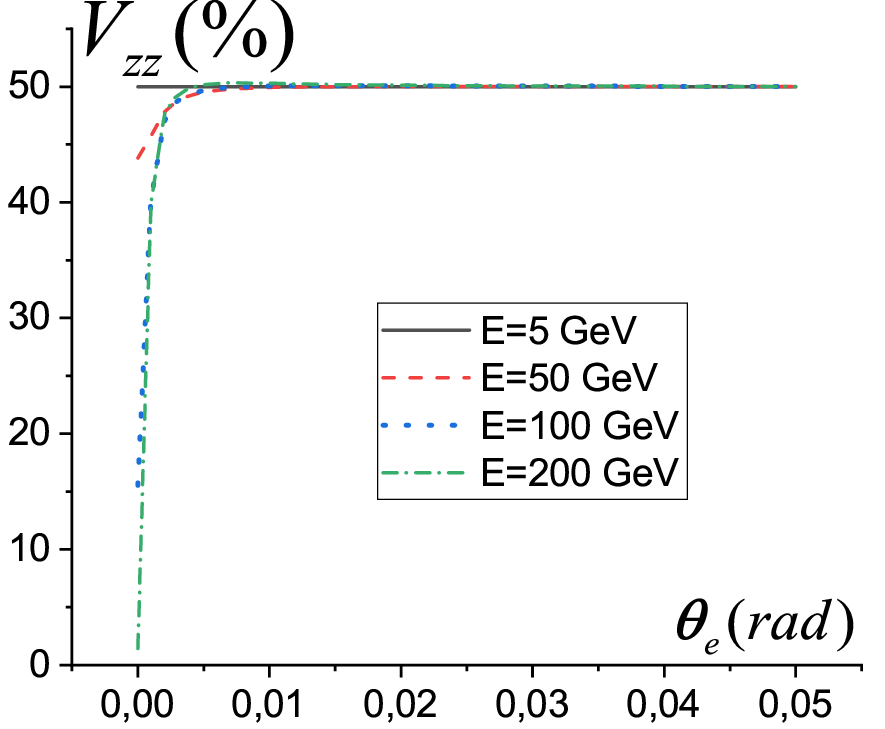}

\vspace{0.2 cm}

\includegraphics[width=0.3\textwidth]{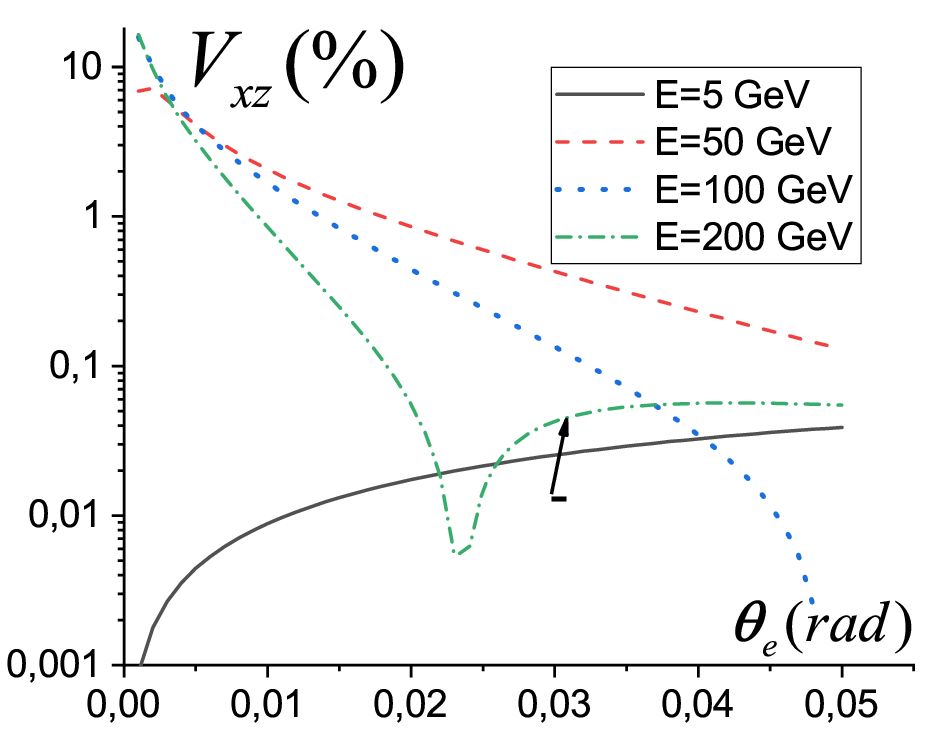}
\includegraphics[width=0.3\textwidth]{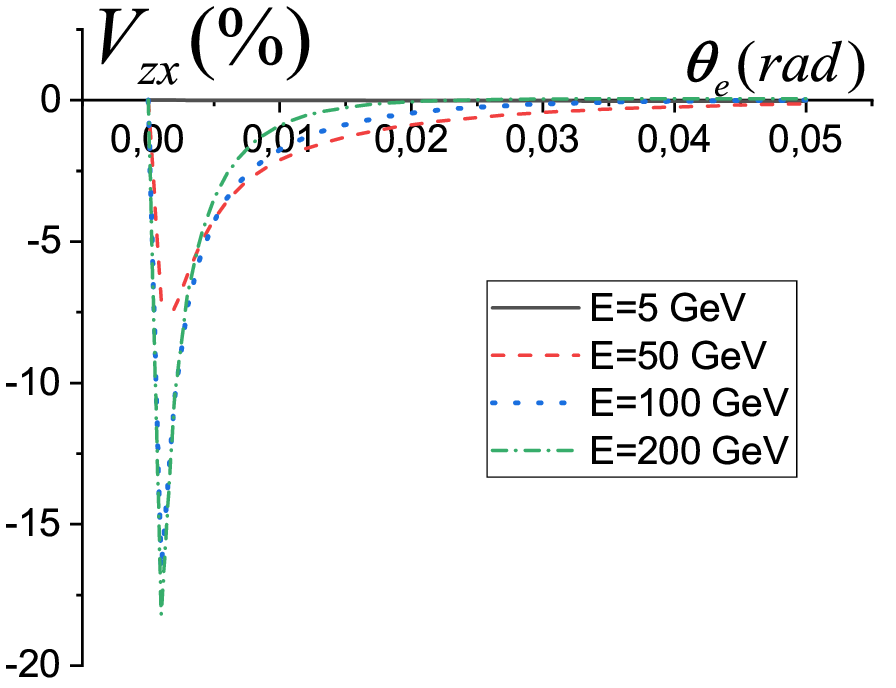}

 \parbox[t]{0.9\textwidth}{\caption{Polarization coefficients which describe the vector polarization transfer from the initial to the scattered deuteron in the reaction $\vec{d}\,^V +e\to \vec{d}\,^V + e $  given by Eqs.\,(\ref{eq:DVij}).}\label{fig.Vij}}
\end{figure}

The scattering of a vector polarized deuteron beam on an
unpolarized electron target is considered here. In this case the scattered deuterons
can be vector polarized. The hadronic tensor which
describes the case of the vector polarized deuteron beam and
scattered deuteron can be written as
\begin{eqnarray}
H_{\mu\nu}(\eta_1, \eta_2)&=&V_1\tilde g_{\mu\nu}+V_2 P_{\mu}P_{\nu}+V_3(\tilde \eta_{1\mu}\tilde \eta_{2\nu}+\tilde\eta_{1\nu}\tilde\eta_{2\mu})+  +V_4(P_{\mu}\tilde \eta_{1\nu}+ \nonumber\\
&&+P_{\nu}\tilde \eta_{1\mu})+V_5(P_{\mu}\tilde \eta_{2\nu}+P_{\nu}\tilde \eta_{2\mu}),  \ \ \tilde
\eta_{i\mu}=\eta_{i\mu}-\frac{k\cdot \eta_i}{k^2}k_{\mu}, 
\label{eq:Heta1eta2}
\end{eqnarray}
where the
structure functions $V_i$ have the following expressions in the
terms of the deuteron electromagnetic FFs
\begin{eqnarray}
\label{eq:Vij}
V_1&=&\frac{1}{2}G_M^2[(1+2\tau )k\cdot \eta_1k\cdot \eta_2+4\tau (1+\tau)M^2\eta_1\cdot \eta_2],\\
V_2&=&-\frac{1}{8M^2}(1+\tau )^{-1}\{(2G_C-\frac{4}{3}\tau G_Q)[(G_C+ \frac{4}{3}\tau G_Q)(k\cdot \eta_1k\cdot \eta_2+2(1+\tau )M^2\eta_1\cdot \eta_2)+ \nonumber\\
&&+(G_Q-G_M)k\cdot \eta_1k\cdot \eta_2]+(1+\tau )G_M^2(k\cdot \eta_1k\cdot \eta_2+4\tau M^2\eta_1\cdot \eta_2)\}, \nonumber\\
V_3&=&-\tau (1+\tau )M^2G_M^2,  \ \ V_4=-\frac{1}{4}k\cdot \eta_2G_M(G_C+\tau G_M-\frac{2}{3}\tau G_Q), \nonumber\\
V_5&=&\frac{1}{4}k\cdot \eta_1G_M(G_C+\tau G_M-\frac{2}{3}\tau G_Q). 
\nonumber 
\end{eqnarray}
For this configuration of the hadron polarizations it is sufficient to have an unpolarized
electron target since the hadronic tensor in this case is
symmetrical over $\mu , \nu $ indices. Thus, the contraction of the
spin independent $L_{\mu\nu}^{(0)}$ and spin dependent hadronic
$H_{\mu\nu}(\eta_1, \eta_2)$ tensors, gives the following expression which
is valid in an arbitrary reference frame
\begin{eqnarray}
C(\eta_1, \eta_2)&=&\frac{4}{M^2}G_M^2\{(k\cdot \eta_1k\cdot \eta_2-k^2\eta_1\cdot \eta_2)[p_1\cdot k_1(2\tau M^2-p_1\cdot k_1)+M^2((1+\tau)m^2-\tau^2M^2)]+\nonumber\\
&&+\tau M^2[m^2-(1+\tau)M^2]k\cdot \eta_1k\cdot \eta s_2+2\tau M^2[(M^2+p_1\cdot k_1)k\cdot \eta_1k\cdot \eta_2+ \nonumber\\
&&+((1+2\tau)M^2-p_1\cdot k_1)k_1\cdot \eta_1k\cdot \eta_2-2(1+\tau )M^2k_1\cdot \eta_1k_1\cdot \eta_2]\}+ \nonumber\\
&&+\frac{8}{(1+\tau)M^2}(G_C-\frac{2}{3}\tau G_Q)\{G_M(\tau M^2-p_1\cdot k_1)[k\cdot \eta_1k\cdot \eta_2(\tau M^2-p_1\cdot k_1)- \nonumber\\
&&-(1+\tau )M^2(k\cdot \eta_1k_1\cdot \eta_2-k_1\cdot \eta_1k\cdot \eta_2)]+[\tau M^4+2\tau M^2p_1\cdot k_1-(p_1\cdot k_1)^2] \nonumber\\
&&[2(1+\tau )M^2(G_C+\frac{4}{3}\tau G_Q)\eta_1\cdot \eta_2+(G_C+(1+\frac{4}{3}\tau )G_Q)k\cdot \eta_1k\cdot \eta_2]\}. 
\label{eq:Ceta1eta2}
\end{eqnarray}
The corresponding differential cross section  is
\begin{equation}\label{eq:Dseta1eta2}
\frac{d \sigma}{d Q^2} = \Big(\frac{d \sigma}{d Q^2}\Big)_{un}\big[1 + V_{xx} S_{1x}S_{2x} +  V_{yy} S_{1y}S_{2y} + V_{zz} S_{1z}S_{2z}
+  V_{xz} S_{1x}S_{2z} +  V_{zx} S_{1z}S_{2x}\big].
\end{equation}
The explicit expressions for $V_{ij}$ are
\begin{eqnarray}
{\cal D} V_{xx}&=& 2h\Big\{\frac{x^2}{(1+\tau)M^3}\big[E-y M(p(p-z)+2(1+\tau)M^2)\big] -2 \Big\}G^{(-)}\,G^{(+)} +\nonumber\\
&&\frac{2x^2(\tau M^2-m E)}{(1+\tau)M^3}\big[y M p(z-p)(\tau M^2-m E) + \tau M^2(E+m) - mp^2\big] G_M\,G^{(-)}+\nonumber\\
&&\big\{Q^2[\tau M^2(\tau M^2-2 m E - m^2) + m^2 p^2] \nonumber\\
&&- x^2 y M \big[\tau M^4(M + (1+2\tau)E)+ mp^2(m(E+M) -2\tau M^2)\big]\big\}\frac{G_M^2}{M^2},\nonumber\\
{\cal D} V_{yy}&=&-4 h G^{(-)}\,G^{(+)} +[\tau M^2(\tau M^2-2 m E - m^2) + m^2 p^2]\tau G_M^2,    \nonumber  \\
{\cal D} V_{zz}&=& \frac{yM}{m^2p^2}\Big\{\frac{4(E-M) h h_2}{1+\tau}G^{(-)}\,G^{(+)}-4\tau M^4[E+(1+2\tau)M][h +(1+\tau)m^2 M^2] G_M^2 \nonumber\\
&&-2Q^2(E-M)(\tau M^2-m E)[h+(1+\tau)m^2 M^2] \frac{h_1\,G_M\,G^{(-)}}{(1+\tau)M^2}\Big\}, \nonumber\\
{\cal D} V_{xz}&=& \frac{x y M}{m p}\Big\{\frac{2(E-M)}{(1+\tau)M}\big[-2 h h_1 G^{(-)}\,G^{(+)} +h_2(\tau M^2 - m E)G_M\,G^{(-)}\big]\nonumber\\
&& -2\tau M^3(M^2+m E)[E+(1+2\tau)M]G_M^2 \Big\},\nonumber\\
{\cal D} V_{zx}&=& \frac{x y M}{m p}\Big\{\frac{2(E-M)}{(1+\tau)M}\big[2 h h_1 G^{(-)}\,G^{(+)} -h_2(\tau M^2- m E)G_M\,G^{(-)}\big]\nonumber\\
&&-2\tau^2 M^3\big[M^2(M+(1+2\tau)E + 2(1+\tau)m) - m E(E-M)\big]G_M^2\Big\},\label{eq:DVij}
\end{eqnarray}
where we introduced the short notation
\begin{eqnarray}
G^{(-)} &= &G_C-\frac{2 \tau G_Q}{3}, \ \ G^{(+)} = G_C+\frac{4 \tau G_Q}{3}, \nonumber\\
 h& =&\tau M^2(M^2+2m E) - m^2 E^2, \ \ h_1=\tau M^2- m E - (1+\tau)m M, \nonumber\\
 h_2&= & -2\tau M^2(h_1+\tau m E) -m^2\big\{(1+\tau)M[(1+2\tau)M +2 E]+(1-\tau)E^2\big\}.\label{eq:eqshortFF}
\end{eqnarray}
The polarization coefficients $V_{ij}$ describing the vector polarization transfer from the deuteron beam to the scattered deuteron (the deuteron depolarization) are shown in Fig.\,\ref{fig.Vij}.

\section{Discussion and conclusion}
In this work we calculated the differential cross section and some polarization observables for the elastic reaction induced by deuteron scattering
off electrons at rest assuming the one-photon-exchange approximation. We limited the study to  the  estimation of  one-spin effects when one deuteron (initial or scattered) is tensor polarized, and  to double-spin effects when two particles are vector polarized. In the last case,  all possible polarization states are considered. Our analytical and numerical results are obtained under the condition that all components of the 3-vector polarization for every particle in reaction (\ref{eq:1}) are defined in the Lab system, as shown in Fig.\,\ref{fig.1}. The same is required for the  components of the deuteron tensor polarizations. In this respect, it should be noted that  the combination of the form factors, included in $t_{20}$ and measured in $e-d$ scattering \cite{JLABt20:2000uor}, corresponds to the $z$-axis along the unit 3-vector of the momentum transfer \cite{Gakh:2004cc,Gakh:2012hf} in the rest frame of the initial deuteron. Our choice corresponds to the direction opposite to the unit 3-vector of the initial electron 3-momentum (in Lab system, the $z$-axis is just along deuteron 3-momentum). Therefore $A_{zz}$ has the contribution of a term $~G_M\,G_Q$ which is absent in $t_{20}$. 
Along the numerical calculation we used the parametrization of the deuteron electromagnetic form factors suggested in Ref. \cite{Tomasi-Gustafsson:2005dyd} and extrapolate it to the small $Q^2$ region (Fig.~\ref{fig.Gi}). Others form factor parameterizations exhibit very similar behaviour  in this region.

Our result can be applied  to measure the polarization of or to create the polarization of the participating particles. Note, that the unpolarized cross section is large enough
(see Fig.\,\ref{fig.DsBorn}), indicating that the number of events in the  different polarization conditions can be sufficient to perform fairly accurate measurements despite of the fact that 
the corresponding effects are at the percent level.

Our formalism is very general, based on the symmetries of the strong and electromagnetic interactions. The lepton and hadron tensors are obtained in terms of the deuteron electromagnetic current Eq.\,(\ref{eq:JD}) and the density matrices of the initial and scattered deuteron, Eq.\.(\ref{eq:MPD}). All the coefficients, which describe the single- and double-spin effects, are the ratio of the corresponding spin-dependent parts of the matrix element squared by the spin-independent parts, i.e.,  $H_{\mu\nu}\,L^{\mu\nu}/(2\,{\cal D}).$ , according to  the used normalization of the unpolarized cross section (see Eq.\,(\ref{eq:dsQ201})). The additional factor 1/2 in the differential cross sections when the recoil electron polarization is measured (see Eqs.\,(\ref{eq:DSs1s2}),\,(\ref{eq:DSeta2s2}),\, (\ref{eq:Dseta1s2})), is due to the density matrix of the recoil electron. Our main results are illustrated in Figs.\,(4-10) in which we plotted different coefficients as a function of the electron scattering angle $\theta_e$ at fixed values of the deuteron beam energy $E$ and vice versa.

In  the Lab system the tensor asymmetries $A_{ij}$ (Fig.\,\ref{fig.Aij}) and the tensor polarization coefficients $P_{ij}$ (Fig.\,\ref{fig.Pij}) are small, not exceeding the order of percent. Nevertheless this situation leaves room for measurements due to large cross section. The coefficients of the polarization transfer $t_{ij}$ (Fig.\,\ref{fig.tij}) from the target to the recoil electron vary in the range $-1$ to $+1,$ which makes it possible to change the polarization of electrons. The coefficients $C_{ij}$ except $C_{xz}$ (Fig.\,\ref{fig.Cij}) are on the level a few tens, thus the correlation between the vector polarizations of the deuteron beam and the target electrons is large and measurable.

The possibility to create vector polarized deuterons from polarized target electrons is illustrated by the polarization transfer coefficients $T_{ij}$
(Fig.\,\ref{fig.Tij}). They are of the order  of 10\%, except $T_{xz}$, showing a realistic possibility of applications. The correlation between the vector polarization of the final deuterons and electrons is noticeable although not as large as for the initial ones  (Fig.\,\ref{fig.cij}). It is quite unexpected that the vector polarization transfer coefficients $\widetilde T_{ij}$ (Fig.\,\ref{fig.wTij}) from the initial deuterons to the recoil electrons is several times larger than $T_{ij}.$ The  large values of the coefficients $V_{ij}$, which describe the vector polarization transfer from the initial o the scattered deuterons (Fig.\,\ref{fig.Vij}) should also be noted.

Our formalism, being very general, gives the essential formulas for deuteron polariation and polarimetry studies. The specific ingredients of the deuteron structure are contained in the form factors. The sensitivity to different models is expected not to be large, because of the low-$Q^2$ involved and the constrains for static deuteron properties at $Q^2\to 0$.

\section{Acknowledgments}
This work was partly supported (G.I.G, M.I.K. and N.P.M.) by the National
 Academy of Sciences via the program "Participation in the international
 projects in high energy and nuclear physics" (project no.0121U111693).


\begin{thebibliography}{50}%
\makeatletter
\providecommand \@ifxundefined [1]{%
 \@ifx{#1\undefined}
}%
\providecommand \@ifnum [1]{%
 \ifnum #1\expandafter \@firstoftwo
 \else \expandafter \@secondoftwo
 \fi
}%
\providecommand \@ifx [1]{%
 \ifx #1\expandafter \@firstoftwo
 \else \expandafter \@secondoftwo
 \fi
}%
\providecommand \natexlab [1]{#1}%
\providecommand \enquote  [1]{``#1''}%
\providecommand \bibnamefont  [1]{#1}%
\providecommand \bibfnamefont [1]{#1}%
\providecommand \citenamefont [1]{#1}%
\providecommand \href@noop [0]{\@secondoftwo}%
\providecommand \href [0]{\begingroup \@sanitize@url \@href}%
\providecommand \@href[1]{\@@startlink{#1}\@@href}%
\providecommand \@@href[1]{\endgroup#1\@@endlink}%
\providecommand \@sanitize@url [0]{\catcode `\\12\catcode `\$12\catcode
  `\&12\catcode `\#12\catcode `\^12\catcode `\_12\catcode `\%12\relax}%
\providecommand \@@startlink[1]{}%
\providecommand \@@endlink[0]{}%
\providecommand \url  [0]{\begingroup\@sanitize@url \@url }%
\providecommand \@url [1]{\endgroup\@href {#1}{\urlprefix }}%
\providecommand \urlprefix  [0]{URL }%
\providecommand \Eprint [0]{\href }%
\providecommand \doibase [0]{http://dx.doi.org/}%
\providecommand \selectlanguage [0]{\@gobble}%
\providecommand \bibinfo  [0]{\@secondoftwo}%
\providecommand \bibfield  [0]{\@secondoftwo}%
\providecommand \translation [1]{[#1]}%
\providecommand \BibitemOpen [0]{}%
\providecommand \bibitemStop [0]{}%
\providecommand \bibitemNoStop [0]{.\EOS\space}%
\providecommand \EOS [0]{\spacefactor3000\relax}%
\providecommand \BibitemShut  [1]{\csname bibitem#1\endcsname}%
\let\auto@bib@innerbib\@empty
\bibitem [{\citenamefont {Adylov}\ \emph {et~al.}(1974)\citenamefont {Adylov}
  \emph {et~al.}}]{Adylov:1974sg}%
  \BibitemOpen
  \bibfield  {author} {\bibinfo {author} {\bibfnamefont {G.}~\bibnamefont
  {Adylov}} \emph {et~al.},\ }\bibfield  {booktitle} {\emph {\bibinfo
  {booktitle} {{High energy physics. Proceedings, 17th International
  Conference, ICHEP 1974, London, England, July 01-July 10, 1974}}},\ }\href
  {\doibase 10.1016/0370-2693(74)90239-1} {\bibfield  {journal} {\bibinfo
  {journal} {Phys. Lett.}\ }\textbf {\bibinfo {volume} {B51}},\ \bibinfo
  {pages} {402} (\bibinfo {year} {1974})}\BibitemShut {NoStop}%
\bibitem [{\citenamefont {Dally}\ \emph {et~al.}(1977)\citenamefont {Dally}
  \emph {et~al.}}]{Dally:1977vt}%
  \BibitemOpen
  \bibfield  {author} {\bibinfo {author} {\bibfnamefont {E.~B.}\ \bibnamefont
  {Dally}} \emph {et~al.},\ }\href {\doibase 10.1103/PhysRevLett.39.1176}
  {\bibfield  {journal} {\bibinfo  {journal} {Phys. Rev. Lett.}\ }\textbf
  {\bibinfo {volume} {39}},\ \bibinfo {pages} {1176} (\bibinfo {year}
  {1977})}\BibitemShut {NoStop}%
\bibitem [{\citenamefont {Dally}\ \emph {et~al.}(1982)\citenamefont {Dally}
  \emph {et~al.}}]{Dally:1982zk}%
  \BibitemOpen
  \bibfield  {author} {\bibinfo {author} {\bibfnamefont {E.~B.}\ \bibnamefont
  {Dally}} \emph {et~al.},\ }\href {\doibase 10.1103/PhysRevLett.48.375}
  {\bibfield  {journal} {\bibinfo  {journal} {Phys. Rev. Lett.}\ }\textbf
  {\bibinfo {volume} {48}},\ \bibinfo {pages} {375} (\bibinfo {year}
  {1982})}\BibitemShut {NoStop}%
\bibitem [{\citenamefont {Dally}\ \emph {et~al.}(1980)\citenamefont {Dally}
  \emph {et~al.}}]{Dally:1980dj}%
  \BibitemOpen
  \bibfield  {author} {\bibinfo {author} {\bibfnamefont {E.~B.}\ \bibnamefont
  {Dally}} \emph {et~al.},\ }\href {\doibase 10.1103/PhysRevLett.45.232}
  {\bibfield  {journal} {\bibinfo  {journal} {Phys. Rev. Lett.}\ }\textbf
  {\bibinfo {volume} {45}},\ \bibinfo {pages} {232} (\bibinfo {year}
  {1980})}\BibitemShut {NoStop}%
\bibitem [{\citenamefont {Amendolia}\ \emph {et~al.}(1986)\citenamefont
  {Amendolia} \emph {et~al.}}]{Amet86}%
  \BibitemOpen
  \bibfield  {author} {\bibinfo {author} {\bibfnamefont {S.~R.}\ \bibnamefont
  {Amendolia}} \emph {et~al.} (\bibinfo {collaboration} {NA7}),\ }\bibfield
  {booktitle} {\emph {\bibinfo {booktitle} {{Proceedings, 23RD International
  Conference on High Energy Physics, JULY 16-23, 1986, Berkeley, CA}}},\ }\href
  {\doibase 10.1016/0550-3213(86)90437-2} {\bibfield  {journal} {\bibinfo
  {journal} {Nucl. Phys.}\ }\textbf {\bibinfo {volume} {B277}},\ \bibinfo
  {pages} {168} (\bibinfo {year} {1986})}\BibitemShut {NoStop}%
\bibitem [{\citenamefont {Amendolia}\ \emph {et~al.}(1984)\citenamefont
  {Amendolia} \emph {et~al.}}]{Amet84}%
  \BibitemOpen
  \bibfield  {author} {\bibinfo {author} {\bibfnamefont {S.~R.}\ \bibnamefont
  {Amendolia}} \emph {et~al.},\ }\href {\doibase 10.1016/0370-2693(84)90655-5}
  {\bibfield  {journal} {\bibinfo  {journal} {Phys. Lett.}\ }\textbf {\bibinfo
  {volume} {B146}},\ \bibinfo {pages} {116} (\bibinfo {year}
  {1984})}\BibitemShut {NoStop}%
\bibitem [{\citenamefont {Abbiendi}\ \emph {et~al.}(2016)\citenamefont
  {Abbiendi} \emph {et~al.}}]{Abbiendi:2016xup}%
  \BibitemOpen
  \bibfield  {author} {\bibinfo {author} {\bibfnamefont {G.}~\bibnamefont
  {Abbiendi}} \emph {et~al.},\ }\href@noop {} {\  (\bibinfo {year} {2016})},\
  \Eprint {http://arxiv.org/abs/1609.08987} {arXiv:1609.08987 [hep-ex]}
  \BibitemShut {NoStop}%
\bibitem [{\citenamefont {Gakh}\ \emph {et~al.}(2011)\citenamefont {Gakh},
  \citenamefont {Dbeyssi}, \citenamefont {Marchand}, \citenamefont
  {Tomasi-Gustafsson},\ and\ \citenamefont {Bytev}}]{Gakh:2011sh}%
  \BibitemOpen
  \bibfield  {author} {\bibinfo {author} {\bibfnamefont {G.}~\bibnamefont
  {Gakh}}, \bibinfo {author} {\bibfnamefont {A.}~\bibnamefont {Dbeyssi}},
  \bibinfo {author} {\bibfnamefont {D.}~\bibnamefont {Marchand}}, \bibinfo
  {author} {\bibfnamefont {E.}~\bibnamefont {Tomasi-Gustafsson}}, \ and\
  \bibinfo {author} {\bibfnamefont {V.}~\bibnamefont {Bytev}},\ }\href
  {\doibase 10.1103/PhysRevC.84.015212} {\bibfield  {journal} {\bibinfo
  {journal} {Phys.Rev.}\ }\textbf {\bibinfo {volume} {C84}},\ \bibinfo {pages}
  {015212} (\bibinfo {year} {2011})},\ \Eprint {http://arxiv.org/abs/1103.2540}
  {arXiv:1103.2540 [nucl-th]} \BibitemShut {NoStop}%
\bibitem [{\citenamefont {Pohl}\ \emph {et~al.}(2010)\citenamefont {Pohl},
  \citenamefont {Antognini}, \citenamefont {Nez}, \citenamefont {Amaro},
  \citenamefont {Biraben} \emph {et~al.}}]{Pohl:2010zza}%
  \BibitemOpen
  \bibfield  {author} {\bibinfo {author} {\bibfnamefont {R.}~\bibnamefont
  {Pohl}}, \bibinfo {author} {\bibfnamefont {A.}~\bibnamefont {Antognini}},
  \bibinfo {author} {\bibfnamefont {F.}~\bibnamefont {Nez}}, \bibinfo {author}
  {\bibfnamefont {F.~D.}\ \bibnamefont {Amaro}}, \bibinfo {author}
  {\bibfnamefont {F.}~\bibnamefont {Biraben}},  \emph {et~al.},\ }\href
  {\doibase 10.1038/nature09250} {\bibfield  {journal} {\bibinfo  {journal}
  {Nature}\ }\textbf {\bibinfo {volume} {466}},\ \bibinfo {pages} {213}
  (\bibinfo {year} {2010})}\BibitemShut {NoStop}%
\bibitem [{\citenamefont {Bernauer}\ \emph {et~al.}(2010)\citenamefont
  {Bernauer} \emph {et~al.}}]{Bernauer:2010wm}%
  \BibitemOpen
  \bibfield  {author} {\bibinfo {author} {\bibfnamefont {J.}~\bibnamefont
  {Bernauer}} \emph {et~al.} (\bibinfo {collaboration} {A1 Collaboration}),\
  }\href {\doibase 10.1103/PhysRevLett.105.242001} {\bibfield  {journal}
  {\bibinfo  {journal} {Phys.Rev.Lett.}\ }\textbf {\bibinfo {volume} {105}},\
  \bibinfo {pages} {242001} (\bibinfo {year} {2010})},\ \Eprint
  {http://arxiv.org/abs/1007.5076} {arXiv:1007.5076 [nucl-ex]} \BibitemShut
  {NoStop}%
\bibitem [{\citenamefont {Mohr}\ \emph {et~al.}(2012)\citenamefont {Mohr},
  \citenamefont {Taylor},\ and\ \citenamefont {Newell}}]{RevModPhys.84.1527}%
  \BibitemOpen
  \bibfield  {author} {\bibinfo {author} {\bibfnamefont {P.~J.}\ \bibnamefont
  {Mohr}}, \bibinfo {author} {\bibfnamefont {B.~N.}\ \bibnamefont {Taylor}}, \
  and\ \bibinfo {author} {\bibfnamefont {D.~B.}\ \bibnamefont {Newell}},\
  }\href {\doibase 10.1103/RevModPhys.84.1527} {\bibfield  {journal} {\bibinfo
  {journal} {Rev. Mod. Phys.}\ }\textbf {\bibinfo {volume} {84}},\ \bibinfo
  {pages} {1527} (\bibinfo {year} {2012})}\BibitemShut {NoStop}%
\bibitem [{\citenamefont {Gasparian}(2014)}]{Gasparian:2014rna}%
  \BibitemOpen
  \bibfield  {author} {\bibinfo {author} {\bibfnamefont {A.}~\bibnamefont
  {Gasparian}} (\bibinfo {collaboration} {PRad at JLab}),\ }\bibfield
  {booktitle} {\emph {\bibinfo {booktitle} {{Proceedings, 13th International
  Conference on Meson-Nucleon Physics and the Structure of the Nucleon (MENU
  2013): Rome, Italy, September 30-October 4, 2013}}},\ }\href {\doibase
  10.1051/epjconf/20147307006} {\bibfield  {journal} {\bibinfo  {journal} {EPJ
  Web Conf.}\ }\textbf {\bibinfo {volume} {73}},\ \bibinfo {pages} {07006}
  (\bibinfo {year} {2014})}\BibitemShut {NoStop}%
\bibitem [{\citenamefont {Cline}\ \emph {et~al.}(2021)\citenamefont {Cline},
  \citenamefont {Bernauer}, \citenamefont {Downie},\ and\ \citenamefont
  {Gilman}}]{Cline:2021ehf}%
  \BibitemOpen
  \bibfield  {author} {\bibinfo {author} {\bibfnamefont {E.}~\bibnamefont
  {Cline}}, \bibinfo {author} {\bibfnamefont {J.}~\bibnamefont {Bernauer}},
  \bibinfo {author} {\bibfnamefont {E.~J.}\ \bibnamefont {Downie}}, \ and\
  \bibinfo {author} {\bibfnamefont {R.}~\bibnamefont {Gilman}},\ }\href
  {\doibase 10.21468/SciPostPhysProc.5.023} {\bibfield  {journal} {\bibinfo
  {journal} {SciPost Phys. Proc.}\ }\textbf {\bibinfo {volume} {5}},\ \bibinfo
  {pages} {023} (\bibinfo {year} {2021})}\BibitemShut {NoStop}%
\bibitem [{\citenamefont {Xiong}\ \emph {et~al.}(2019)\citenamefont {Xiong}
  \emph {et~al.}}]{Xiong:2019umf}%
  \BibitemOpen
  \bibfield  {author} {\bibinfo {author} {\bibfnamefont {W.}~\bibnamefont
  {Xiong}} \emph {et~al.},\ }\href {\doibase 10.1038/s41586-019-1721-2}
  {\bibfield  {journal} {\bibinfo  {journal} {Nature}\ }\textbf {\bibinfo
  {volume} {575}},\ \bibinfo {pages} {147} (\bibinfo {year}
  {2019})}\BibitemShut {NoStop}%
\bibitem [{\citenamefont {Tiesinga}\ \emph {et~al.}(2021)\citenamefont
  {Tiesinga}, \citenamefont {Mohr}, \citenamefont {Newell},\ and\ \citenamefont
  {Taylor}}]{RevModPhys.93.025010}%
  \BibitemOpen
  \bibfield  {author} {\bibinfo {author} {\bibfnamefont {E.}~\bibnamefont
  {Tiesinga}}, \bibinfo {author} {\bibfnamefont {P.~J.}\ \bibnamefont {Mohr}},
  \bibinfo {author} {\bibfnamefont {D.~B.}\ \bibnamefont {Newell}}, \ and\
  \bibinfo {author} {\bibfnamefont {B.~N.}\ \bibnamefont {Taylor}},\ }\href
  {\doibase 10.1103/RevModPhys.93.025010} {\bibfield  {journal} {\bibinfo
  {journal} {Rev. Mod. Phys.}\ }\textbf {\bibinfo {volume} {93}},\ \bibinfo
  {pages} {025010} (\bibinfo {year} {2021})}\BibitemShut {NoStop}%
\bibitem [{\citenamefont {Sick}\ and\ \citenamefont
  {Trautmann}(1998)}]{Sick:1998cvq}%
  \BibitemOpen
  \bibfield  {author} {\bibinfo {author} {\bibfnamefont {I.}~\bibnamefont
  {Sick}}\ and\ \bibinfo {author} {\bibfnamefont {D.}~\bibnamefont
  {Trautmann}},\ }\href {\doibase 10.1016/S0375-9474(98)00334-0} {\bibfield
  {journal} {\bibinfo  {journal} {Nucl. Phys. A}\ }\textbf {\bibinfo {volume}
  {637}},\ \bibinfo {pages} {559} (\bibinfo {year} {1998})}\BibitemShut
  {NoStop}%
\bibitem [{\citenamefont {Antognini}\ \emph {et~al.}(2021)\citenamefont
  {Antognini}, \citenamefont {Kottmann},\ and\ \citenamefont
  {Pohl}}]{10.21468/SciPostPhysProc.5.021}%
  \BibitemOpen
  \bibfield  {author} {\bibinfo {author} {\bibfnamefont {A.}~\bibnamefont
  {Antognini}}, \bibinfo {author} {\bibfnamefont {F.}~\bibnamefont {Kottmann}},
  \ and\ \bibinfo {author} {\bibfnamefont {R.}~\bibnamefont {Pohl}},\ }\href
  {\doibase 10.21468/SciPostPhysProc.5.021} {\bibfield  {journal} {\bibinfo
  {journal} {SciPost Phys. Proc.}\ ,\ \bibinfo {pages} {021}} (\bibinfo {year}
  {2021})}\BibitemShut {NoStop}%
\bibitem [{\citenamefont {Pacetti}\ and\ \citenamefont
  {Gustafsson}(2016)}]{Pacetti:2016tqi}%
  \BibitemOpen
  \bibfield  {author} {\bibinfo {author} {\bibfnamefont {S.}~\bibnamefont
  {Pacetti}}\ and\ \bibinfo {author} {\bibfnamefont {E.~T.}\ \bibnamefont
  {Gustafsson}},\ }\href@noop {} {\bibfield  {journal} {\bibinfo  {journal} {to
  appear in Phys. Rev. C}\ } (\bibinfo {year} {2016})},\ \Eprint
  {http://arxiv.org/abs/1604.02421} {arXiv:1604.02421 [nucl-th]} \BibitemShut
  {NoStop}%
\bibitem [{\citenamefont {Pacetti}\ and\ \citenamefont
  {Tomasi-Gustafsson}(2020)}]{Pacetti:2018wwk}%
  \BibitemOpen
  \bibfield  {author} {\bibinfo {author} {\bibfnamefont {S.}~\bibnamefont
  {Pacetti}}\ and\ \bibinfo {author} {\bibfnamefont {E.}~\bibnamefont
  {Tomasi-Gustafsson}},\ }\href {\doibase 10.1140/epja/s10050-020-00076-1}
  {\bibfield  {journal} {\bibinfo  {journal} {Eur. Phys. J. A}\ }\textbf
  {\bibinfo {volume} {56}},\ \bibinfo {pages} {74} (\bibinfo {year} {2020})},\
  \Eprint {http://arxiv.org/abs/1812.04444} {arXiv:1812.04444 [nucl-th]}
  \BibitemShut {NoStop}%
\bibitem [{\citenamefont {Glavanakov}\ \emph {et~al.}(1996)\citenamefont
  {Glavanakov}, \citenamefont {Krechetov}, \citenamefont {Potylitsyn},
  \citenamefont {Radutsky}, \citenamefont {Tabachenko},\ and\ \citenamefont
  {Nurushev}}]{Glavanakov:96}%
  \BibitemOpen
  \bibfield  {author} {\bibinfo {author} {\bibfnamefont {I.~V.}\ \bibnamefont
  {Glavanakov}}, \bibinfo {author} {\bibfnamefont {{\relax Yu}.~F.}\
  \bibnamefont {Krechetov}}, \bibinfo {author} {\bibfnamefont {A.~P.}\
  \bibnamefont {Potylitsyn}}, \bibinfo {author} {\bibfnamefont {G.~M.}\
  \bibnamefont {Radutsky}}, \bibinfo {author} {\bibfnamefont {A.~N.}\
  \bibnamefont {Tabachenko}}, \ and\ \bibinfo {author} {\bibfnamefont {S.~B.}\
  \bibnamefont {Nurushev}},\ }\href {\doibase 10.1016/S0168-9002(96)00714-0}
  {\bibfield  {journal} {\bibinfo  {journal} {Nucl. Instrum. Meth.}\ }\textbf
  {\bibinfo {volume} {A381}},\ \bibinfo {pages} {275} (\bibinfo {year}
  {1996})}\BibitemShut {NoStop}%
\bibitem [{\citenamefont {Reifarth}\ and\ \citenamefont
  {Litvinov}(2014)}]{Reifarth:2013bta}%
  \BibitemOpen
  \bibfield  {author} {\bibinfo {author} {\bibfnamefont {R.}~\bibnamefont
  {Reifarth}}\ and\ \bibinfo {author} {\bibfnamefont {Y.~A.}\ \bibnamefont
  {Litvinov}},\ }\href {\doibase 10.1103/PhysRevSTAB.17.014701} {\bibfield
  {journal} {\bibinfo  {journal} {Phys. Rev. ST Accel. Beams}\ }\textbf
  {\bibinfo {volume} {17}},\ \bibinfo {pages} {014701} (\bibinfo {year}
  {2014})},\ \Eprint {http://arxiv.org/abs/1312.3714} {arXiv:1312.3714
  [nucl-ex]} \BibitemShut {NoStop}%
\bibitem [{\citenamefont {Taggart}\ \emph {et~al.}(2019)\citenamefont {Taggart}
  \emph {et~al.}}]{Taggart:2019ump}%
  \BibitemOpen
  \bibfield  {author} {\bibinfo {author} {\bibfnamefont {M.~P.}\ \bibnamefont
  {Taggart}} \emph {et~al.},\ }\href {\doibase 10.1016/j.physletb.2019.134894}
  {\bibfield  {journal} {\bibinfo  {journal} {Phys. Lett. B}\ }\textbf
  {\bibinfo {volume} {798}},\ \bibinfo {pages} {134894} (\bibinfo {year}
  {2019})},\ \Eprint {http://arxiv.org/abs/1910.00870} {arXiv:1910.00870
  [nucl-ex]} \BibitemShut {NoStop}%
\bibitem [{\citenamefont {Williams}\ \emph {et~al.}(2020)\citenamefont
  {Williams} \emph {et~al.}}]{Williams:2019wxo}%
  \BibitemOpen
  \bibfield  {author} {\bibinfo {author} {\bibfnamefont {M.}~\bibnamefont
  {Williams}} \emph {et~al.},\ }\href {\doibase 10.1103/PhysRevC.102.035801}
  {\bibfield  {journal} {\bibinfo  {journal} {Phys. Rev. C}\ }\textbf {\bibinfo
  {volume} {102}},\ \bibinfo {pages} {035801} (\bibinfo {year} {2020})},\
  \Eprint {http://arxiv.org/abs/1910.01698} {arXiv:1910.01698 [nucl-ex]}
  \BibitemShut {NoStop}%
\bibitem [{\citenamefont {Phuc}\ \emph {et~al.}(2019)\citenamefont {Phuc},
  \citenamefont {Yoshida},\ and\ \citenamefont {Ogata}}]{Phuc:2019sur}%
  \BibitemOpen
  \bibfield  {author} {\bibinfo {author} {\bibfnamefont {N.~T.~T.}\
  \bibnamefont {Phuc}}, \bibinfo {author} {\bibfnamefont {K.}~\bibnamefont
  {Yoshida}}, \ and\ \bibinfo {author} {\bibfnamefont {K.}~\bibnamefont
  {Ogata}},\ }\href {\doibase 10.1103/PhysRevC.100.064604} {\bibfield
  {journal} {\bibinfo  {journal} {Phys. Rev. C}\ }\textbf {\bibinfo {volume}
  {100}},\ \bibinfo {pages} {064604} (\bibinfo {year} {2019})},\ \Eprint
  {http://arxiv.org/abs/1908.00667} {arXiv:1908.00667 [nucl-th]} \BibitemShut
  {NoStop}%
\bibitem [{\citenamefont {Holl}\ \emph {et~al.}(2019)\citenamefont {Holl} \emph
  {et~al.}}]{R3B:2019gix}%
  \BibitemOpen
  \bibfield  {author} {\bibinfo {author} {\bibfnamefont {M.}~\bibnamefont
  {Holl}} \emph {et~al.} (\bibinfo {collaboration} {R3B}),\ }\href {\doibase
  10.1016/j.physletb.2019.06.069} {\bibfield  {journal} {\bibinfo  {journal}
  {Phys. Lett. B}\ }\textbf {\bibinfo {volume} {795}},\ \bibinfo {pages} {682}
  (\bibinfo {year} {2019})}\BibitemShut {NoStop}%
\bibitem [{\citenamefont {A.I.}\ and\ \citenamefont {Rekalo}(1977)}]{ARbook}%
  \BibitemOpen
  \bibfield  {author} {\bibinfo {author} {\bibnamefont {A.I.}}\ and\ \bibinfo
  {author} {\bibfnamefont {M.}~\bibnamefont {Rekalo}},\ }\href@noop {} {\emph
  {\bibinfo {title} {Hadron Electrodynamics(in russian)}}}\ (\bibinfo
  {publisher} {Naukova Dumka},\ \bibinfo {address} {Kiev},\ \bibinfo {year}
  {1977})\BibitemShut {NoStop}%
\bibitem [{\citenamefont {Mohr}\ and\ \citenamefont
  {Taylor}(2000)}]{Mohr:2000ie}%
  \BibitemOpen
  \bibfield  {author} {\bibinfo {author} {\bibfnamefont {P.~J.}\ \bibnamefont
  {Mohr}}\ and\ \bibinfo {author} {\bibfnamefont {B.~N.}\ \bibnamefont
  {Taylor}},\ }\href {\doibase 10.1103/RevModPhys.72.351} {\bibfield  {journal}
  {\bibinfo  {journal} {Rev. Mod. Phys.}\ }\textbf {\bibinfo {volume} {72}},\
  \bibinfo {pages} {351} (\bibinfo {year} {2000})}\BibitemShut {NoStop}%
\bibitem [{\citenamefont {Ericson}\ and\ \citenamefont
  {Rosa-Clot}(1983)}]{Ericson:1982ei}%
  \BibitemOpen
  \bibfield  {author} {\bibinfo {author} {\bibfnamefont {T.~E.~O.}\
  \bibnamefont {Ericson}}\ and\ \bibinfo {author} {\bibfnamefont
  {M.}~\bibnamefont {Rosa-Clot}},\ }\href {\doibase
  10.1016/0375-9474(83)90516-X} {\bibfield  {journal} {\bibinfo  {journal}
  {Nucl. Phys. A}\ }\textbf {\bibinfo {volume} {405}},\ \bibinfo {pages} {497}
  (\bibinfo {year} {1983})}\BibitemShut {NoStop}%
\bibitem [{\citenamefont {Gourdin}\ and\ \citenamefont
  {C.A.}(1964)}]{Gourdin:1963ub}%
  \BibitemOpen
  \bibfield  {author} {\bibinfo {author} {\bibfnamefont {M.}~\bibnamefont
  {Gourdin}}\ and\ \bibinfo {author} {\bibfnamefont {P.}~\bibnamefont {C.A.}},\
  }\href {\doibase 10.1007/BF02812613} {\bibfield  {journal} {\bibinfo
  {journal} {Nuovo Cim.}\ }\textbf {\bibinfo {volume} {32}},\ \bibinfo {pages}
  {1137} (\bibinfo {year} {1964})}\BibitemShut {NoStop}%
\bibitem [{\citenamefont {Schildknecht}(1964)}]{SCHILDKNECHT1964254}%
  \BibitemOpen
  \bibfield  {author} {\bibinfo {author} {\bibfnamefont {D.}~\bibnamefont
  {Schildknecht}},\ }\href {\doibase
  https://doi.org/10.1016/0031-9163(64)90189-1} {\bibfield  {journal} {\bibinfo
   {journal} {Physics Letters}\ }\textbf {\bibinfo {volume} {10}},\ \bibinfo
  {pages} {254} (\bibinfo {year} {1964})}\BibitemShut {NoStop}%
\bibitem [{\citenamefont {Ferro-Luzzi}\ \emph {et~al.}(1998)\citenamefont
  {Ferro-Luzzi} \emph {et~al.}}]{Ferro-Luzzi:1997sqo}%
  \BibitemOpen
  \bibfield  {author} {\bibinfo {author} {\bibfnamefont {M.}~\bibnamefont
  {Ferro-Luzzi}} \emph {et~al.},\ }\href {\doibase
  10.1016/S0375-9474(98)00023-2} {\bibfield  {journal} {\bibinfo  {journal}
  {Nucl. Phys. A}\ }\textbf {\bibinfo {volume} {631}},\ \bibinfo {pages} {190C}
  (\bibinfo {year} {1998})}\BibitemShut {NoStop}%
\bibitem [{\citenamefont {Dmitriev}\ \emph {et~al.}(1985)\citenamefont
  {Dmitriev}, \citenamefont {Nikolenko}, \citenamefont {Popov}, \citenamefont
  {Rachek}, \citenamefont {Shatunov}, \citenamefont {Toporkov}, \citenamefont
  {Tsentalovich}, \citenamefont {Ukraintsev}, \citenamefont {Voitsekhovsky},\
  and\ \citenamefont {Zelevinsky}}]{Dmitriev:1985us}%
  \BibitemOpen
  \bibfield  {author} {\bibinfo {author} {\bibfnamefont {V.~F.}\ \bibnamefont
  {Dmitriev}}, \bibinfo {author} {\bibfnamefont {D.~M.}\ \bibnamefont
  {Nikolenko}}, \bibinfo {author} {\bibfnamefont {S.~G.}\ \bibnamefont
  {Popov}}, \bibinfo {author} {\bibfnamefont {I.~A.}\ \bibnamefont {Rachek}},
  \bibinfo {author} {\bibfnamefont {Y.~M.}\ \bibnamefont {Shatunov}}, \bibinfo
  {author} {\bibfnamefont {D.~K.}\ \bibnamefont {Toporkov}}, \bibinfo {author}
  {\bibfnamefont {E.~P.}\ \bibnamefont {Tsentalovich}}, \bibinfo {author}
  {\bibfnamefont {Y.~G.}\ \bibnamefont {Ukraintsev}}, \bibinfo {author}
  {\bibfnamefont {B.~B.}\ \bibnamefont {Voitsekhovsky}}, \ and\ \bibinfo
  {author} {\bibfnamefont {V.~G.}\ \bibnamefont {Zelevinsky}},\ }\href
  {\doibase 10.1016/0370-2693(85)91534-5} {\bibfield  {journal} {\bibinfo
  {journal} {Phys. Lett. B}\ }\textbf {\bibinfo {volume} {157}},\ \bibinfo
  {pages} {143} (\bibinfo {year} {1985})}\BibitemShut {NoStop}%
\bibitem [{\citenamefont {Gilman}\ \emph {et~al.}(1990)\citenamefont {Gilman}
  \emph {et~al.}}]{Gilman:1990vg}%
  \BibitemOpen
  \bibfield  {author} {\bibinfo {author} {\bibfnamefont {R.~A.}\ \bibnamefont
  {Gilman}} \emph {et~al.},\ }\href {\doibase 10.1103/PhysRevLett.65.1733}
  {\bibfield  {journal} {\bibinfo  {journal} {Phys. Rev. Lett.}\ }\textbf
  {\bibinfo {volume} {65}},\ \bibinfo {pages} {1733} (\bibinfo {year}
  {1990})}\BibitemShut {NoStop}%
\bibitem [{\citenamefont {Ferro-Luzzi}\ \emph {et~al.}(1996)\citenamefont
  {Ferro-Luzzi} \emph {et~al.}}]{Ferro-Luzzi:1996znh}%
  \BibitemOpen
  \bibfield  {author} {\bibinfo {author} {\bibfnamefont {M.}~\bibnamefont
  {Ferro-Luzzi}} \emph {et~al.},\ }\href {\doibase 10.1103/PhysRevLett.77.2630}
  {\bibfield  {journal} {\bibinfo  {journal} {Phys. Rev. Lett.}\ }\textbf
  {\bibinfo {volume} {77}},\ \bibinfo {pages} {2630} (\bibinfo {year}
  {1996})}\BibitemShut {NoStop}%
\bibitem [{\citenamefont {Bouwhuis}\ \emph {et~al.}(1999)\citenamefont
  {Bouwhuis}, \citenamefont {Alarcon}, \citenamefont {Botto}, \citenamefont
  {van~den Brand}, \citenamefont {Bulten}, \citenamefont {Dolfini},
  \citenamefont {Ent}, \citenamefont {Ferro-Luzzi}, \citenamefont
  {Higinbotham}, \citenamefont {de~Jager}, \citenamefont {Lang}, \citenamefont
  {de~Lange}, \citenamefont {Papadakis}, \citenamefont {Passchier},
  \citenamefont {Poolman}, \citenamefont {Six}, \citenamefont {Steijger},
  \citenamefont {Vodinas}, \citenamefont {de~Vries},\ and\ \citenamefont
  {Zhou}}]{PhysRevLett.82.3755}%
  \BibitemOpen
  \bibfield  {author} {\bibinfo {author} {\bibfnamefont {M.}~\bibnamefont
  {Bouwhuis}}, \bibinfo {author} {\bibfnamefont {R.}~\bibnamefont {Alarcon}},
  \bibinfo {author} {\bibfnamefont {T.}~\bibnamefont {Botto}}, \bibinfo
  {author} {\bibfnamefont {J.~F.~J.}\ \bibnamefont {van~den Brand}}, \bibinfo
  {author} {\bibfnamefont {H.~J.}\ \bibnamefont {Bulten}}, \bibinfo {author}
  {\bibfnamefont {S.}~\bibnamefont {Dolfini}}, \bibinfo {author} {\bibfnamefont
  {R.}~\bibnamefont {Ent}}, \bibinfo {author} {\bibfnamefont {M.}~\bibnamefont
  {Ferro-Luzzi}}, \bibinfo {author} {\bibfnamefont {D.~W.}\ \bibnamefont
  {Higinbotham}}, \bibinfo {author} {\bibfnamefont {C.~W.}\ \bibnamefont
  {de~Jager}}, \bibinfo {author} {\bibfnamefont {J.}~\bibnamefont {Lang}},
  \bibinfo {author} {\bibfnamefont {D.~J.~J.}\ \bibnamefont {de~Lange}},
  \bibinfo {author} {\bibfnamefont {N.}~\bibnamefont {Papadakis}}, \bibinfo
  {author} {\bibfnamefont {I.}~\bibnamefont {Passchier}}, \bibinfo {author}
  {\bibfnamefont {H.~R.}\ \bibnamefont {Poolman}}, \bibinfo {author}
  {\bibfnamefont {E.}~\bibnamefont {Six}}, \bibinfo {author} {\bibfnamefont
  {J.~J.~M.}\ \bibnamefont {Steijger}}, \bibinfo {author} {\bibfnamefont
  {N.}~\bibnamefont {Vodinas}}, \bibinfo {author} {\bibfnamefont
  {H.}~\bibnamefont {de~Vries}}, \ and\ \bibinfo {author} {\bibfnamefont
  {Z.-L.}\ \bibnamefont {Zhou}},\ }\href {\doibase 10.1103/PhysRevLett.82.3755}
  {\bibfield  {journal} {\bibinfo  {journal} {Phys. Rev. Lett.}\ }\textbf
  {\bibinfo {volume} {82}},\ \bibinfo {pages} {3755} (\bibinfo {year}
  {1999})}\BibitemShut {NoStop}%
\bibitem [{\citenamefont {Nikolenko}\ \emph
  {et~al.}(2003{\natexlab{a}})\citenamefont {Nikolenko} \emph
  {et~al.}}]{Nikolenko:2003zq}%
  \BibitemOpen
  \bibfield  {author} {\bibinfo {author} {\bibfnamefont {D.~M.}\ \bibnamefont
  {Nikolenko}} \emph {et~al.},\ }\href {\doibase 10.1103/PhysRevLett.90.072501}
  {\bibfield  {journal} {\bibinfo  {journal} {Phys. Rev. Lett.}\ }\textbf
  {\bibinfo {volume} {90}},\ \bibinfo {pages} {072501} (\bibinfo {year}
  {2003}{\natexlab{a}})}\BibitemShut {NoStop}%
\bibitem [{\citenamefont {Nikolenko}\ \emph
  {et~al.}(2003{\natexlab{b}})\citenamefont {Nikolenko} \emph
  {et~al.}}]{Nikolenko:2003fjj}%
  \BibitemOpen
  \bibfield  {author} {\bibinfo {author} {\bibfnamefont {D.~M.}\ \bibnamefont
  {Nikolenko}} \emph {et~al.},\ }\href {\doibase 10.1016/S0375-9474(03)01084-4}
  {\bibfield  {journal} {\bibinfo  {journal} {Nucl. Phys. A}\ }\textbf
  {\bibinfo {volume} {721}},\ \bibinfo {pages} {C409} (\bibinfo {year}
  {2003}{\natexlab{b}})}\BibitemShut {NoStop}%
\bibitem [{\citenamefont {Schulze}\ \emph {et~al.}(1984)\citenamefont {Schulze}
  \emph {et~al.}}]{Schulze:1984ms}%
  \BibitemOpen
  \bibfield  {author} {\bibinfo {author} {\bibfnamefont {M.~E.}\ \bibnamefont
  {Schulze}} \emph {et~al.},\ }\href {\doibase 10.1103/PhysRevLett.52.597}
  {\bibfield  {journal} {\bibinfo  {journal} {Phys. Rev. Lett.}\ }\textbf
  {\bibinfo {volume} {52}},\ \bibinfo {pages} {597} (\bibinfo {year}
  {1984})}\BibitemShut {NoStop}%
\bibitem [{\citenamefont {Garcon}\ \emph {et~al.}(1994)\citenamefont {Garcon}
  \emph {et~al.}}]{Garcon:1993vm}%
  \BibitemOpen
  \bibfield  {author} {\bibinfo {author} {\bibfnamefont {M.}~\bibnamefont
  {Garcon}} \emph {et~al.},\ }\href {\doibase 10.1103/PhysRevC.49.2516}
  {\bibfield  {journal} {\bibinfo  {journal} {Phys. Rev. C}\ }\textbf {\bibinfo
  {volume} {49}},\ \bibinfo {pages} {2516} (\bibinfo {year}
  {1994})}\BibitemShut {NoStop}%
\bibitem [{\citenamefont {Abbott}\ \emph
  {et~al.}(2000{\natexlab{a}})\citenamefont {Abbott} \emph
  {et~al.}}]{JLABt20:2000uor}%
  \BibitemOpen
  \bibfield  {author} {\bibinfo {author} {\bibfnamefont {D.}~\bibnamefont
  {Abbott}} \emph {et~al.} (\bibinfo {collaboration} {JLAB t(20)}),\ }\href
  {\doibase 10.1103/PhysRevLett.84.5053} {\bibfield  {journal} {\bibinfo
  {journal} {Phys. Rev. Lett.}\ }\textbf {\bibinfo {volume} {84}},\ \bibinfo
  {pages} {5053} (\bibinfo {year} {2000}{\natexlab{a}})},\ \Eprint
  {http://arxiv.org/abs/nucl-ex/0001006} {arXiv:nucl-ex/0001006} \BibitemShut
  {NoStop}%
\bibitem [{\citenamefont {Haftel}\ \emph {et~al.}(1980)\citenamefont {Haftel},
  \citenamefont {Mathelitsch},\ and\ \citenamefont {Zingl}}]{Haftel:1980zz}%
  \BibitemOpen
  \bibfield  {author} {\bibinfo {author} {\bibfnamefont {M.~I.}\ \bibnamefont
  {Haftel}}, \bibinfo {author} {\bibfnamefont {L.}~\bibnamefont {Mathelitsch}},
  \ and\ \bibinfo {author} {\bibfnamefont {H.~F.~K.}\ \bibnamefont {Zingl}},\
  }\href {\doibase 10.1103/PhysRevC.22.1285} {\bibfield  {journal} {\bibinfo
  {journal} {Phys. Rev. C}\ }\textbf {\bibinfo {volume} {22}},\ \bibinfo
  {pages} {1285} (\bibinfo {year} {1980})}\BibitemShut {NoStop}%
\bibitem [{\citenamefont {Alexa}\ \emph {et~al.}(1999)\citenamefont {Alexa}
  \emph {et~al.}}]{JeffersonLabHallA:1998xrv}%
  \BibitemOpen
  \bibfield  {author} {\bibinfo {author} {\bibfnamefont {L.~C.}\ \bibnamefont
  {Alexa}} \emph {et~al.} (\bibinfo {collaboration} {Jefferson Lab Hall A}),\
  }\href {\doibase 10.1103/PhysRevLett.82.1374} {\bibfield  {journal} {\bibinfo
   {journal} {Phys. Rev. Lett.}\ }\textbf {\bibinfo {volume} {82}},\ \bibinfo
  {pages} {1374} (\bibinfo {year} {1999})},\ \Eprint
  {http://arxiv.org/abs/nucl-ex/9812002} {arXiv:nucl-ex/9812002} \BibitemShut
  {NoStop}%
\bibitem [{\citenamefont {Bosted}\ \emph {et~al.}(1990)\citenamefont {Bosted}
  \emph {et~al.}}]{Bosted:1989hy}%
  \BibitemOpen
  \bibfield  {author} {\bibinfo {author} {\bibfnamefont {P.~E.}\ \bibnamefont
  {Bosted}} \emph {et~al.},\ }\href {\doibase 10.1103/PhysRevC.42.38}
  {\bibfield  {journal} {\bibinfo  {journal} {Phys. Rev. C}\ }\textbf {\bibinfo
  {volume} {42}},\ \bibinfo {pages} {38} (\bibinfo {year} {1990})}\BibitemShut
  {NoStop}%
\bibitem [{\citenamefont {Abbott}\ \emph
  {et~al.}(2000{\natexlab{b}})\citenamefont {Abbott} \emph
  {et~al.}}]{JLABt20:2000qyq}%
  \BibitemOpen
  \bibfield  {author} {\bibinfo {author} {\bibfnamefont {D.}~\bibnamefont
  {Abbott}} \emph {et~al.} (\bibinfo {collaboration} {JLAB t20}),\ }\href
  {\doibase 10.1007/PL00013629} {\bibfield  {journal} {\bibinfo  {journal}
  {Eur. Phys. J. A}\ }\textbf {\bibinfo {volume} {7}},\ \bibinfo {pages} {421}
  (\bibinfo {year} {2000}{\natexlab{b}})},\ \Eprint
  {http://arxiv.org/abs/nucl-ex/0002003} {arXiv:nucl-ex/0002003} \BibitemShut
  {NoStop}%
\bibitem [{\citenamefont {Kobushkin}\ and\ \citenamefont
  {Syamtomov}(1995)}]{Kobushkin:1994ed}%
  \BibitemOpen
  \bibfield  {author} {\bibinfo {author} {\bibfnamefont {A.~P.}\ \bibnamefont
  {Kobushkin}}\ and\ \bibinfo {author} {\bibfnamefont {A.~I.}\ \bibnamefont
  {Syamtomov}},\ }\href@noop {} {\bibfield  {journal} {\bibinfo  {journal}
  {Phys. Atom. Nucl.}\ }\textbf {\bibinfo {volume} {58}},\ \bibinfo {pages}
  {1477} (\bibinfo {year} {1995})},\ \Eprint
  {http://arxiv.org/abs/hep-ph/9409411} {arXiv:hep-ph/9409411} \BibitemShut
  {NoStop}%
\bibitem [{\citenamefont {Iachello}\ \emph {et~al.}(1973)\citenamefont
  {Iachello}, \citenamefont {Jackson},\ and\ \citenamefont
  {Lande}}]{Iachello:1972nu}%
  \BibitemOpen
  \bibfield  {author} {\bibinfo {author} {\bibfnamefont {F.}~\bibnamefont
  {Iachello}}, \bibinfo {author} {\bibfnamefont {A.~D.}\ \bibnamefont
  {Jackson}}, \ and\ \bibinfo {author} {\bibfnamefont {A.}~\bibnamefont
  {Lande}},\ }\href {\doibase 10.1016/0370-2693(73)90266-9} {\bibfield
  {journal} {\bibinfo  {journal} {Phys. Lett. B}\ }\textbf {\bibinfo {volume}
  {43}},\ \bibinfo {pages} {191} (\bibinfo {year} {1973})}\BibitemShut
  {NoStop}%
\bibitem [{\citenamefont {Bijker}\ and\ \citenamefont
  {Iachello}(2004)}]{Bijker:2004yu}%
  \BibitemOpen
  \bibfield  {author} {\bibinfo {author} {\bibfnamefont {R.}~\bibnamefont
  {Bijker}}\ and\ \bibinfo {author} {\bibfnamefont {F.}~\bibnamefont
  {Iachello}},\ }\href {\doibase 10.1103/PhysRevC.69.068201} {\bibfield
  {journal} {\bibinfo  {journal} {Phys. Rev. C}\ }\textbf {\bibinfo {volume}
  {69}},\ \bibinfo {pages} {068201} (\bibinfo {year} {2004})},\ \Eprint
  {http://arxiv.org/abs/nucl-th/0405028} {arXiv:nucl-th/0405028} \BibitemShut
  {NoStop}%
\bibitem [{\citenamefont {Tomasi-Gustafsson}\ \emph {et~al.}(2006)\citenamefont
  {Tomasi-Gustafsson}, \citenamefont {Gakh},\ and\ \citenamefont
  {Adamuscin}}]{Tomasi-Gustafsson:2005dyd}%
  \BibitemOpen
  \bibfield  {author} {\bibinfo {author} {\bibfnamefont {E.}~\bibnamefont
  {Tomasi-Gustafsson}}, \bibinfo {author} {\bibfnamefont {G.~I.}\ \bibnamefont
  {Gakh}}, \ and\ \bibinfo {author} {\bibfnamefont {C.}~\bibnamefont
  {Adamuscin}},\ }\href {\doibase 10.1103/PhysRevC.73.045204} {\bibfield
  {journal} {\bibinfo  {journal} {Phys. Rev. C}\ }\textbf {\bibinfo {volume}
  {73}},\ \bibinfo {pages} {045204} (\bibinfo {year} {2006})},\ \Eprint
  {http://arxiv.org/abs/nucl-th/0512039} {arXiv:nucl-th/0512039} \BibitemShut
  {NoStop}%
\bibitem [{\citenamefont {Gakh}\ and\ \citenamefont
  {Merenkov}(2004)}]{Gakh:2004cc}%
  \BibitemOpen
  \bibfield  {author} {\bibinfo {author} {\bibfnamefont {G.~I.}\ \bibnamefont
  {Gakh}}\ and\ \bibinfo {author} {\bibfnamefont {N.~P.}\ \bibnamefont
  {Merenkov}},\ }\href {\doibase 10.1134/1.1767552} {\bibfield  {journal}
  {\bibinfo  {journal} {J. Exp. Theor. Phys.}\ }\textbf {\bibinfo {volume}
  {98}},\ \bibinfo {pages} {853} (\bibinfo {year} {2004})}\BibitemShut
  {NoStop}%
\bibitem [{\citenamefont {Gakh}\ \emph {et~al.}(2012)\citenamefont {Gakh},
  \citenamefont {Konchatnij},\ and\ \citenamefont {Merenkov}}]{Gakh:2012hf}%
  \BibitemOpen
  \bibfield  {author} {\bibinfo {author} {\bibfnamefont {G.~I.}\ \bibnamefont
  {Gakh}}, \bibinfo {author} {\bibfnamefont {M.~I.}\ \bibnamefont
  {Konchatnij}}, \ and\ \bibinfo {author} {\bibfnamefont {N.~P.}\ \bibnamefont
  {Merenkov}},\ }\href {\doibase 10.1134/S1063776112070060} {\bibfield
  {journal} {\bibinfo  {journal} {J. Exp. Theor. Phys.}\ }\textbf {\bibinfo
  {volume} {115}},\ \bibinfo {pages} {212} (\bibinfo {year} {2012})},\ \Eprint
  {http://arxiv.org/abs/1202.2225} {arXiv:1202.2225 [hep-ph]} \BibitemShut
  {NoStop}%
\end{thebibliography}
%

\end{document}